\documentclass[%
 reprint,
 amsmath,amssymb,
 superscriptaddress,
 aps,
 pre,
]{revtex4-2}

\usepackage[utf8]{inputenc}
\usepackage{graphicx}
\usepackage{dcolumn}
\usepackage{bm}
\usepackage{amssymb}
\usepackage{amsmath}
\usepackage{tabularx}
\usepackage{hyperref} 

\usepackage{blindtext}

\begin{document}
\title{Quantum Chaos in Triangular Billiards}

\author{\v Crt Lozej}
\affiliation{Department of Physics, Faculty of Mathematics and Physics, University of Ljubljana, Jadranska 19, SI-1000 Ljubljana, Slovenia}
\author{Giulio Casati}
\affiliation{Center for Nonlinear and Complex Systems, Dipartimento di Scienza e Alta Tecnologia, Universita` degli Studi dell’Insubria, via Valleggio 11, 22100 Como, Italy}
\author{Toma\v z Prosen}
\affiliation{Department of Physics, Faculty of Mathematics and Physics, University of Ljubljana, Jadranska 19, SI-1000 Ljubljana, Slovenia}

\begin{abstract}
We present an extensive numerical study of spectral statistics and eigenfunctions of quantized triangular billiards. We compute two million consecutive eigenvalues for six representative cases of triangular billiards, three with generic angles with irrational ratios with $\pi$, whose classical dynamics is presumably mixing, and three with exactly one angle rational with $\pi$, which are presumably only weakly mixing or even non-ergodic in case of right-triangles. We find excellent agreement of short and long range spectral statistics with the Gaussian orthogonal ensemble of random matrix theory for the most irrational generic triangle, while the other cases show small but significant deviations which are attributed either to scarring or super-scarring mechanism. This result, which extends the quantum chaos conjecture to systems with dynamical mixing in the absence of hard (Lyapunov) chaos, has been corroborated by analysing distributions of phase-space localisation measures of eigenstates and inspecting the structure of characteristic typical and atypical eigenfunctions.

\end{abstract}

\maketitle

\section{Introduction}
\label{Intro}

\begin{figure*}
  \centering
  \includegraphics[]{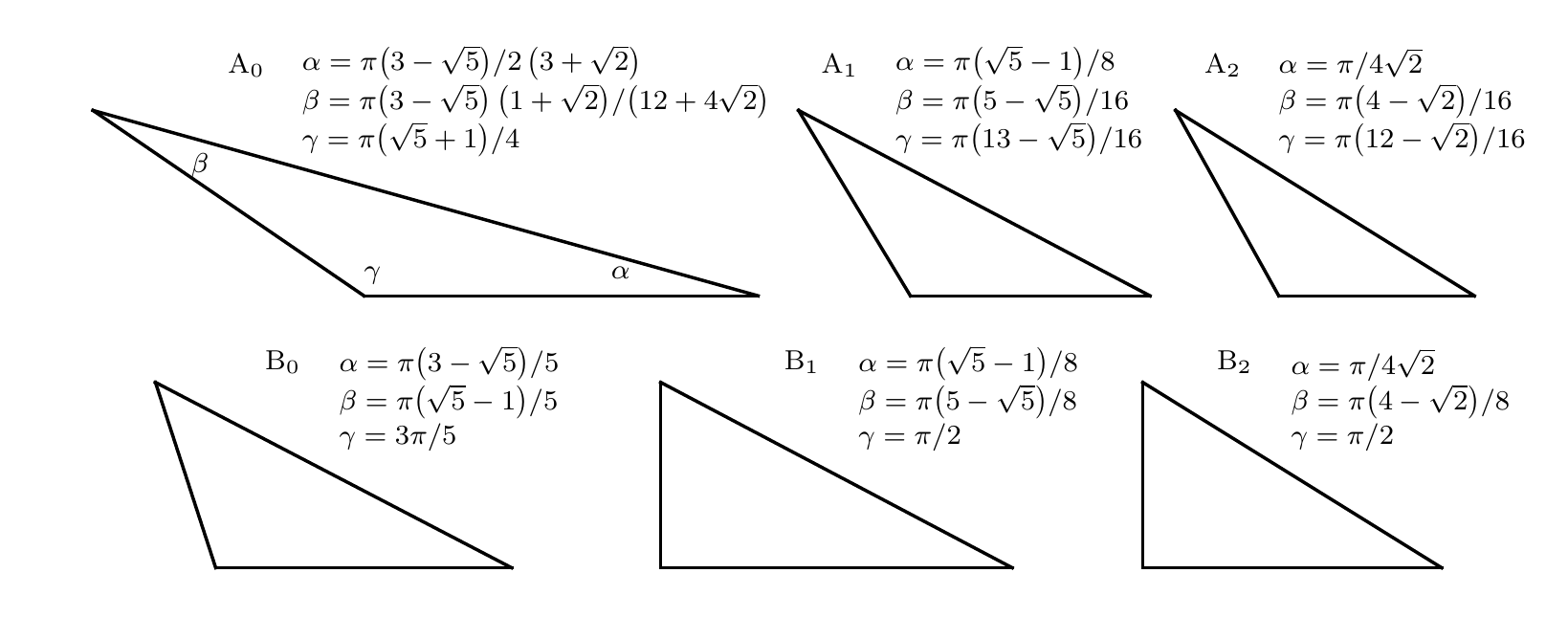}
  \caption{The geometry of the triangle billiards considered in the numerical study. The generic triangles of class (A) are shown in the top and those of class (B) with one angle rational with $\pi$ on the bottom row. The exact values of the angles are listed next to each triangle. All triangles have the same height $h=0.75$, the position of the angles is defined on the triangle $\mathrm{A_0}$. } 
  \label{triangles}
\end{figure*}

Classical ergodic theory provides a fairly satisfactory classification of statistical properties of classical dynamical systems. Even though it is very difficult to determine to which category (K-property, mixing, weak mixing, ergodic etc.) a given dynamical system belongs, the existence of a rigorous classification of different properties proved to be very useful, for instance, for connecting to macroscopic physical behavior such as transport. 

The situation is much less clear for quantum dynamical systems with a few (finite number of) degrees of freedom. Due to the quasi-periodic nature of time evolution in finite quantum systems, such systems can not have positive Kolmogorov-Sinai entropy neither can they be dynamically mixing in the strict sense. One instead looks for definite signatures of classical ergodic behavior on the statistical properties of quantum spectra, eigenfunctions, etc.
The Quantum Chaos conjecture \cite{BGS,CVG} states that quantum systems with chaotic classical dynamics should have spectral statistics (or quantum statistical properties in general) described by an appropriate ensemble of random matrix theory (RMT), which is determined solely by unitary and anti-unitary (say time-reversal) symmetries of the system. For example, time-reversal invariant systems without (half-integer) spin correspond to Gaussian orthogonal ensemble (GOE) of random real symmetric matrices.
Although the conjecture is not proven in a mathematical sense, a heuristic proof that was initiated by Sieber and Richter \cite{Sieber2001} and later developed by the group of Haake~\cite{Mueller2004a,Mueller2004b,Mueller2005} clearly relates random matrix spectral correlations to correlations among classical unstable (hyperbolic) orbits.
Yet, this mechanism requires (uniform) hyperbolicity of classical dynamics, i.e. essentially all periodic orbits need to be exponentially unstable. It has remained unclear what happens in systems with weaker ergodic properties. In the other extreme case of completely integrable classical dynamics, Berry and Tabor conjectured Poisson statistics of energy levels \cite{BerryTabor}, which, however, despite corroborated by a vast amount of data has not been proven as well.

A possibility to approach this problem is to consider classical systems
with very definite statistical properties. Billiards are very convenient
models in this respect. Completely integrable and completely chaotic
quantum billiards have been studied in detail. On the other hand,
triangular billiards (or polygonal billiards in general) are very interesting for this purpose, since they may possess a minimal ergodic property for (quantum) statistical behaviour:
they are not chaotic (linear separation of nearby trajectories, zero KS-entropy), and yet they may be ergodic and mixing, hence they occupy a special place in the {\em ergodic hierarchy} \cite{EH}.
 
Yet, investigating the classical properties of billiards in polygons is notoriously difficult (see \cite{gutkin1986, gutkin1996} and references therein). Even basic properties like the existence of periodic orbits are hard to prove (see e.g. Ref.~\cite{schwartz2009} for triangles). The ergodic properties of the triangular billiards depend sensitively on the number theoretic aspects of the angles. In this paper, we consider representatives of two classes of triangular billiards with, according to current understanding, the strongest ergodic properties:

(A) \emph{Generic triangles} with all angles irrational with respect to $\pi$. It has been demonstrated numerically \cite{CasPro1999tri} that such triangular billiards are ergodic and mixing (with clear power-law
decay of correlations $1/t^{\sigma}$, and typically $\sigma \approx 1$) and
have much stronger properties than class (B) below.

(B)  Only \emph{one} angle is \emph{rational} with respect to $\pi$. 
Such are, for example, \emph{generic right triangles}, which have been extensively numerically studied in \cite{ArtCasGua1997} {where evidence of weak-mixing (but not mixing) has been found. Later \cite{wang2014nonergodicity},
right triangular billiards have been revisited suggesting that generic right triangles are not even ergodic as the invariant measure may be localized in the direction space and even more recently \cite{huang} extremely 'slow' (logarithmic) difussion has been demonstrated. Yet, as the right triangular billiard is always arbitrary close to an ergodic billiard, we will refer to them as {\em pseudo-ergodic} in this paper}~\footnote{We define the system as pseudo-ergodic, if a fixed typical trajectory (for almost any initial condition) comes arbitrary close to any point in phase space, but the corresponding phase-space measure is not flat (as observed in ~\cite{wang2014nonergodicity}).}.  

In our numerical study, we consider six triangles (three in each class) which we label $\mathrm{C}_i$, where $\mathrm{C}\ = \mathrm{A}, \mathrm{B}$ denotes the class and $i = 0, 1, 2$ is the index of the triangle. The representatives were chosen with the following considerations. In class (A) the triangle $\mathrm{A_0}$ is chosen to have highly irrational angles, implying the strongest mixing properties. The obtuse angle has a one half golden ratio with $\gamma = \pi{\left(\sqrt{5}+1\right)}/{4}$ and the ratio between  the remaining angles is $(1+\sqrt{2})/2$, that is one half of the silver ratio. The triangles  $\mathrm{A_1}$ and  $\mathrm{A_2}$ are chosen to have increasingly less noble ratios between the angles and $\pi$ and presumably weaker mixing properties. In class (B) the triangle $\mathrm{B_0}$ has an obtuse angle of $\gamma = {3 \pi}/{5}$ with a golden ratio between the other angles. Triangles  $\mathrm{B_1}$ and  $\mathrm{B_2}$ are representatives of generic right triangles $\gamma = {\pi}/{2}$ and share the angle $\alpha$ with the generic triangle with the same index. Incidentally, the top angle $\beta$ of the right triangle is twice that of the generic one with the same index. 

The third class of triangles with all angles rational with $\pi$ are not considered in this study. Triangle billiards of this class belong to the class of so-called \textit{pseudointegrable} systems and have been extensively studied in other works (see \cite{bogomolny2004, bogomolny2021} and references therein). Their classical trajectories belong to 2-dimensional surfaces of finite genus defined by the angles \cite{richens1981} and hence cannot be ergodic on 3-dimensional energy surfaces. Their quantum spectral properties belong to neither the chaotic nor integrable universality classes, but have potentially less universal intermediate spectral statistics. The eigenstates of pseudointegrable triangles are also known to form super-scars produced by the series of diffractions off singular points in the triangle corners. The singular scatterers effectively produce channels of propagating plane waves inside parts of the billiard containing short classical periodic orbits. The effect becomes more pronounced in the semiclassical limit as opposed to the usual scarred states in chaotic systems where scarring is diminished. The super-scarring effect is well known also in barrier billiards \cite{bogomolny2021}, and is conjectured to exist in a more general class of polygonal billiards.

In the quantum billiard problem, we consider a quantum particle trapped inside a (in our case triangular) region $\mathcal{B} \subset \mathbb{R}^2$ referred to as the billiard table. The eigenfuncitons $\psi_n$ are given by the solutions of the Helmholtz equation 
\begin{equation}
    \left(\nabla^{2}+k_n^{2}\right)\psi_n=0, 
\end{equation}
and Dirichlet b.c. $\psi_n|_{\partial \mathcal{B}}=0$, with eigenenergies $E_n=k_n^2$, where $k_n$ is the wavenumber of the $n$-th eigenstate. Here and in the following, we use a system of units where $\hbar =1$, and the mass of the particle is $m = 1/2$.  Using the very efficient scaling method of Vergini and Saraceno \cite{VerSer1995, BarnettPHD} with a corner adapted Fourier-Bessel basis \cite{BarBet2007} (the implementation is available as part of \cite{QuantumBilliards}) we computed more than $2\cdot10^6$ levels for each triangle. The most recent study of spectral statistics in generic triangular quantum billiards \cite{lima2013} considered spectra of up to $1.5 \cdot 10^5$ levels in a family of class (A) billiards, where in most cases intermediate level statistics were observed. 

Using several short-range and long-range measures of spectral statistics, such as level spacings, spacing ratios, number variance, spectral form factor, and mode fluctuations, we confirm that the class A triangle billiards conform to GOE universality. The deviations are statistically insignificant for billiard $\mathrm{A}_0$, while for billiards $\mathrm{A}_{1,2}$ we find excellent agreement with GOE for short range statistics and a very small spectral compressibility on long energy ranges.{ In class B the non-right-angled triangle $\mathrm{B}_{0}$ shows similar statistics to class A,} but for the right-triangle billiards we find some significant deviations from GOE at finite energies which seem to persist when increasing the energy (level number), in particular we find a finite spectral compressibility. Our results on spectral statistics are corroborated with a study of statistics of localization measures of eignenstates, and illustrated with the galleries of eigenfunctions in configuration-space, Poincar\' e-Husimi and nodal-domain representations.

The paper is organized as follows. The analysis of the spectral statistics is presented in Sec. \ref{sec:Spectral statistics}. The subsections are devoted to the spectral staircase function and the mode fluctuations \ref{sub:Spectral staircase}, the level spacing distributions and ratios \ref{sub:Level spacings}, the number variance \ref{sub:Number variance}.  and  to the spectral form factor \ref{sub:sff}. The eigenstates are analyzed in Sec. \ref{sec:Eigenstates}. In subsection \ref{sub:PH representation} we introduce the Poincaré-Husimi representation that maps the eigenstates onto the classical phase space. In \ref{sub: Localization measures} we analyze the distributions of localization measures of ensembles of eigenstates, and in \ref{sub: Gallery of states} we show and analyze particular characteristic examples of eigenfunctions. The results are concluded and discussed in Sec. \ref{sec:Conclusion}.

\section{Spectral statistics}\label{sec:Spectral statistics}
The main question we wish to address in this section is, whether the classical property of mixing rather than Lyapunov chaos is enough for RMT statistics of the spectra. We will examine the spectral statistics of the triangle billiards in view of the most commonly used spectral statistics. These include the spectral staircase function and the mode-fluctuations, the level spacing distributions and ratios, the number variance and finally, the spectral form factor. The spectral samples contain in excess of $2\cdot10^6$ levels for each triangle, which is, to the best of our knowledge, by a good margin the largest number of modes considered in the literature so far.

\subsection{Spectral staircase function and mode-fluctuations}\label{sub:Spectral staircase} 
\begin{figure*}
  \centering
  \includegraphics[]{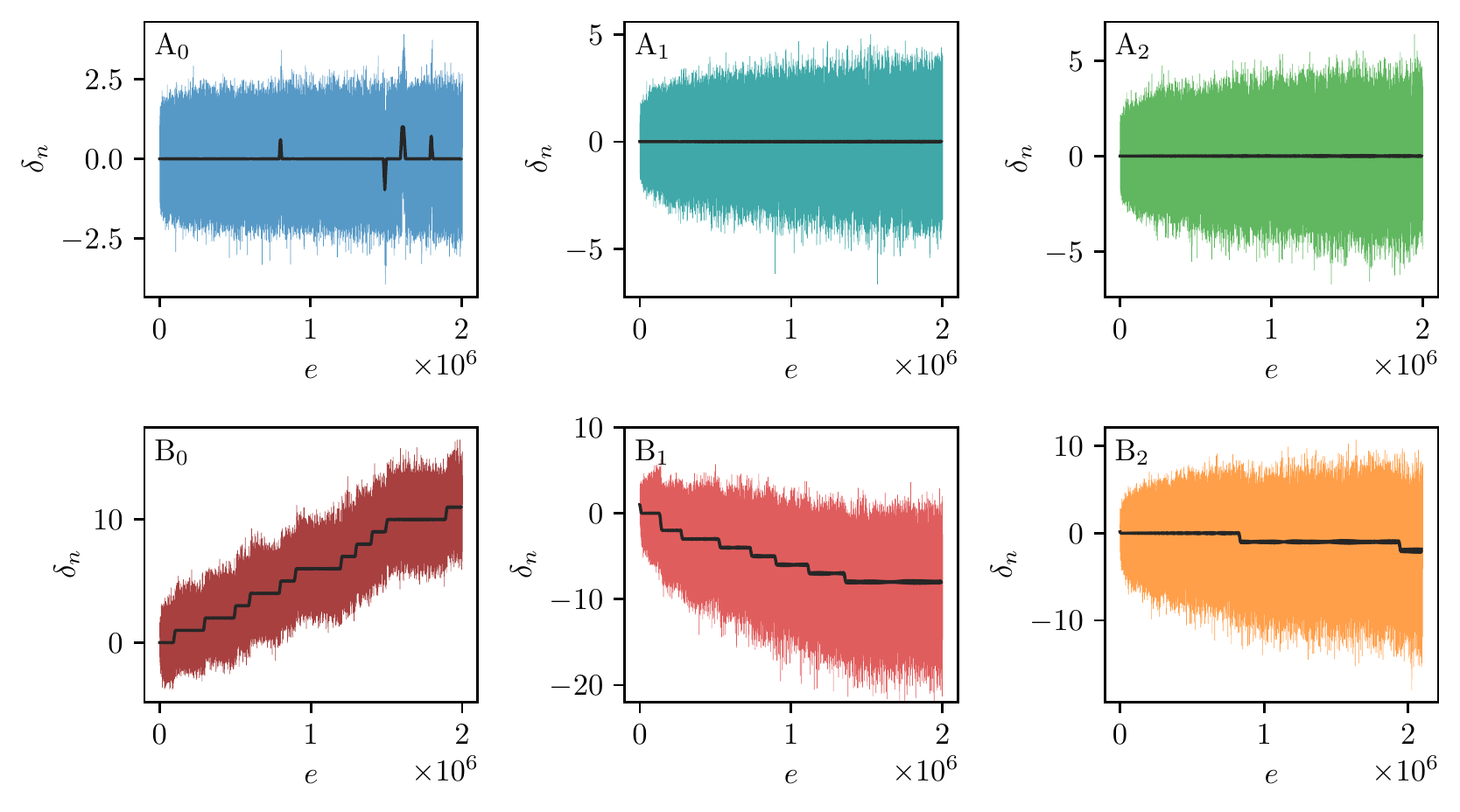}
  \caption{The fluctuating part of the spectral staircase as a function of the unfolded energy in each triangle. The colored lines represent the fluctuations at each eigenstate wavenumber. The black lines are moving averages over $10^4$ consecutive eigenstates, revealing the approximate location of the spurious or missing levels.} 
  \label{fig:stateCheck}
\end{figure*}

The spectral staircase function counts the number of eigenstates (or modes) up to some energy 
\begin{equation}
    N(E) := \#\{n|E_n<E\}.
\end{equation}
The asymptotic mean of the spectral staircase is given by the well known generalized Weyl formula \cite{Hilf}
\begin{equation}
    N_{\mathrm{Weyl}}(E) = (\mathcal{A}E-\mathcal{L}\sqrt{E})/4\pi + \sum_i{\frac{\pi^2-{\varphi_i}^2}{24 \pi {\varphi_i} }},
\end{equation}
where $\mathcal{A}$ is the area, $\mathcal{L}$ the perimeter of the billiard and the last term is a constant corner correction term summing over the triangle corners $\varphi_i \in \{\alpha, \beta, \gamma\}$. 
This formula is already specialized to polygonal billiards, dropping the vanishing curvature term.
Despite being an asymptotic formula, the result evidently holds from the ground state, as has been observed in many numerical studies. The spectral staircase typically fluctuates around the mean value $N(E)= N_{\mathrm{Weyl}}(E) + N_{\mathrm{fluct}}(E)$. In order to compare the universal aspects of spectra of different triangles, we unfold the spectra. This is most commonly done by evaluating the Weyl formula at each point of the computed discrete spectrum, giving the unfolded energy levels 
\begin{equation}
    e_n := N_{\mathrm{Weyl}}(E_n),
\end{equation}
with unit mean level spacings. The Weyl formula also provides a means of checking whether the computed spectrum is complete. The scaling method computes the states in some small, finite spectral interval. The final spectral sample is a composite of many small overlapping spectral samples, where we try to identify which of the levels in the overlap interval belong to the same eigenstates. Because of the finite precision and numerical errors in the computation of the individual levels this is not always possible, and some levels are missed, while some may be counted twice. By considering the fluctuating part of the spectral staircase at the points of the eigenenergies i.e.
\begin{equation}
    \delta_n= N_{\mathrm{fluct}}(E_n) = n - \frac{1}{2} - e_n,  
\end{equation}
known as the \textit{mode-fluctuation}, we obtain a sensitive probe for determining the completeness of the spectrum. This discrete function should fluctuate around 0 if the spectrum is complete. Any steps in the constant value indicate missing or spurious levels. However, the amplitude of the fluctuations increases as we progress higher in the spectrum, and the exact point of the missing level is nearly impossible to determine. In Fig. \ref{fig:stateCheck} we show $\delta_n$ as a function of the unflolded energy $e$ for each triangle. The amplitude of the fluctuations is quickly larger than 1, obscuring errors in the spectral computation. The errors are more easily locatable if we compute a moving average over $10^4$ consecutive eigenstates. We thus observe that spurious levels are compensated by missing levels on four occasions (eight errors) in the spectrum of $\mathrm{A_0}$, while the spectra of $\mathrm{A_1}$ and $\mathrm{A_2}$ appear complete and without errors. The spectrum of $\mathrm{B_0}$ contains about eleven spurious levels, $\mathrm{B_1}$ nine missing levels and $\mathrm{B_2}$ two to three missing levels. Because the number of errors is tiny compared to the overall number of levels, they should have no effect on the statistical properties of the spectra.

\begin{figure*}
  \centering
  \includegraphics[]{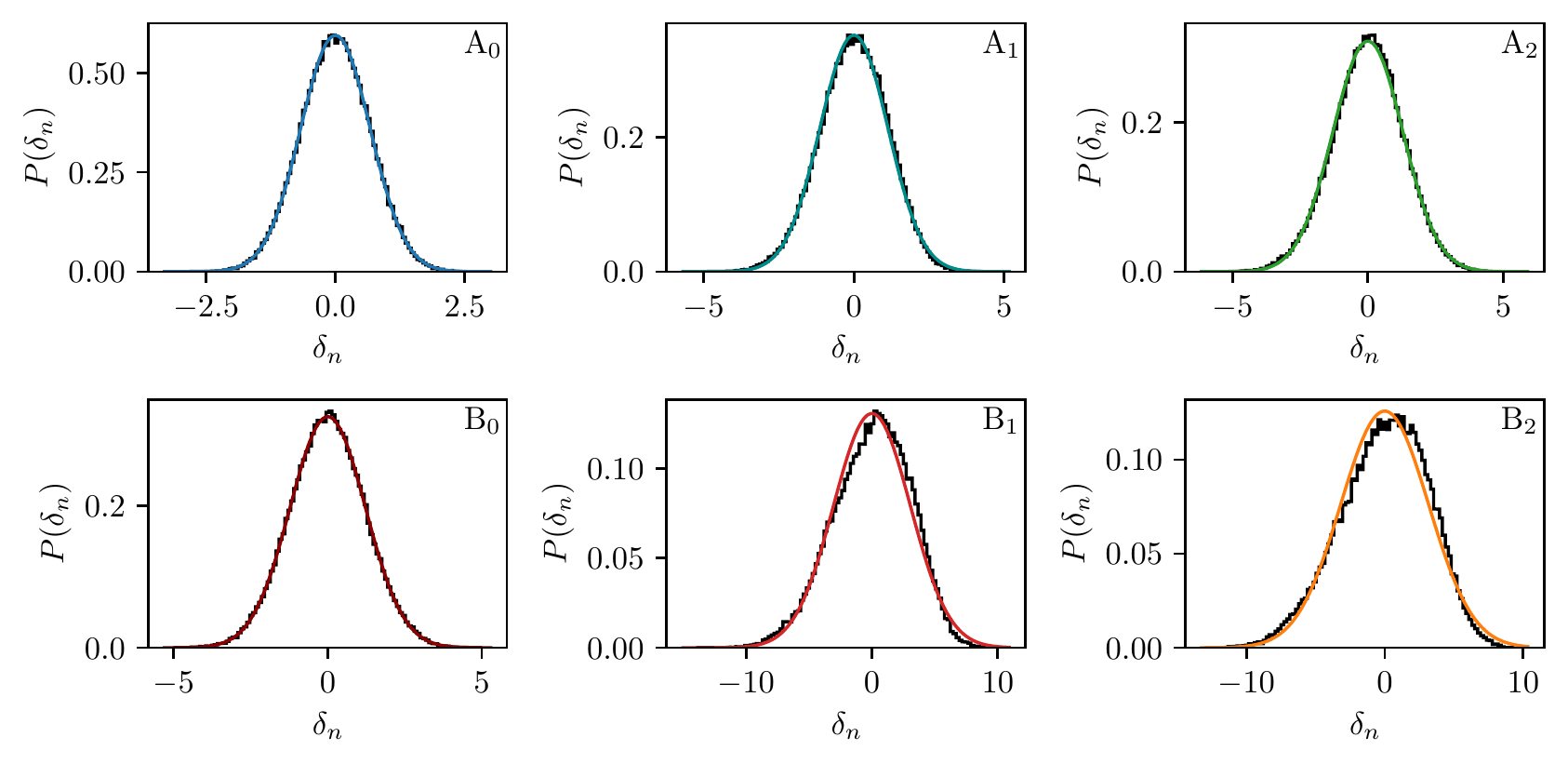}
  \caption{The histograms of the mode-fluctuation distributions for $10^5$ consecutive eigenstates starting from $e = 10^6$. The best fitting Gaussian distributions are shown with colored lines and the values of the central moments are shown in Tab. \ref{tab:CentralMoments}. }
  \label{fig:stateHist}
\end{figure*}

\begin{figure*}
  \centering
  \includegraphics[]{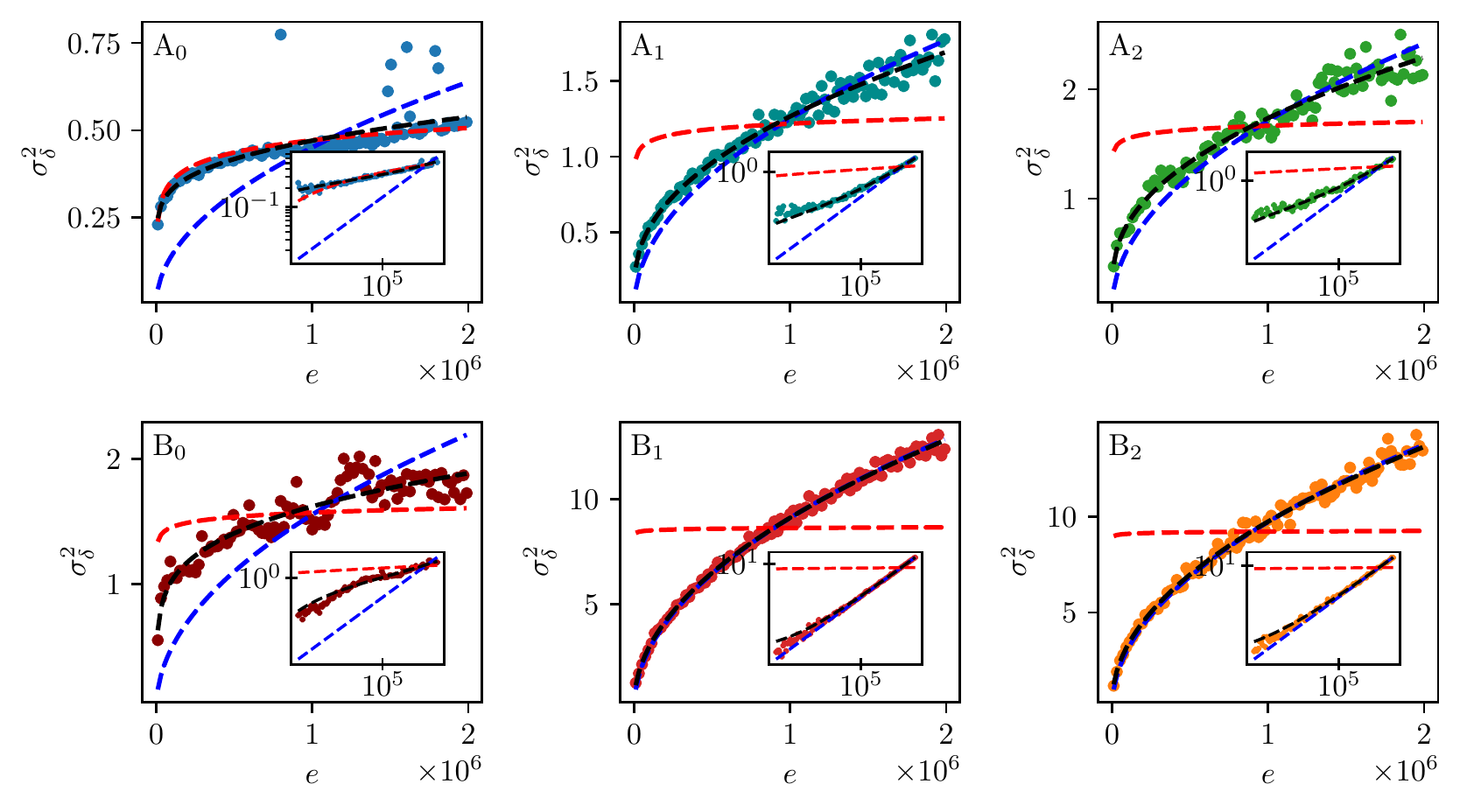}
  \caption{The variance of the mode-fluctuation distributions as a function of the unfolded energy. The insets show the data in the log-log scale. The dashed lines show the best fitting theoretical curves \eqref{eq:VarCha} in red, \eqref{eq:VarInt} in blue, and a linear superposition of both \eqref{eq:VarSuper} in black, the fitting parameters are given in Tab. \ref{tab:FitParams}.} 
  \label{fig:stateVar}
\end{figure*}

The distributions of the mode-fluctuations were conjectured \cite{aurich1994,aurich1997} to be distinct in chaotic and integrable systems. The limiting distribution in the semiclassical limit $e \rightarrow \infty$ is expected to be universally Gaussian in chaotic systems. In integrable systems, the distribution is expected to be system specific. The variance of the mode-fluctuations in some finite spectral interval is related to the saturation level (see Ref. \cite{aurich1997} for a detailed analysis) of some long range spectral statistics, specifically the \textit{spectral rigidity} $\Delta_3(L)$ and \textit{number variance } $\Sigma^2(L)$, with $\sigma^2_{\delta} = \Delta_3(\infty) \simeq\Sigma^2(\infty)/2 $. The number variance will be thoroughly investigated in subsection \ref{sub:Number variance}. Based on Berry's semiclassical analysis \cite{berry1985} (see also \cite{backer2002autocorrelation}), one would expect 
\begin{equation}\label{eq:VarCha}
    \sigma^2_{\delta}(e) = \frac{1}{2\pi^2} \mathrm{ln}e + a, 
\end{equation}    
in chaotic systems with time reversal symmetry and
\begin{equation}\label{eq:VarInt}
    \sigma^2_{\delta}(e) = b \sqrt{e}, 
\end{equation}   
for integrable systems, where $a$ and $b$ are system dependent constants. Before further analyzing the mode-fluctuations, the steps in the data because of the errors in the spectral computation have been carefully removed by subtracting the appropriate integer part of the local mean value. In Fig. \ref{fig:stateHist} the high-lying mode-fluctuation distributions are shown for each of the triangles. The distributions agree with the Gaussian in all cases except in the two right triangles  $\mathrm{B_1}$ and $\mathrm{B_2}$, where the distributions are skewed slightly. This is caused either by the (possibly super-scarred) bouncing ball modes that are most prominent in the right triangles, or by the non-ergodicity of the latter.
This type of effect has been observed experimentally in semiconductor microwave billiards in Ref. \cite{alt1998mode}. The scarring mechanisms and the bouncing ball modes will be further discussed in Sec. \ref{sub: Gallery of states}. It is worth noting that the distribution is Gaussian even in the $\mathrm{B_0}$ triangle. In Fig. \ref{fig:stateVar} we plot the variance of the mode-fluctuations as a function of the unfolded energy. The energy spectra of $2\cdot 10^6$ unfolded levels were divided into 100 equally spaced and logarithmically spaced intervals, and the mode-fluctuation variances computed. Even though there is significant scattering of the data, some conclusions may be made. Relying on the semiclassical analysis, one would expect the triangles with stronger mixing to be closer to the prediction for chaotic systems Eq. \eqref{eq:VarCha}. The energy dependence $\sigma^2_{\delta}(e)$ in the most irrational triangle $\mathrm{A_0}$ does indeed coincide quite closely with the predicted scaling. However, the scaling in the right triangles coincides with the integrable case Eq. \eqref{eq:VarInt}. In general, the data is best described by a linear superposition of both scaling laws 
\begin{equation}\label{eq:VarSuper}
    \sigma^2_{\delta}(e) =  \frac{c}{2\pi^2} \mathrm{ln}e + a + b \sqrt{e}. 
\end{equation} 
This is most likely caused by bouncing ball modes along marginally stable classical periodic trajectories~\footnote{Thereby generalizing the term `bouncing ball' which is usually used only in reference to orbits bouncing between parallel walls}, which generally behave like regular modes.
These modes (eigenstates) can be as well referred to as scars, even though the latter term is typically reserved to effects of (weakly) unstable classical periodic orbits, thus we prefer to use the term bouncing ball modes.
Since the scaling of variance for the bouncing ball modes is much stronger compared to the generic modes, this contribution is significant even if the relative fraction of the bouncing ball modes is small, as for example in the generic triangles $\mathrm{A_1}$ and $\mathrm{A_2}$. In particular, the data is far from both the chaotic and integrable prediction in the triangle $\mathrm{B_0}$, but is still well described by the superposition. The causes may be twofold. Firstly, the system has weaker mixing properties and secondly, there exist heavily scarred bouncing ball-modes. It is difficult to say which of the causes is more significant. 
The mode-fluctuation distributions provide the first piece of evidence that mixing is enough for a correspondence with RMT. However, weaker mixing and bouncing ball modes result in deviations from the RMT predictions.

\begin{table}[]

\centering\caption{Central moments of the distributions shown in Fig. \ref{fig:stateHist}, computed directly from the mode fluctuations of the $10^5$ sampled consecutive states. }{\label{tab:CentralMoments}}
\begin{tabular}{>{\centering}p{1.5cm}>{\centering}p{1.5cm}>{\centering}p{1.5cm}>{\centering}p{1.5cm}>{\centering}p{1.5cm}}
\multicolumn{5}{p{8.6cm}}{}\tabularnewline
\hline 
\hline 
label & $\mu$ & $\sigma$ & $\mathrm{skew}$ & $\mathrm{kurt}$\tabularnewline
\hline 
\hline 
$\mathrm{A_0}$ & 0.00 & 0.67 &  -0.014 &  -0.041 \tabularnewline
$\mathrm{A_1}$ & 0.00 & 1.13 & -0.129 &  -0.048 \tabularnewline
$\mathrm{A_2}$ & 0.00 &  1.29 & -0.093 & 0.061 \tabularnewline
$\mathrm{B_0}$ & 0.00 & 1.22 & -0.005 & -0.009 \tabularnewline
$\mathrm{B_1}$ & 0.00 & 3.05 & -0.285 & -0.137 \tabularnewline
$\mathrm{B_2}$ & -0.01 &  3.17 & -0.295 & -0.191 \tabularnewline

\hline 
\end{tabular}

\end{table}

\begin{table}[]

\centering\caption{Fitting parameters of Eq. \eqref{eq:VarSuper} fitted to the variances of the mode-fluctuations shown in Fig. \ref{fig:stateVar}.}{\label{tab:FitParams}}
\begin{tabular}{>{\centering}p{1.2cm}>{\centering}p{2.2cm}>{\centering}p{2.5cm}>{\centering}p{2cm}}
\multicolumn{4}{>{\centering}p{8.6cm}}{}\tabularnewline
\hline 
\hline 
label & $a$ & $b$ & $c$\tabularnewline
\hline 
\hline 
$\mathrm{A_0}$ &-0.071 $\pm$ 0.088  &(8.4 $\pm$ 2.9)E-5  &0.64 $\pm$ 0.17 \tabularnewline
$\mathrm{A_1}$ &-0.057 $\pm$ 0.072  &(9.8 $\pm$ 0.3)E-4  &0.49 $\pm$ 0.14 \tabularnewline
$\mathrm{A_2}$ &-0.18 $\pm$ 0.14  &(1.2 $\pm$ 0.057)E-3  &0.95 $\pm$ 0.27 \tabularnewline
$\mathrm{B_0}$ &-0.78 $\pm$ 0.21  &(3.7 $\pm$ 0.75)E-4  &2.9 $\pm$ 0.4 \tabularnewline
$\mathrm{B_1}$ &1.1 $\pm$ 0.22  &(9.3 $\pm$ 0.11)E-3  &-1.8 $\pm$ 0.46 \tabularnewline
$\mathrm{B_2}$ &-0.26 $\pm$ 0.32  &(9.2 $\pm$ 0.16)E-3  &1.1 $\pm$ 0.66 \tabularnewline

\hline 
\end{tabular}

\end{table}

\subsection{Level spacings}\label{sub:Level spacings} 
\begin{figure}
  \centering
  \includegraphics[width=8.6cm]{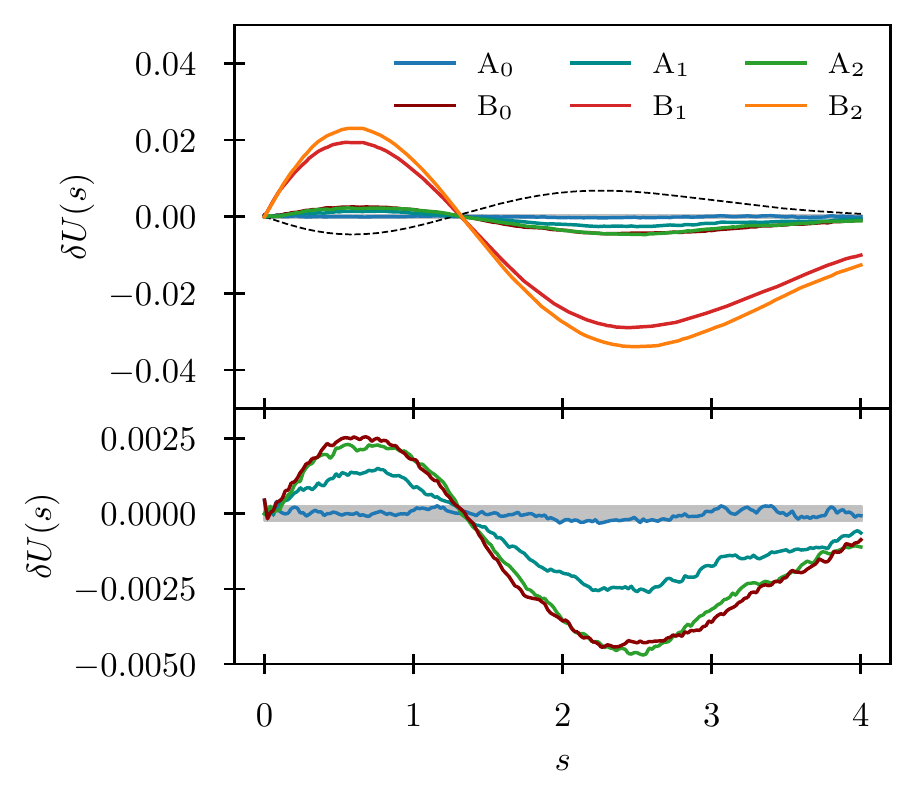}
  \caption{Deviations of the level spacings for the triangle billiards compared to GOE $\delta U = U_{\mathrm{data}}-U_{\mathrm{GOE}}$. The extent of expected statistical fluctuations is shown by the gray band. The top panel shows only the data for all triangles. The WD distribution is plotted with a dashed line. In the lower panel, we show a magnification of the results, omitting the right triangles for clarity. }
  \label{fig:LevelSpacing}
\end{figure}

Nearest neighbor level spacing distributions are the most widely used indicator of quantum chaos. The level spacing is defined as the difference in energy between two consecutive levels in the unfolded spectrum $s_i=e_{i+1}-e_i$. The unfolding procedure guarantees that the mean level spacing is unity. The main object of interest is the \textit{level spacing distribution} i.e. the probability density $P(s)$ or its cumulative density $W(s)=\int_{0}^{s} P(s) \,ds$. Furthermore, it is useful to perform the following nonlinear transformation of the cumulative level spacing distribution,
    \begin{equation}
    U(s) := \frac{2}{\pi}\arccos\sqrt{1-W(s)}, \label{eq:TILS}
    \end{equation}
since the expected statistical fluctuations $\sigma_U=1/\pi \sqrt{N}$ are independent of $s$ and depend only on the sample size $N$ (see \cite{batistic2013dynamical} for a short derivation). In chaotic systems with time reversal symmetries, the level spacing statistics are expected to coincide with the GOE random matrix ensemble, well approximated by the Wigner-Dyson (WD) distribution
\begin{equation}
    P_{WD}(s)=\frac{\pi}{2} s \exp{(-\frac{\pi}{4} s^2)}, \label{eq:WD}
\end{equation}
based on the two-dimensional random matrix approximation. However, far more accurate formulae for the level spacing distributions based on asymptotic expansions around $s \rightarrow 0$ and $s \rightarrow \infty$ exist and may be compounded and numerically evaluated to obtain a very accurate approximation of the infinite dimensional GOE result (see Ref. \cite{dietz1990taylor} for details). We will refer to these numerically evaluated distributions with the subscript GOE, for instance $U_\mathrm{GOE}(s)$. Since the level spacing distributions in the selected triangle billiards are all very close to the GOE result, we will mainly consider the deviation $\delta U(s) = U(s) - U_\mathrm{GOE}(s)$, which we show in Fig. \ref{fig:LevelSpacing}. The deviations are largest in the two right triangles $\mathrm{B_1}$ and $\mathrm{B_2}$, and decrease systematically when we consider more irrational triangles. We may note that the deviations go in the opposite direction to those of the WD distribution. Curiously, the deviations in $\mathrm{A_2}$ and $\mathrm{B_0}$ largely overlap even though the triangles belong to different classes. The deviations in $\mathrm{A_0}$ fall almost entirely within the expected statistical fluctuations for the sample size, thereby presenting strong evidence that strongly mixing systems conform to RMT statistics. The deviations in the other triangles are not statistical. They are caused by the bouncing ball scarred states. The level spacing distributions are mostly stable with respect to increasing the energy range, even if we compare only the $10^5$ highest or lowest lying states. This fact indicates that our results have essentially converged to the semiclassical limit. In spite of the expectation of quantum ergodicity, it is known that bouncing ball scarred states may contribute to the level statistics significantly at any energy \cite{backer1997number}. 

\begin{figure}
  \centering
  \includegraphics[width=8.6cm]{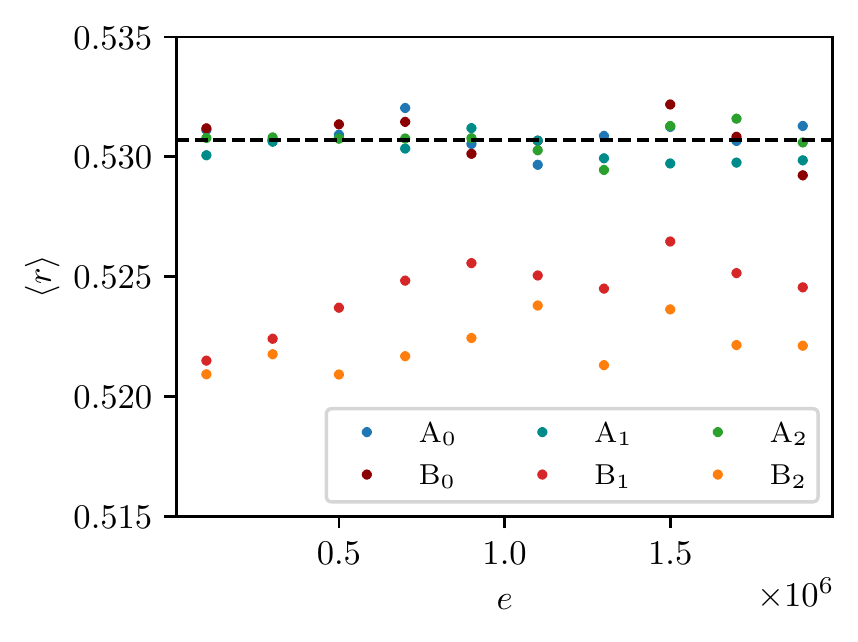}
  \caption{The averaged level spacing ratio for consecutive samples of $2.5\cdot10^5$ levels in the unfolded energy spectra.}  
  \label{fig:LevelSpacingRatio}
\end{figure}

To further corroborate the results, we computed a second commonly used short range statistic, namely the \textit{level spacing ratio} (LSR) \cite{atas2013}. This is defined as  
\begin{equation}
    r_i = \frac{\mathrm{min}(s_i,s_{i-1})}{\mathrm{max}(s_i,s_{i-1})},
\end{equation}
where $s_i=e_{i+1}-e_i$ is the level spacing. The level spacing ratio provides the benefit that it is not necessary to unfold the spectra, as it is independent of the local density of states. For GOE random matrices, the mean LSR is $\langle r \rangle_{\mathrm{GOE}} = 0.5307$ and $\langle r \rangle_{\mathrm{Poiss}} = 0.3863$  for Poissonian level statistics. In Fig. \ref{fig:LevelSpacingRatio} we show the mean LSR computed for consecutive intervals of $2.5\cdot10^5$ levels. Even though unfolding is not required, it is easiest to compare the statistics in the unfolded energies. The mean LSR values fluctuate around $\langle r \rangle_{\mathrm{GOE}} = 0.5307$. Again, the two right triangles are the exception, where the mean LSR fluctuates around slightly lower values after saturation, namely $\langle r \rangle_{\mathrm{B}_1} \approx 0.5255$ and $\langle r \rangle_{\mathrm{B}_2} \approx 0.5223$. The results on the mean LRS thus corroborate the previously presented results on the level spacing distributions. However, the mean LRS does not distinguish between the different non-right-angled triangles. One might wish to consider the distributions of the LRS instead, but we have already gained all the relevant information from the level spacing distributions.        
   
\subsection{Number Variance}\label{sub:Number variance} 
When studying medium- and long-range statistics of the spectra, two statistics are most commonly considered, namely the \textit{spectral rigidity} $\Delta_3$ and the \textit{number variance} (NV) $\Sigma^2$ (both were mentioned already in Sec. \ref{sub:Spectral staircase}).  The two are related via a simple integral transformation (see for instance \cite{aurich1997}), and we will therefore focus only on the number variance. The NV is defined as
\begin{equation}
    \Sigma^2(L,e):=\left\langle (n(L,x)-L)^2 \right\rangle_{e,w}, \, L>0,
\end{equation}
that is the local variance of the number $n(L,x) = N(x+L/2)-N(x-L/2)$ of unfolded energy levels in the interval $e_n\in[x-L/2,\,x+L/2]$. The brackets $\langle ... \rangle_{e,w}$ denote a local average around the central energy $e$ and window width $w$, so that $x\in[e-w/2,\,e+w/2]$.
The RMT result for the GOE case is the following,
\begin{multline*}
    \Sigma^2_\mathrm{GOE}(L) = \frac{2}{\pi^2}\Bigg\{\mathrm{ln}(2\pi L)+\gamma+1 
    \\+\frac{1}{2}\mathrm{Si}^2(\pi L)- \frac{\pi}{2}\mathrm{Si}(\pi L) - \cos{(2\pi L)} \\
     - \mathrm{Ci}(2\pi L)
    +\pi^2L\left(1-\frac{2}{\pi}\mathrm{Si}(2\pi L)\right)\Bigg\},
\end{multline*}
where $\gamma=0.5772...$ is Euler's constant and $\mathrm{Si}(x)$ and $\mathrm{Ci}(x)$ are the sine and cosine integral respectively. In the Poissonian case, we have $\Sigma^2(L)=L$. Moreover, the short range behavior of the NV $\Sigma^2(L)=L+O(L^2)$ is fixed by the fact that the spectra are unfolded. It is well known that the medium- and long-range statistics measured by the number variance are strongly influenced by (non-universal) short periodic orbits (see \cite{backer1995spectral} and references therein). The universality regime in which RMT spectral statistics are expected in real chaotic systems is restricted to short correlation lengths $L\ll L_\mathrm{max}$, where following semiclassical arguments $L_\mathrm{max}\propto \sqrt{e}$. For $L> L_\mathrm{max}$, the NV oscillates around its saturation plateau $\Sigma^2(\infty)$.

\begin{figure*}
  \centering
  \includegraphics[]{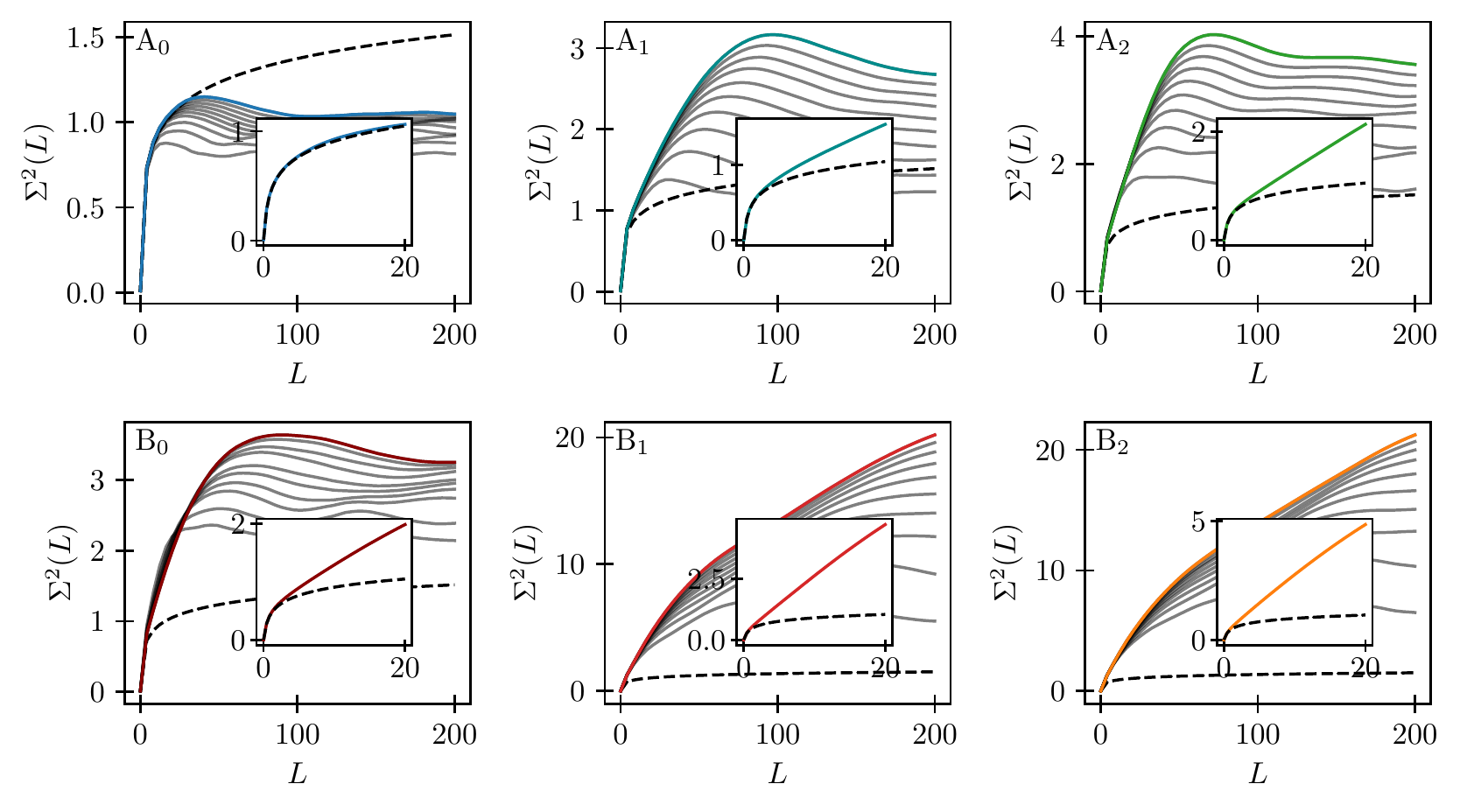}
  \caption{Number variance of cumulative spectral samples, adding $2\cdot 10^5$ levels at each step (grey lines) up to $e=2 \cdot 10^6$ (coloured lines). The insets show the short range behavior of the number variance. The dashed line is the GOE prediction.} 
  \label{numberVariance}
\end{figure*}
    In Fig. \ref{numberVariance} we show $\Sigma^2(L,e)$. The curves show the NV for cumulative spectral samples, adding $2\cdot 10^5$ levels at each step (grey lines) up to $e=2 \cdot 10^6$. We observe the curves saturate at ever higher values as we progress deeper into the semiclassical limit. The progressive curves form an envelope approximating the asymptotic result. The insets show the small $L$ behavior of the envelope function. This is well described by GOE up to the saturation in the $\mathrm{A_0}$ triangle. In the other cases, the short range behavior is followed by a linear regime.  The proportionality coefficient in the linear regime is known as the \emph{spectral compressibility}. In the right-triangles, we observe multiple linear regimes with varying spectral compressibility, until saturation. The qualitative shapes of the NV are similar for the triangles $\mathrm{A_1}$, $\mathrm{A_2}$ and $\mathrm{B_0}$ and also within the subclass of right-triangles $\mathrm{B_1}$ and $\mathrm{B_2}$. The linear regimes are caused by the bouncing ball scarred modes. It is evident that the bouncing ball contributions may become dominant even at fairly short correlation lengths. In the most irrational triangle, the NV follows the GOE prediction up to saturation, confirming the medium-range statistics conform to random matrix theory.     
    
\subsection{Spectral form factor} \label{sub:sff}
As a final test of the long-range spectral statistics, we computed the spectral form factor (SFF). The SFF is loosely defined as the Fourier transform of the spectral two point correlation function and may be written as
\begin{equation}
    K(t) = \left\langle \left|\sum_n  \mathrm{exp}(2 \pi i e_n t) \right|^2\right\rangle,
\end{equation} \label{eq:sff}
where the sum goes over the unfolded energy levels, and $\langle\cdots\rangle$ represents an average over an ensemble of similar systems or a moving time average as discussed below.
The time $t$ is measured in units of the Heisenberg time $t_H =1$, defined as $t_H = 2\pi \hbar/\langle s \rangle$ in full units. The Heisenberg time is the typical timescale after which discreetness of the spectrum is resolved. In a monumental feat of semiclassical analysis, spanning many years of research \cite{Sieber2001, Mueller2004a, Mueller2004b, Mueller2005, Heustler2007, Mueller2009},  the periodic orbit contributions to SFF in chaotic systems have been fully identified and shown to conform to the RMT statistics. This is also supported by an extensive amount of numerical evidence. Recently, the RMT statistics of the SFF have been analytically shown to hold in a kicked Ising spin system as a minimal model of many-body quantum chaos (without a meaningful classical limit) \cite{bertini2018exact} and other systems represented with dual-unitary quantum circuits \cite{bertini2021cmp}. In the infinite dimensional GOE case, the SFF has the following analytical form,
\begin{equation}
K_\mathrm{GOE}(t) =\begin{cases} 2t-t\mathrm{ln}(2t+1) &t < 1 \\
                                2-t\mathrm{ln}(\frac{2t+1}{2t-1}) &t > 1 \end{cases}.
\end{equation}
It is worth mentioning that the SFF is related to the NV studied in the previous section by the integral transformation $\Sigma^2(L)=\int_0^\infty dt K(t) \left[\sin(\frac{\pi L t}{\pi t})\right]^2$. 
The behavior of the NV in long-range limit is given by the SFF at short times. If the SFF remains finite in the limit $t \rightarrow 0$, then $\Sigma^2(L)= L \lim_{L \rightarrow \infty} K(\frac{1}{L})$. A finite limit of the SFF at $t \rightarrow 0$, the so-called {\em spectral compressibility} $K(0)$, thus results in the linear behavior of the NV, which we observed in most triangles (except in the `most mixing' case $\mathrm{A}_0$). 

\begin{figure}
  \centering
  \includegraphics[width=8.6cm]{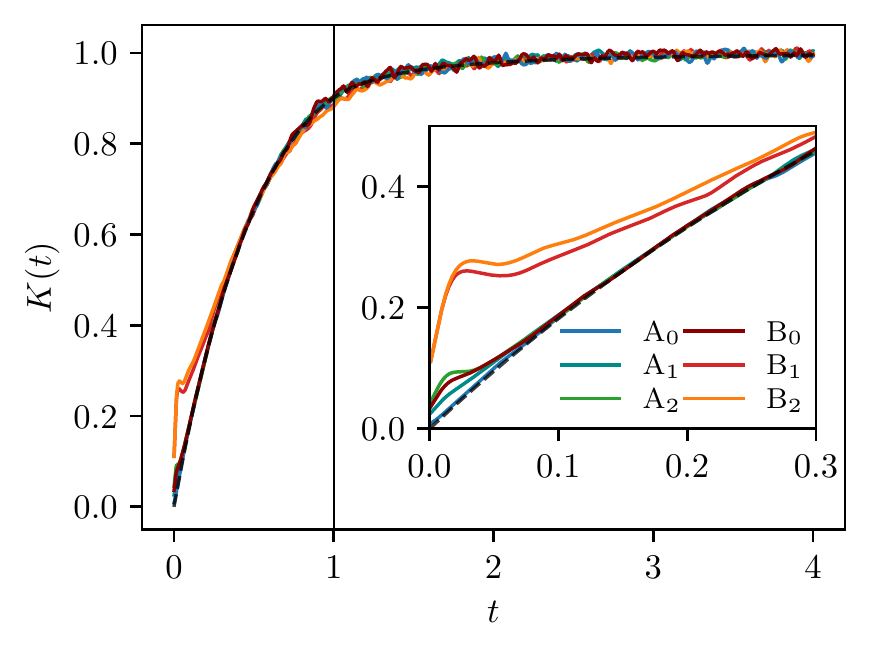}
  \caption{The connected spectral form factor of the triangle billiards computed for the entire data sets of $2\cdot 10^6$ levels. Since the SFF is not self-averaging, a moving-time average is performed to smooth the data. The non-universal disconnected part (the delta like peak at times $t<10^{-5}$) is removed from the data before the averaging. The inset shows a magnification of the short time behavior. The vertical line shows the Heisenberg timescale $t_{\rm H}=1$. The black dashed line is the GOE prediction.} 
  \label{fig:SFF}
\end{figure}

\begin{figure*}
  \centering
  \includegraphics[]{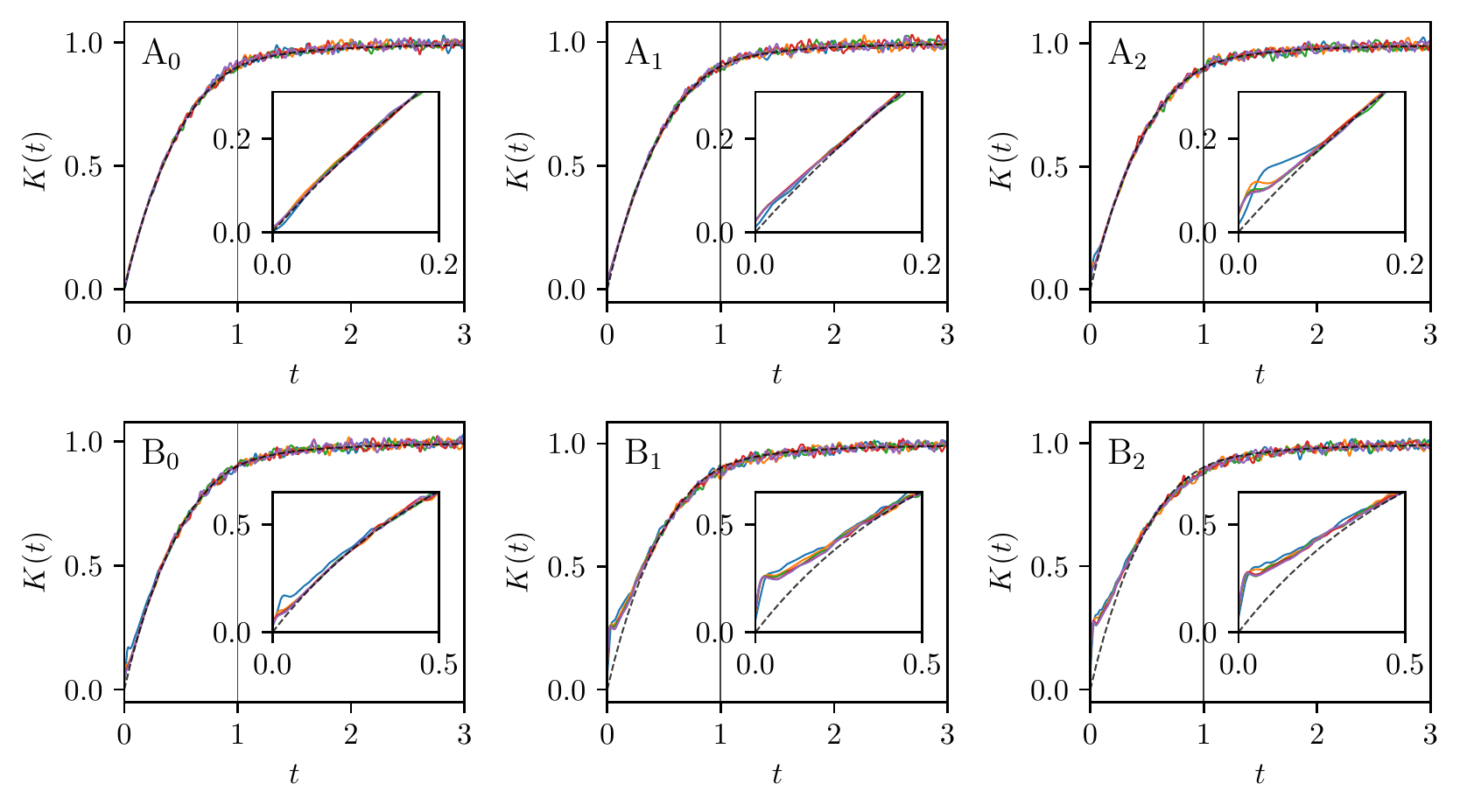}
  \caption{The connected spectral form factor of the triangle billiards for $2 \cdot 10^5$ consecutive levels starting at different energies $e_0$, with $e_0 = 0$ (blue),  $e_0 = 4 \cdot 10^5$ (orange), $e_0 = 8 \cdot 10^5$ (green) $e_0 = 12 \cdot 10^5$ (red), $e_0 = 16 \cdot 10^5$ (violet). The black dashed line is the GOE prediction.} 
  \label{fig:SFFcompare}
\end{figure*}

The main practical drawback of the SFF is the fact that it is not a self averaging quantity \cite{prange1997spectral}, meaning that the typical value at some time may be far from the average value over some short time interval. This problem is traditionally circumvented by performing an average over an ensemble of systems, for example different realizations of disordered systems, random matrices etc. This approach is not suitable in our case, where we want to understand the properties of very specific dynamical systems. 
We will thus opt for a different approach, namely employing a moving average in time to smooth the results. This is done by convolving the time dependent SFF data with a Gaussian kernel in time. We thus introduce an additional numerical parameter to the calculation, the width of the Gaussian kernel $\delta_t$. The SFF exhibits a delta-like peak at short times $t<10^{-5}$ (inversely proportional to the number of states), with an amplitude proportional to the square of the number of spectral levels, which would obscure the results of the time smoothing. We therefore consider only the so-called connected part of the SFF by removing this initial peak from the data. In Fig. \ref{fig:SFF} we show the connected spectral form factor, giving us a time resolved view of the spectrum. The entire computed spectral samples were used for the calculation, and the width of the time smoothing kernel is $\delta_t = 0.01$.  In keeping with the previously presented spectral statistics, the SFF in the most irrational triangle $\mathrm{A_0}$ conforms perfectly to the GOE prediction. The other triangles show deviations at short times (particularly visible in the inset) as one would expect based on the results on the NV. The qualitative behavior is similar in all cases. {The characteristic time at which the SFF starts to overlap the GOE result is related to the transport time of the classical system. The relevance of this timescale will be further discussed in Sec. \ref{sub: Localization measures}.}  Again, this short time behavior is most likely due to the bouncing ball orbits, which feature most prominently in the right triangles. The SFF does not change, if we omit the first $10^5$ levels to diminish any possible non-generic effect associated with the low-lying part spectra. { Moreover, we have also compared the SFF computed for intervals of consecutive $2\cdot 10^5$ levels, starting at different energy, as shown in Fig. \ref{fig:SFFcompare}. Since, the density of states depends on the energy and the time units are fixed by the local Heisenberg time we observe a weak dependence of the short time behavior on the energy i.e. the transport time is rescaled slightly. However, because the density of states becomes nearly constant at high energies, the different curves quickly start to overlap even in the short time regime.}  

Finally, we need to admit that it is not evident if the systematic discrepancy from GOE statistics, which seem not to decrease with increasing the excitation energies
for the right triangle billiards, can be solely attributed to abundance of bouncing ball modes due to marginally stable periodic orbits. We stress that even the precise counting of the number of periodic orbit manifolds in triangular billiards is an open mathematical problem. The discrepancy from GOE could also be related to observed and suggested~\cite{wang2014nonergodicity} lack of ergodicity in generic right triangular billiards, or, alternatively~\cite{kaplan1998,horvat,huang}, extremely `slow' (logarithmic) ergodicity (or pseudo-ergodicity). We note that even with state of the art numerical experiments, it seems impossible to draw a definite conclusion.

\section{Eigenstates}\label{sec:Eigenstates} 
The triangle billiards considered in this work belong to a general class of quantized ergodic systems, although ergodicity has been rigorously proven only for a dense subset of triangular billiards~\cite{gutkin2003review} {and as previously stated the generic right triangles may not be ergodic but are only pseudo-ergodic in this sense}. Due to the quantum ergodicity theorem \cite{shnirel1974ergodic,zelditch1996ergodicity} (see also \cite{barnett2006asymptotic} and references therein) one expects that the wavefunctions will be spatially close to uniform in the semiclassical limit. The wavefunctions in the bulk of the system (far away from the boundary) are statistically similar to random superpositions of plane waves \cite{Berry1977wf}. However, there are well known exceptions, for example, the states scared by periodic orbits \cite{Heller1984}, or dynamically localized states (see \cite{LozejPHD} for many examples in billiards). We will examine the eigenstates of the triangles in terms of their statistical properties, as well as some recently developed methodology based on the localization measures in the Poincaré-Husimi representation. 

\subsection{Poincaré-Husimi representation}\label{sub:PH representation} 
When considering eigenstates of any quantum system in the semiclassical limit, it is often useful to interpret them as (quasi)probability distributions in terms of the classical canonical coordinates $(q,p)$. Since there is no strict equivalent to the classical phase space in the quantum realm (due to the uncertainty principle) various representations are employed. Most commonly, the eigenstate is represented with the Wigner \cite{Wig1932,Hillery1984} or Husimi \cite{Hus1940} function, the latter being a Gaussian-smoothed equivalent of the former. In contrast to the Wigner function, the Husimi functions are strictly non-negative and may therefore be interpreted, in a vague sense, as probability density functions of quantum states in classical phase space. 
Here we will give a short description of how to define the Husimi functions for billiard systems, following the construction outlined in Refs. \cite{tualle1995, Baecker2004hus}. 

The classical billiard dynamics may be reduced to a two-dimensional discrete mapping. We first fix the speed of the particle and employ the Poincaré surface of section (SOS) method, using the boundary of the billiard table as the SOS to discretize the dynamics. The mapping is commonly described in the Poincaré-Birkhoff (PB) coordinates, where we take the arc-length of the billiard boundary $q$ as the spatial coordinate and $p=\sin \alpha$, as the canonical momentum, where $\alpha$ is the angle of reflection. The phase space is a cylinder $\mathcal{M}=\left[0,\,\mathcal{L}\right]\cdot(-1,\,1)$,
where we take $s$ to be periodic with a period equal to the total
length of the billiard boundary $\mathcal{L}$. For a more detailed description, see Ref. \cite{LozejPHD}. 

As we see, all the relevant information about the classical dynamics is contained on the boundary of the billiard table. Similarly, the wavefunction inside the quantum billiard is fully determined by its normal derivative at the boundary, called the boundary function
\begin{equation}
u\left(q\right):=\boldsymbol{n}\cdot\nabla_{\boldsymbol{r}}\psi\left(\boldsymbol{r}\left(q\right)\right),\label{eq:Boundary functions}
\end{equation}
where $\boldsymbol{n}$ is the outward normal unit vector at the boundary
position $\boldsymbol{r}\left(q\right)$. The wavefunction is obtained by the boundary integral
\begin{equation}
\psi_{k}\left(\boldsymbol{r}\right)=-\oint_{\partial B}\mathrm{d}lu_{k}\left(l\right)G\left(\boldsymbol{r},\boldsymbol{r}\left(l\right)\right),\label{eq:Wavefunction Green}
\end{equation}
where $\boldsymbol{r}\left(l\right)$ is a position on the billiard
boundary and $\boldsymbol{r}$ is a position inside the billiard table, and
\begin{equation}
G\left(\boldsymbol{r},\boldsymbol{r}^{\prime}\right)=-\frac{\mathrm{i}}{4}H_{0}^{\left(1\right)}\left(k\left|\boldsymbol{r}-\boldsymbol{r}^{\prime}\right|\right),\label{eq:Green's function}
\end{equation}
is the free particle Green's functions satisfying $(\nabla^{2}+k^{2})G\left(\boldsymbol{r},\boldsymbol{r}^{\prime}\right)=\delta\left(\boldsymbol{r}-\boldsymbol{r}^{\prime}\right),
$ and $H_{0}^{\left(1\right)}\left(x\right)$ is the zero-order Hankel function of the first kind.

The \textit{Pouncaré-Husimi} (PH) functions are a representation of the quantum states of billiards
as probability distributions in the phase space using the classical (PB) coordinates. The basic idea is to use coherent states on the boundary $\partial\mathcal{B}$, that are localized
at $\left(q,p\right)\in\left[0,\mathcal{\,L}\right]\cdot\left[-1,\,1\right]$
and are periodic with a period of $\mathcal{L}$, onto which we project
the boundary functions $u(s)$. We define the coherent state as
\begin{multline*}
c_{\left(q,p\right),k}\left(l\right):=\sum_{m\in\mathbb{Z}}\exp\left[\mathrm{i}kp\left(l-q+m\mathcal{L}\right)\right]\\
\cdot\exp\left[-\frac{k}{2}\left(l-q+m\mathcal{L}\right)^{2}\right].\label{eq:coherent state}    
\end{multline*}

The sum over $m$ ensures the coherent states are periodic, with a period of $\mathcal{L}$.
We omit all normalization factors because we will normalize the PH
functions at the end. Let $u_{n}(q)$ be the boundary function of
the $n$-th billiard eigenstate with the wavenumber $k_{n}$.
The Poincaré-Husimi function of this state is defined as
\begin{equation}
H_{n}\left(q,p\right):=\frac{1}{A_{n}}\left|\oint_{\partial B}c_{\left(q,p\right),k_{n}}\left(l\right)u_{n}\left(l\right)\mathrm{d}l\right|^{2},
\end{equation}
where $A_{n}$ is a normalization factor. We see that the PH functions
are positive definite by construction and would like them
to represent probability distributions in the phase space. We therefore
fix the normalization factor so that 
\begin{equation}
\intop_{0}^{\mathcal{L}}\mathrm{d}q\intop_{-1}^{1}\mathrm{d}pH_{n}\left(q,p\right)=1.\label{eq:HusimiNorm}
\end{equation}
The PH functions are an invaluable tool for interpreting the quantum-classical correspondence in billiards and will help us to distinguish generic eigenstates from bouncing ball states. Many examples are shown in section \ref{sub: Gallery of states}. 

\subsection{Localization measures}\label{sub: Localization measures} 
In ergodic systems, we expect the generic states to be close to uniformly extended over the entire phase space (in the PH representation). On the other hand, the states scarred by marginally stable periodic orbits are localized on the part of the phase space corresponding to the classical periodic orbits. The extent of the localization is one criterion by which the bouncing ball states may be distinguished from the generic ones. Following recently developed methodology, that has been used to describe the dynamical localization in fully chaotic \cite{batistic2020distribution} and mixed-type billiards \cite{batistic2013dynamical,BatLozRob2019,lozej2021effects}, we will define a localization measure based on the entropy of the PH functions of the eigenstates. We interpret the PH function as a probability distribution and define its \textit{information entropy}
\begin{equation}
S_{n}:=-\intop_{0}^{\mathcal{L}}\mathrm{d}q\intop_{-1}^{1}\mathrm{d}pH_{n}\left(q,p\right)\ln\left(H_{n}\left(q,p\right)\right).
\end{equation}
The entropy localization measure (ELM) is then defined as
\begin{equation}
l_{n}:=\frac{\exp\left(S_{n}\right)}{\mathrm{Vol}(\mathcal{M})},
\end{equation}
where $\mathrm{Vol}(\mathcal{M}) = 2 \mathcal{L}$ is the volume (surface area in our case) of the classical phase space.  The ELM is minimized by a state localized within one Planck's cell of the phase space, giving $l\rightarrow0$. Conversely, the ELM is maximized by a uniform distribution in the phase space, giving $l=1$.  However, pure states cannot produce a completely uniform distribution, so in practice there is some upper bound for the ELM. Empirically and based on some numerical results using Berry's random wave ansatz, we find $l_{max}\approx0.7$ \cite{LozejPHD}. Let us mention that traditionally (see for instance Ref. \cite{Izr1990} for a review of results in the quantum kicked rotor), similar localization measures were defined by considering the occupation of the basis vectors in some natural basis (which is not the eigenbasis) for the particular problem. In this sense, the ELM can be thought of as the localization length.

The distributions of the ELMs have recently been studied in several families of billiards. Results from the ergodic stadium \cite{batistic2020distribution}, cardioid \cite{batistic2020distribution} and special cases of the lemon billiards \cite{LozejPHD} show that the ELMs of a sequence of consecutive eigenstates are distributed according to a common form, which is empirically well described by the beta probability distribution. This was corroborated by special cases of the lemon billiards with a mixed-type phase space but no stickiness \cite{LozejPHD,lozej2021nostick}, as well as the Dicke model \cite{wang2020dicke} in the chaotic regime, which describes a series of atoms coupled to a single elctro-magnetic cavity mode. We thus expect the same shape of the distribution of the ELM in the `most mixing' triangle billiards, namely
\begin{equation}
P(l)=\frac{1}{C}l^{a-1}(l_{max}-l)^{b-1}, \label{eq:betaDist}
\end{equation}
where the normalization constant is given by $C = l_{max}^{a+b-1}B(a,b)$, where $B(x,y)$ is the Beta function, the namesake of the distribution. While the shape of the distribution seems to be universal for ergodic systems and even non-sticky components of mixed-type systems, the shape parameters $a$ and $b$ vary from case to case and spectral interval, depending on the level of localization of the eigenstates. 
This is only one possible choice of localization measure in a wider class of Rényi occupation measures \cite{gnutzmann2001renyi, villasenor2021quantum}, defined as the exponentials of the Rényi entropies of the Husimi functions
\begin{equation}
l^{\alpha}_{n}:=\left\langle{H_n}^{\alpha}\right\rangle^{\frac{1}{1-\alpha}}
\end{equation}
where $\alpha$ is the order of the measure and the angled brackets denote the phase space average (we note that the PH functions are already correctly normalized, so no additional normalization factor is needed). In the limit $\alpha \rightarrow 1$ we obtain the ELM. We will also consider the $\alpha = 2$ measure, i.e. the normalized inverse participation ratio (IPR). Higher order Rényi occupation measures are more sensitive to larger values of the Husimi function and therefore suitable for detecting highly localized or scarred states, as recently demonstrated in Refs. \cite{pilatowsky2021ubiquitous, pilatowsky2021identification} for the Dicke model. Analytical calculation based on the random wave approximation (see \cite{pilatowsky2021identification} and references therein) give the estimate $l^{\alpha}_{max} \approx \Gamma(1+\alpha)^{1/(1-\alpha)}$ for maximally extended pure states. This gives us $l_{max} = l^{1}_{max}\approx 0.66 $ and $l^{2}_{max}\approx0.5 $. In practice, the PH functions were evaluated on a grid of $M_q\cdot M_p$ points in the phase space, with $M_p\approx 5k \mathcal{L}/(2\pi)$ and $M_q = \mathcal{L}M_p$. The PH functions were then considered as discrete probability distributions and the phase space averages evaluated as sums. 

Dynamically localized states are expected to appear if the \textit{classical transport time} $t_T$ (the typical timescale of classical diffusion) is longer than the Heisenberg time. We may define the localization control parameter $\alpha = t_H/t_T$. The Heisenberg time is proportional to the mean density of states, which is given by the Weyl formula $\rho(E)=\mathrm{d}N/\mathrm{d}E = \frac{\mathcal{A}}{4\pi}-\frac{\mathcal{L}}{8\pi\sqrt{E}}$. We may manipulate the local Heisenberg time by considering only the states with wavenumbers close to some $k_0=\sqrt{E_0}$. Interestingly, we may estimate the classical transport time from the SFF. As we saw in Sec. \ref{sub:sff} the SFF initially deviates from the GOE result, but after some time the two curves start to overlap. The point of overlap is the classical transport time. From Fig. \ref{fig:SFF} we see $t_T\ll t_H $ in all cases except for the right triangles, where  $t_T \approx 0.4 t_H$.  We thus expect no significant dynamical localization, with the possible exception of some small effects in the right triangles.   

\begin{figure*}
  \centering
  \includegraphics[]{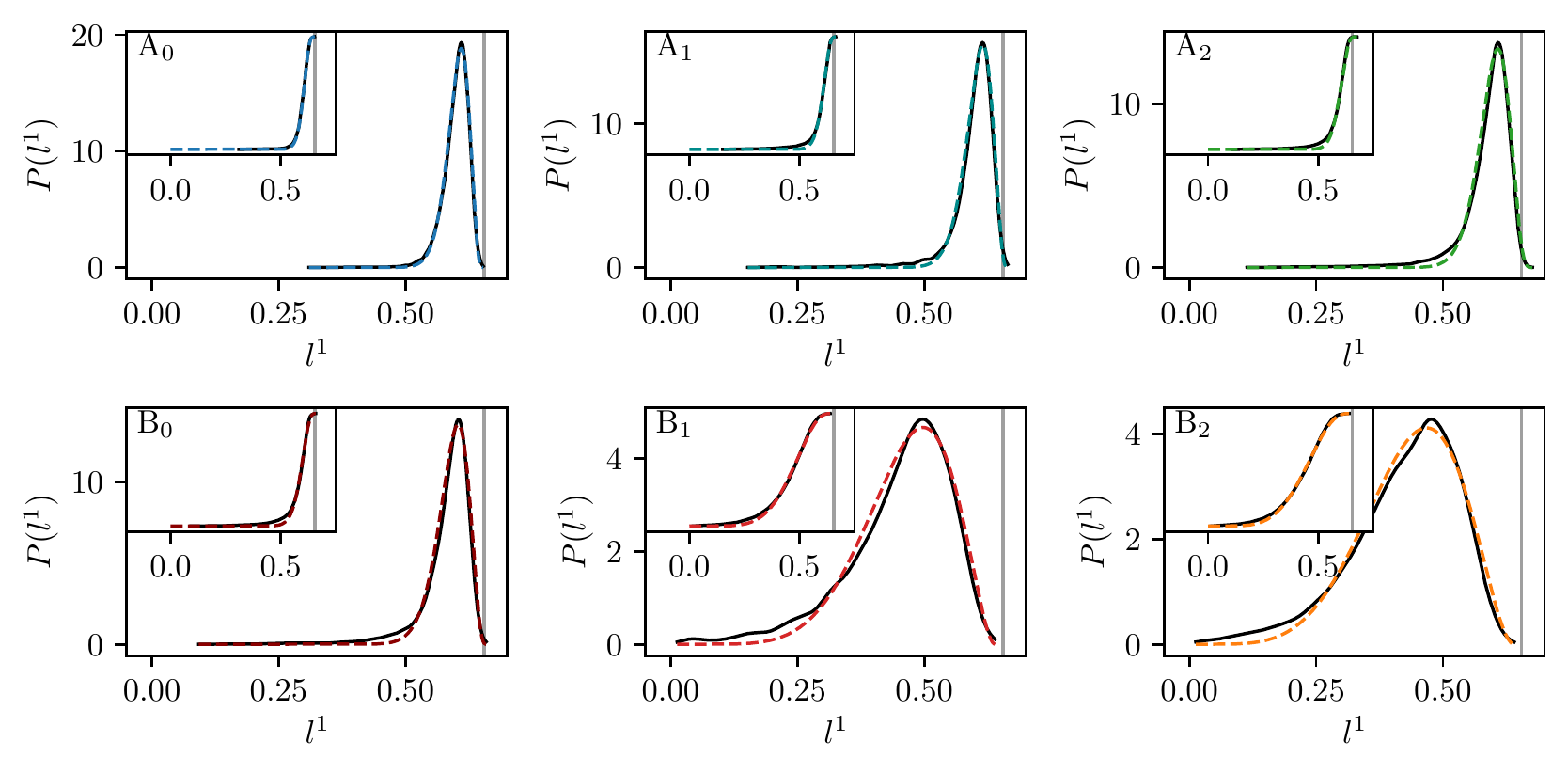}
  \caption{Distributions of the ELMs of $10^4$ consecutive eigenstates, starting form $e_0=1.5\cdot 10^6$. The main figures show the kernel density estimation of the probability density function, and the insets, the cumulative distribution function. The colored dashed lines show the best fitting beta distribution Eq. \eqref{eq:betaDist}. The vertical gray line indicates ${l^1}_{max} = 0.66$ from the RWM.} 
  \label{fig:measures1}
\end{figure*}

\begin{figure*}
  \centering
  \includegraphics[]{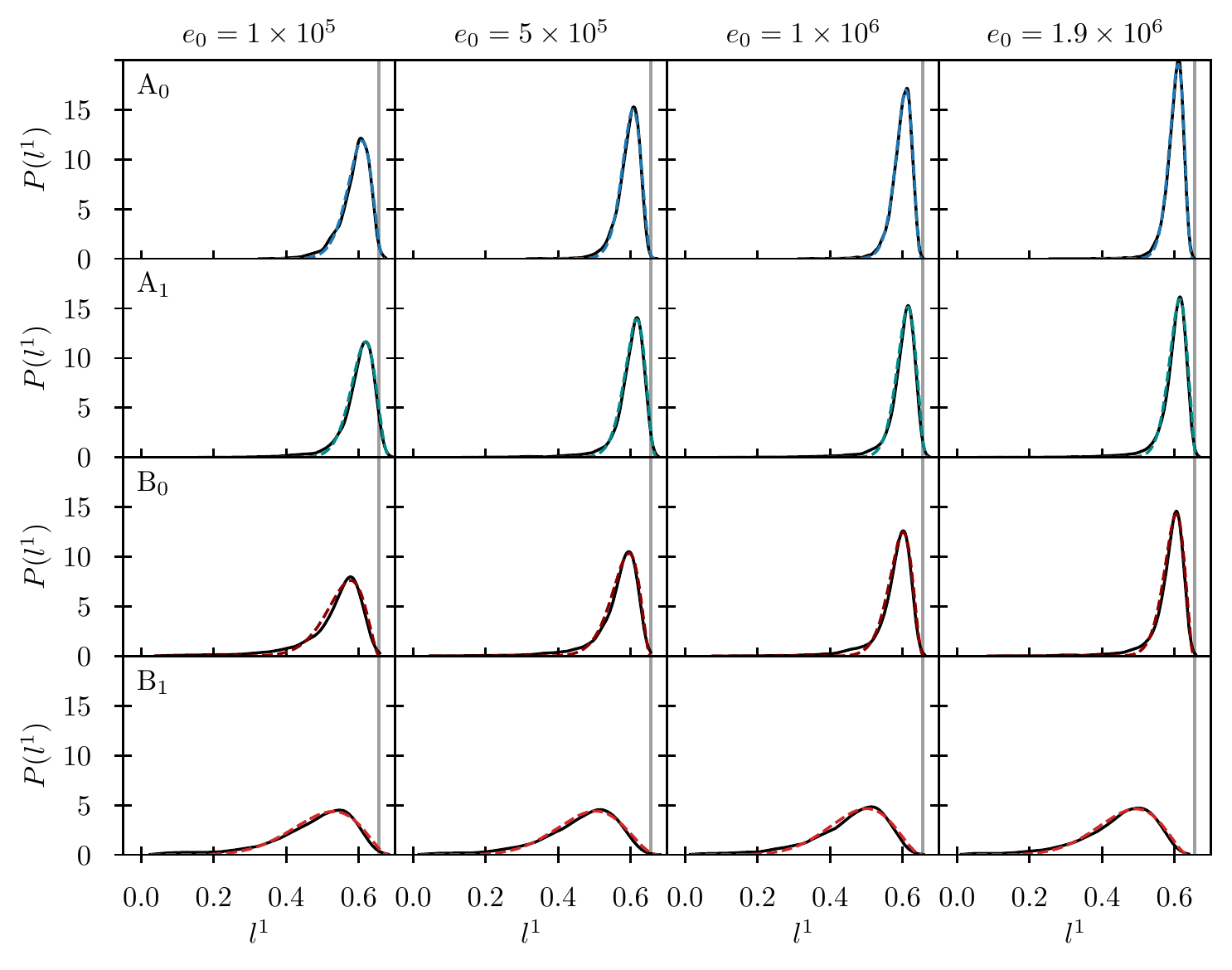}
  \caption{Distributions of the ELMs of $10^4$ consecutive eigenstates, as a function of the unfolded energy $e_0$ (indicated at the top of each column). The colored lines show the best fitting beta distribution Eq. \eqref{eq:betaDist}.} 
  \label{fig:measures_dep1}
\end{figure*}

In Fig. \ref{fig:measures1} we show the distributions of the ELMs of $10^4$ consecutive states, starting from the unfolded energy $e_0=1.5\cdot 10^6$ in each triangle. In all cases, the distribution is skewed heavily towards the extended regime. Especially in the class (A) triangles, the ELMs are narrowly distributed towards the upper bound $l_{max} = l^{1}_{max}\approx 0.7 $. The theoretical expectation $l_{max} \approx 0.66$ is slightly exceeded in some cases, which we attribute to finite size effects not considered in the derivation. The distributions are wider in the class (B) triangles (especially in the right-triangles), but still centered on the extended regime. As expected, there is no significant dynamical localization. The empirical beta distribution \eqref{eq:betaDist} fits the data reasonably well. The fit is almost perfect in the $A_0$ triangle. However, we see slight deviations in the lower tail of the distribution, which are even more pronounced in the class (B) triangles. These deviations are caused by the bouncing ball scarred states. The bouncing ball states are severely localized and thus produce small localization measures. The distributions of the measures show that severely localized states are more probable than expected in a uniformly ergodic system, even in the strongly-mixing triangles, but especially in the weakly-mixing and pseudo-ergodic ones. As already indicated by the spectral statistics, this is especially noticeable in the right-triangles. In Fig. \ref{fig:measures_dep1} we show the dependence of the ELM distributions as function of $e_0$, thus varying the local Heisenberg time. One immediately notices the distributions narrow as we increase $e_0$. As expected from the quantum ergodicty theorem, all states become increasingly delocalized and ever closer to uniformly extended states. However, this is almost not noticeable for the right-triangle due to the much longer transport time. Furthermore, we may see some scarred states persist well into the high-energy parts of the spectrum and may remain even in the semiclassical limit. This has been extensively studied in chaotic billiards \cite{backer1997number}. Using the WKB approximation, one can estimate how the proportion of bouncing ball states scales with $k$. In principle, this may be arbitrarily close to $N_{bb}\propto k^2$ that is the first term of the Weyl formula, giving a non-vanishing contribution in the semiclassical limit. The scarred states are identifiable by having a very small ELM. It is quite possible that the bouncing ball states are actually super-scared states, similar to those found in pseudointegrable triangle billiards (see Ref. \cite{bogomolny2021} and references therein).

\begin{figure*}
  \centering
  \includegraphics[]{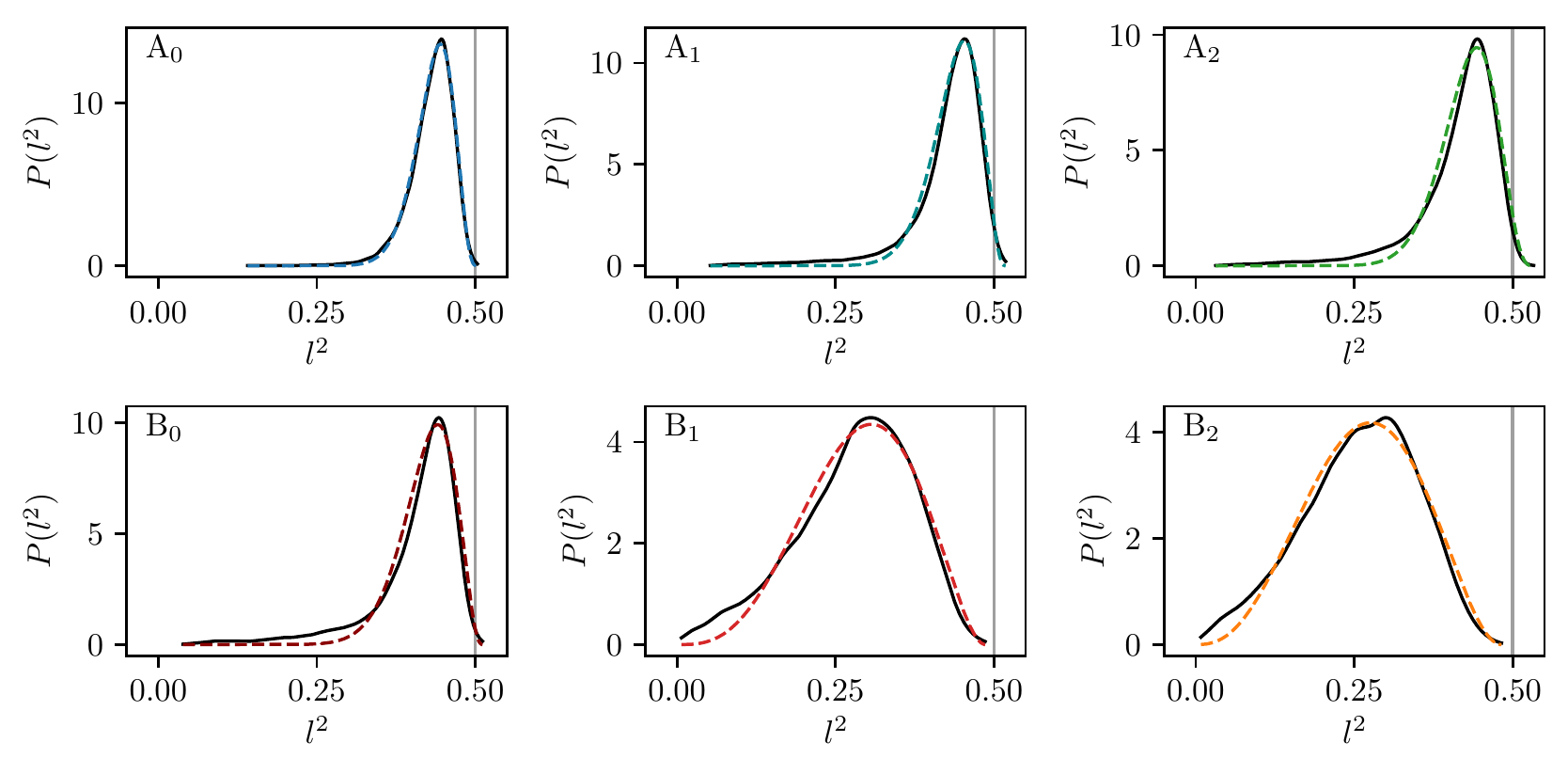}
  \caption{Distributions of the IPRs of $10^4$ consecutive eigenstates, starting form $e_0=1.5\cdot 10^6$. The figures show the kernel density estimation of the probability density function. The colored dashed lines show the best fitting beta distribution Eq. \eqref{eq:betaDist}. The vertical gray line indicates ${l^2}_{max} = 0.5$ from the RWM.} 
  \label{fig:measures2}
\end{figure*}

\begin{figure*}
  \centering
  \includegraphics[]{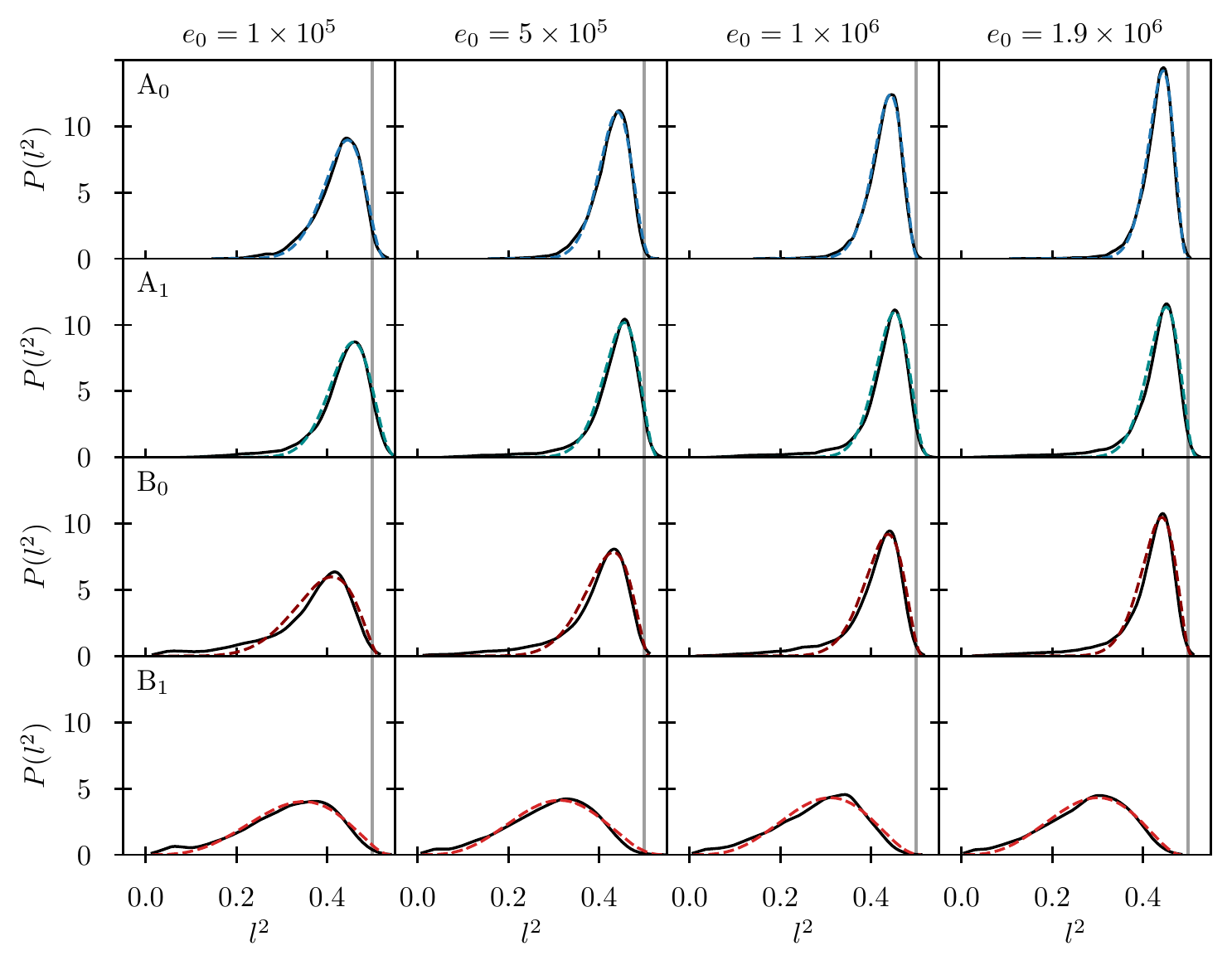}
  \caption{Distributions of the IPRs of $10^4$ consecutive eigenstates, as a function of the unfolded energy $e_0$ (indicated at the top of each column). The colored lines show the best fitting beta distribution Eq.\eqref{eq:betaDist}.} 
  \label{fig:measures_dep2}
\end{figure*}

To corroborate this results, we consider also the IPR as an alternative localization measure. In Fig. \ref{fig:measures2} we show the distributions of the IPRs of $10^4$ consecutive states, starting form $e_0=1.5\cdot 10^6$ in each triangle. Results from chaotic and mixed-type billiards \cite{BatLozRob2019,batistic2020distribution,LozejPHD} show the mean ELM and IPR in small spectral samples are linearly dependent. This leads us to expect that the distributions of the IPRs should also be generally similar to the distributions of the ELMs. Indeed, we find the empirical beta distribution Eq. \eqref{eq:betaDist} fits the distributions of the IPR reasonably well when the parameters are suitably adjusted, that is $l^{2}_{max}\approx0.5 $. The maximum value from the data is close to this, value but is slightly exceeded in most cases, except for the right-triangles where it is slightly lower. The distributions are generally wider than those of the ELMs (compare with Fig. \ref{fig:measures1}) and the bouncing ball scarred states are more prominent and thus more easily detectable as a slight bulge in the lower tail of the distribution. This may also be observed in Fig. \ref{fig:measures_dep2} where we show the energy dependence of the distributions of the IPRs. The bouncing ball scarred states are again noticeable in the class B triangles and persist into the high-energy regime. Since the beta distribution is only an empirical estimate, it is difficult to make any rigorous statements about the proportion of scarred states that persist in the semiclassical limit. We may only state that the deviations from the best fitting beta distribution (taking for instance the Kolmogorov-Smirnov test) do not systematically diminish with progressing energy, but remain roughly the same and relatively small.

\subsection{Gallery of states}\label{sub: Gallery of states} 
\begin{figure*}
  \centering
  \includegraphics[]{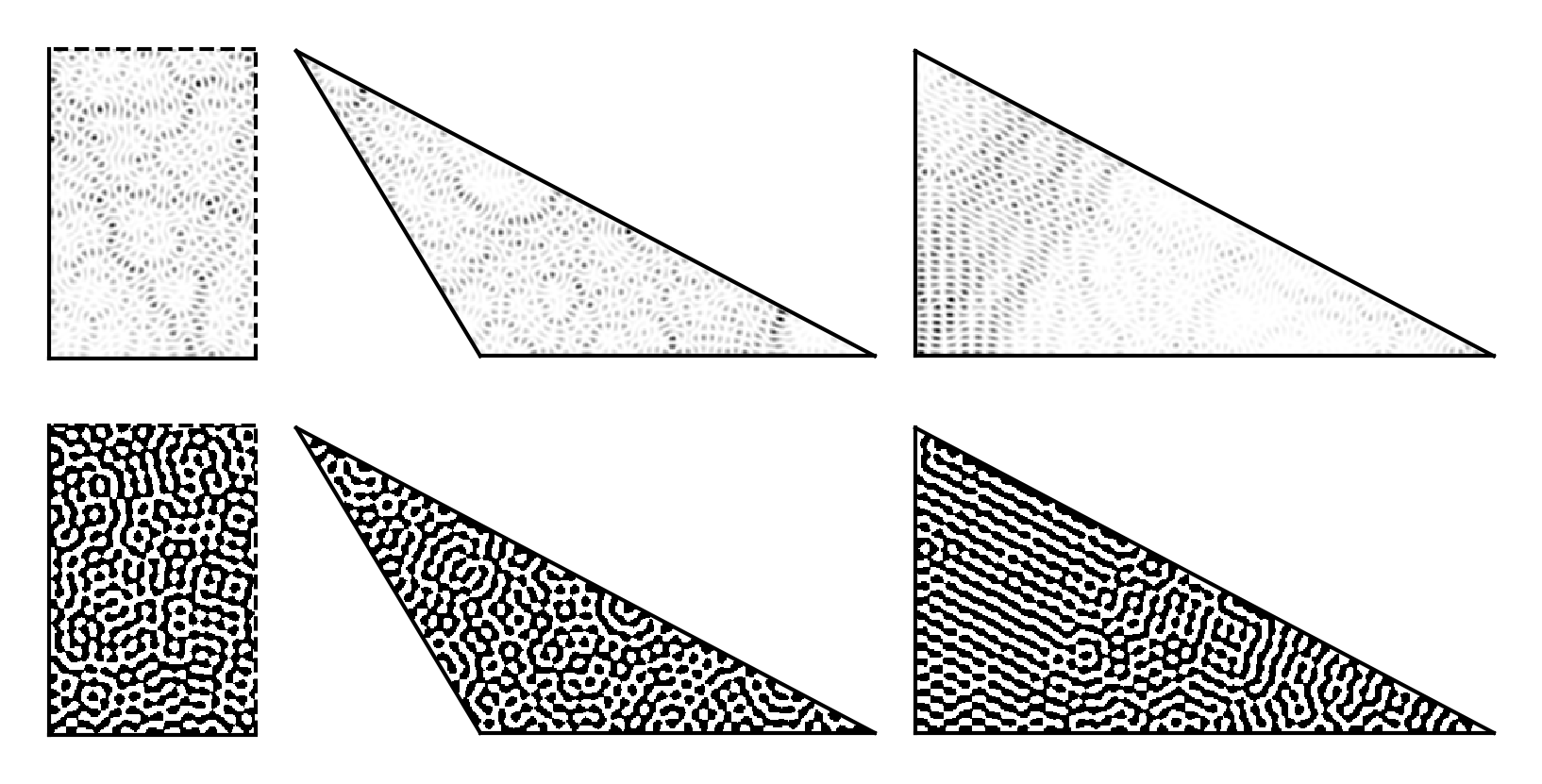}
  \caption{Top row, probability distributions of (from left to right) a Gaussian random superposition of plane waves (that satisfy Dirichlet b.c. only on the bottom and left boundary) with $k = 191.870$, an eigenstate of the $\mathrm{A_1}$ billiard with $k = 191.870$ and a heavily scarred eigenstate of the $\mathrm{B_1}$ billiard with $k = 191.854$. The bottom row shows the corresponding nodal patterns. The wavefunction is real and positive in the black regions and negative in the white regions. Observe the similarity and differences between the nodal patterns of the random waves and the eigenfunctions.} 
  \label{fig:rpw}
\end{figure*}

Thus far, we have studied the collective properties of ensembles of eigenstates. We will now examine the properties of some representative eigenstates themselves. The main goal is to assess if the eigenstates of the triangle billiards comply with the well known results on quantum ergodic states. As conjectured by Berry, with the random wave model (RWM), typical eigenstates in ergodic systems (far away from the boundaries) are expected to be statistically similar to superpositions of plane waves with random phases. The universal properties of such states have been extensively studied (see \cite{urbina2013random} for an overview) and famously form the basis of the eigenstate thermalization hypothesis \cite{srednicki1994eth}. An abundance of numerical evidence in chaotic billiards is available to support the RWM. The similarity with RWM states is apparent already upon visually inspecting the probability distributions of chaotic billiard eigenstates $|\psi|^2$ in the configuration space. The wavefunctions form typical nodal patterns that are very similar to RWM states. Apart from the visual similarity, other statistical properties of the ergodic eigenstates closely coincide with those of RWM states. It is well known that the values of the wavefunction of the RWM states follow a Gaussian distribution \cite{urbina2013random}. The spatial correlations of the wavefunctions in chaotic billiards also decay as a Bessel function of the distance, as predicted by the RWM. Even though the triangle billiards under our investigation are not chaotic, we expect very similar properties to those of RWM states, since the RWM only requires ergodicity. However, scarred states first found by Heller \cite{Heller1984}, which by definition violate the RWM assumptions, are ubiquitous in billiard systems. The most well-known examples are the bouncing ball states of the stadium billiard. Because of the straight boundaries, one may expect to find very similar scarring in the triangle billiards. Furthermore, super-scarred states are known to exist in pseudointegrable triangles \cite{bogomolny2004, bogomolny2021}. In Fig. \ref{fig:rpw} we show a comparison between a RPW state and examples of a generic and heavily scarred triangle eigenstate. 

\begin{figure*}
  \centering
  \includegraphics[]{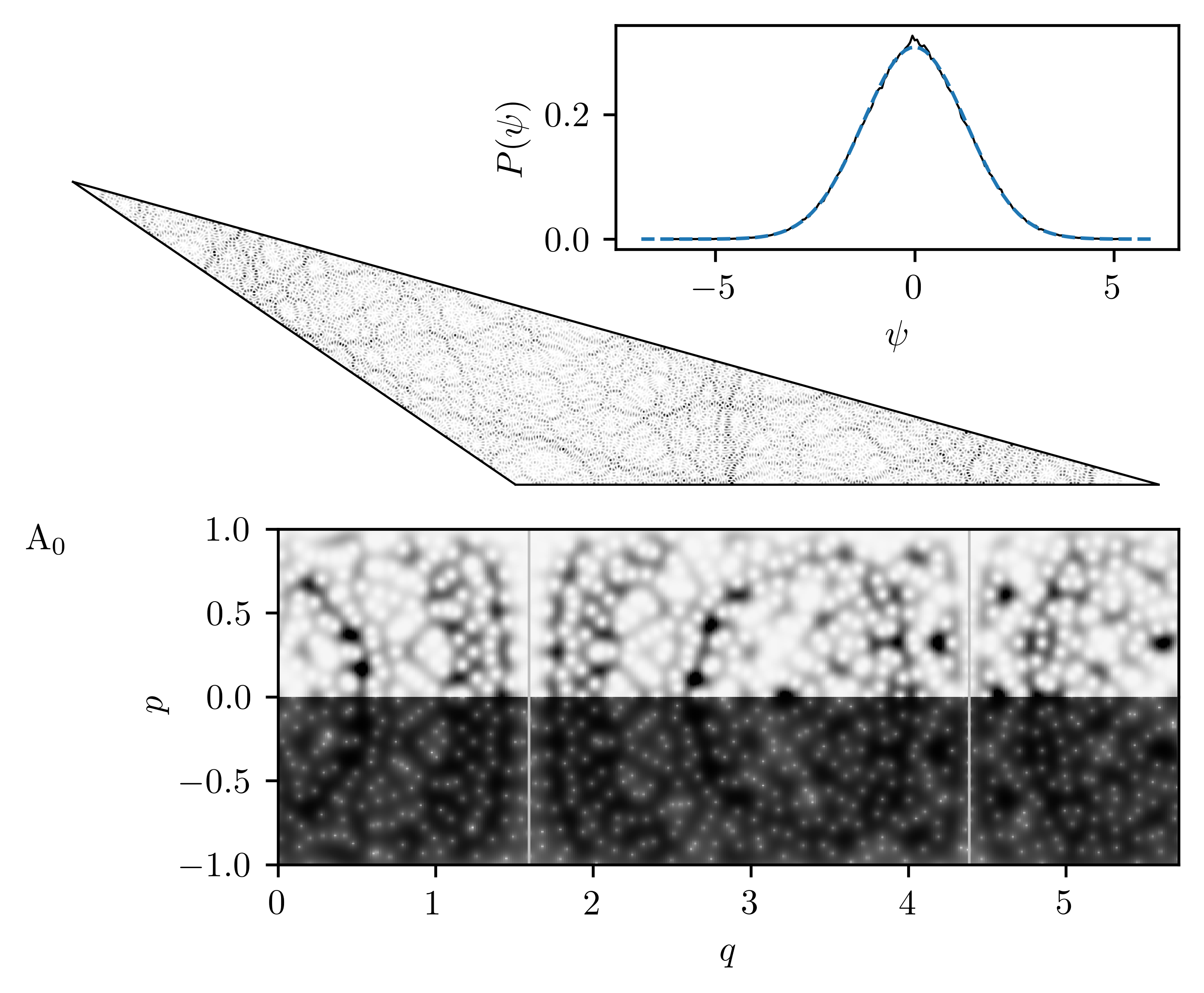}
  \caption{Typical state of the $\mathrm{A_0}$ triangle billiard, with $k =500.1714 $ and $l^{1} = 0.602$, $l^{2} = 0.445 $.(upper part) The probability distribution in real space and the histogram of the (real) wavefunction values fitted by a Gaussian distribution (colored line). (lower part) PH function probability (positive $p$) and stellar representation (negative $p$) in the classical phase space coordinates. The vertical lines show the positions of the corners. The stellar representation is achieved by plotting the PH function in the logarithmic scale. The white dots show the areas where $H_n(p,q)<10^{-16}$.  } 
  \label{fig:WfA0_typ}
\end{figure*}

\begin{figure*}
  \centering
  \includegraphics[]{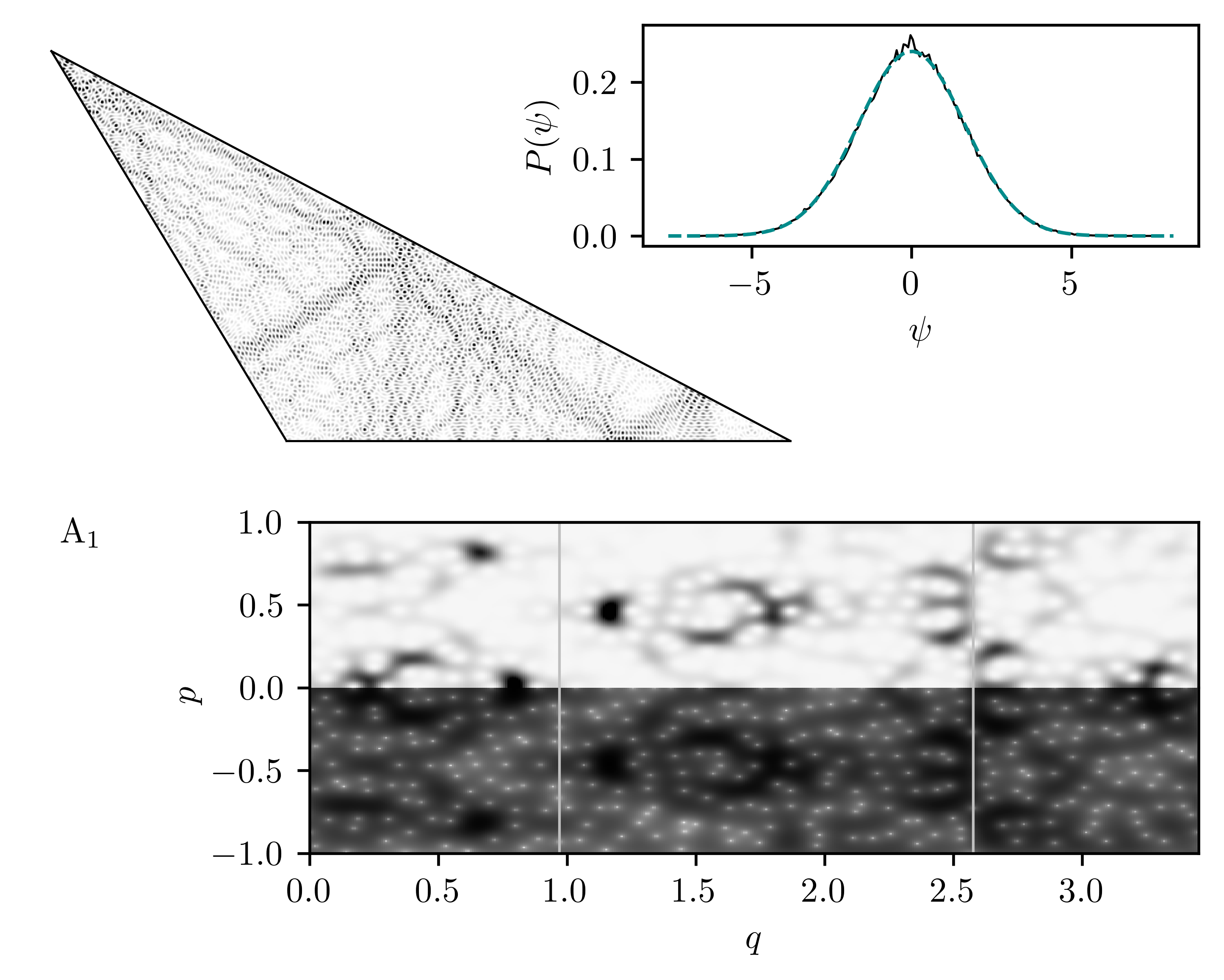}
  \caption{Moderately scarred boucing ball state of the $\mathrm{A_1}$ triangle billiard, with $k =510.3577 $ and $l^{1} = 0.406$, $l^{2} = 0.246$. See Fig. \ref{fig:WfA0_typ} for description.} 
  \label{fig:WfA1_typ}
\end{figure*}

\begin{figure*}
  \centering
  \includegraphics[]{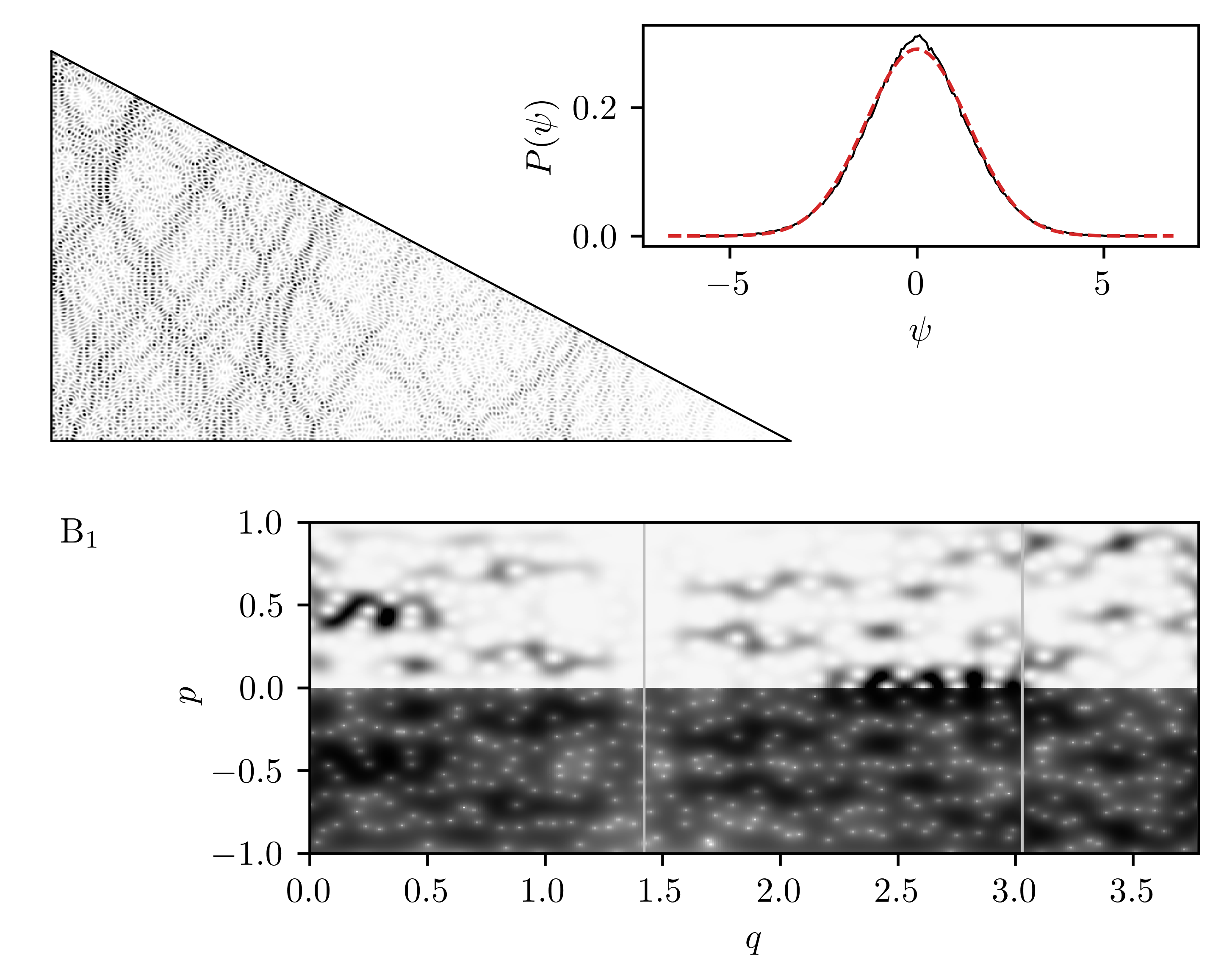}
  \caption{Typical state of the $\mathrm{B_1}$ triangle billiard, with $k = 500.0795$ and $l^{1} =0.401 $, $l^{2} =0.251 $. See Fig. \ref{fig:WfA0_typ} for description.} 
  \label{fig:WfB1_typ}
\end{figure*}

\begin{figure*}
  \centering
  \includegraphics[]{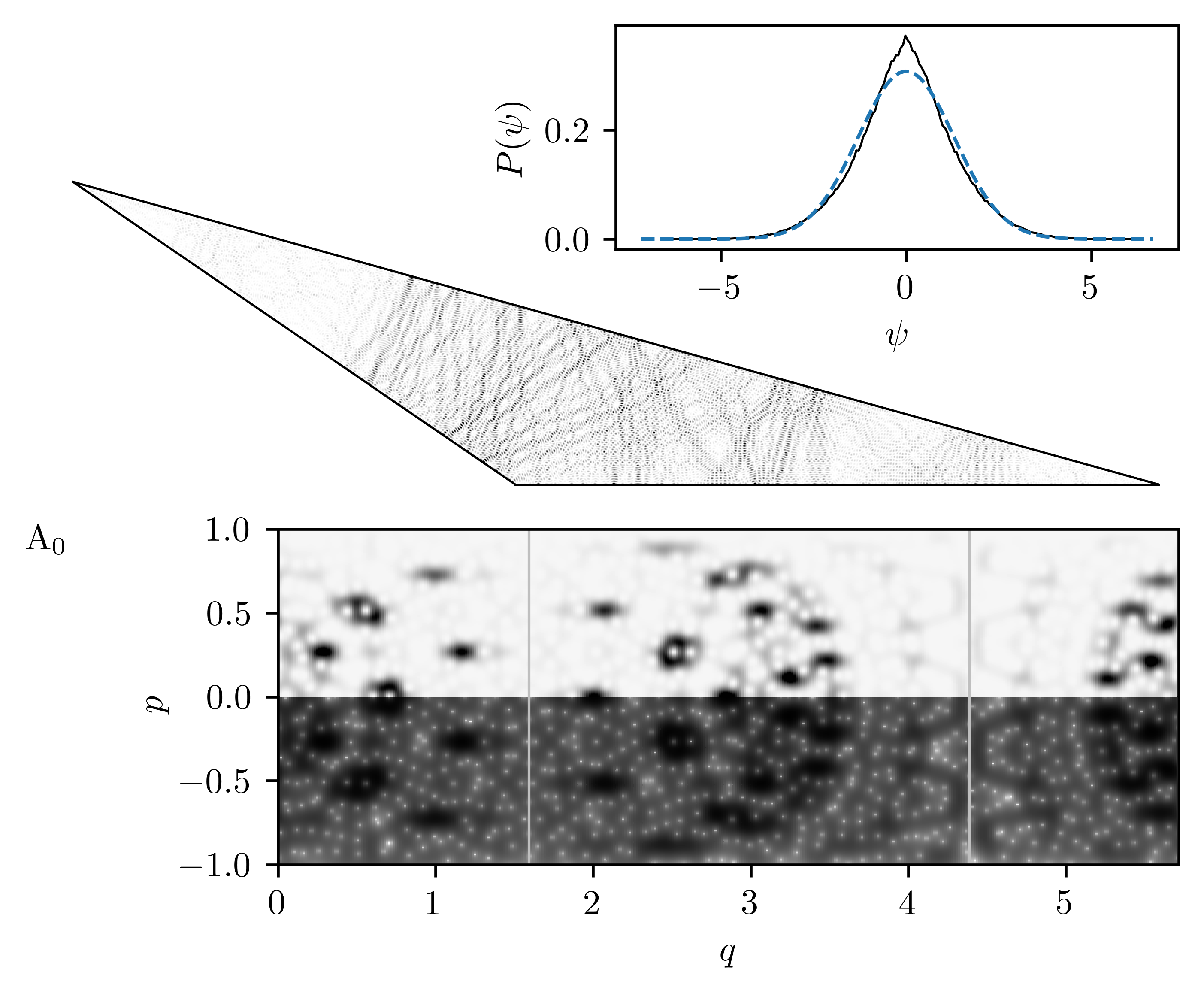}
  \caption{Severely scarred bouncing ball state of the $\mathrm{A_0}$ triangle billiard, with $k = 572.3722$ and $l^{1} =0.309 $, $l^{2} = 0.179$. See Fig. \ref{fig:WfA0_typ} for description.} 
  \label{fig:WfA0_scarr}
\end{figure*}

\begin{figure*}
  \centering
  \includegraphics[]{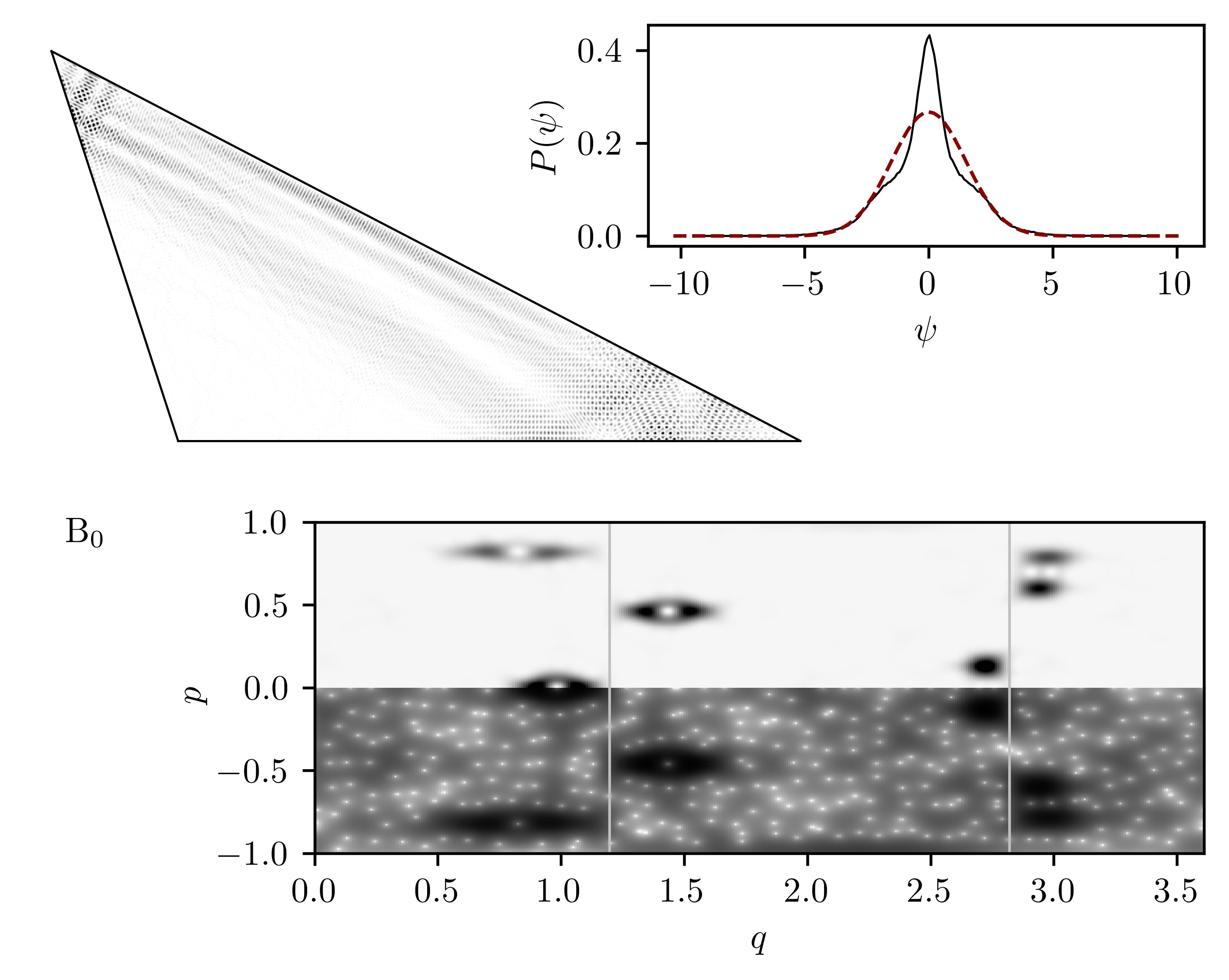}
  \caption{Severely scarred bouncing ball state of the $\mathrm{B_0}$ triangle billiard, with $k = 523.6165$ and $l^{1} =0.108 $, $l^{2} =0.0648 $. See Fig. \ref{fig:WfA0_typ} for description.} 
  \label{fig:WfB0_scarr}
\end{figure*}

\begin{figure*}
  \centering
  \includegraphics[]{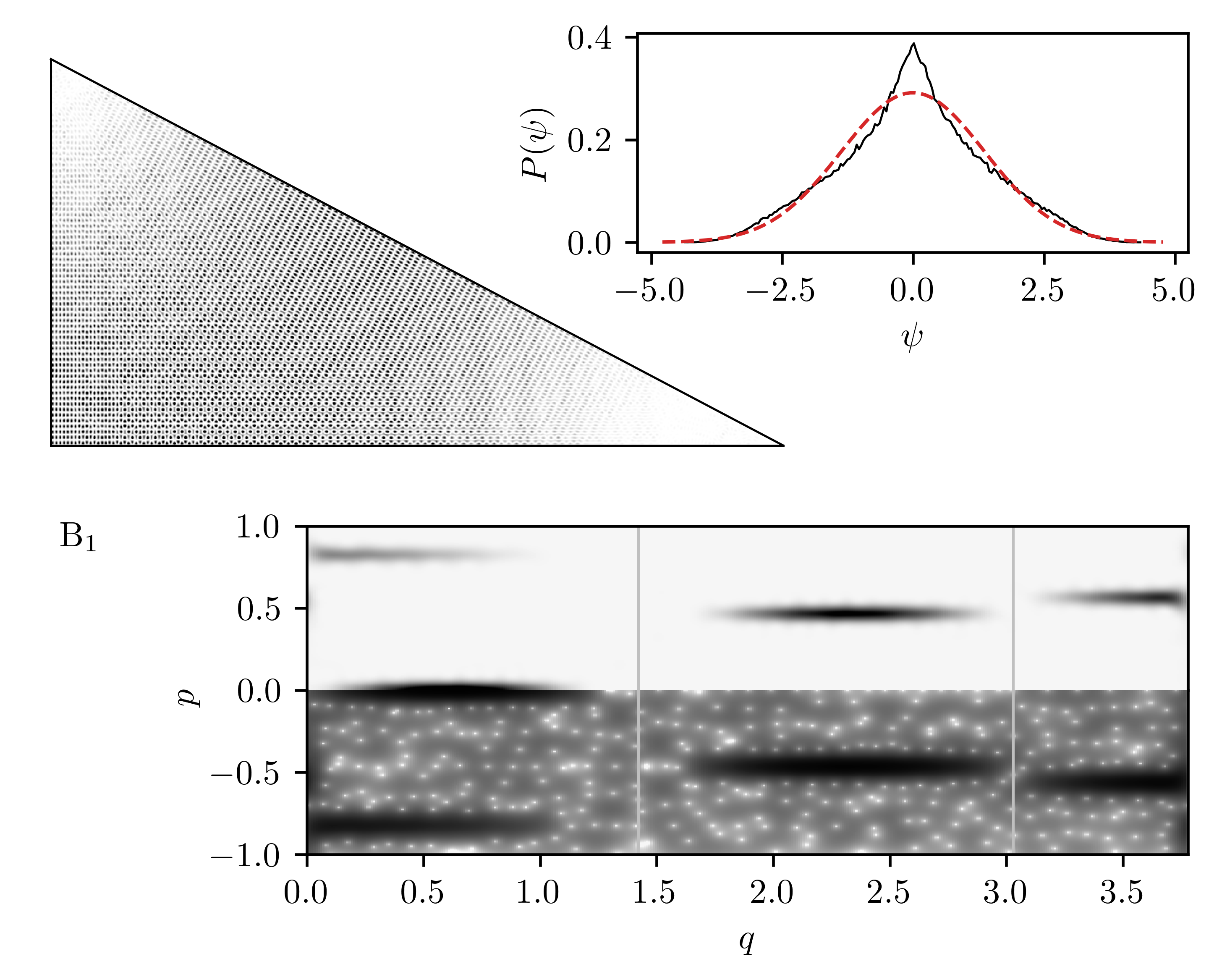}
  \caption{Severely scarred bouncing ball state of the $\mathrm{B_1}$ triangle billiard, with $k =515.3609 $ and $l^{1} =0.110 $, $l^{2} = 0.073$. See Fig. \ref{fig:WfA0_typ} for description.} 
  \label{fig:WfB1_scarr}
\end{figure*}

\begin{figure*}
  \centering
  \includegraphics[]{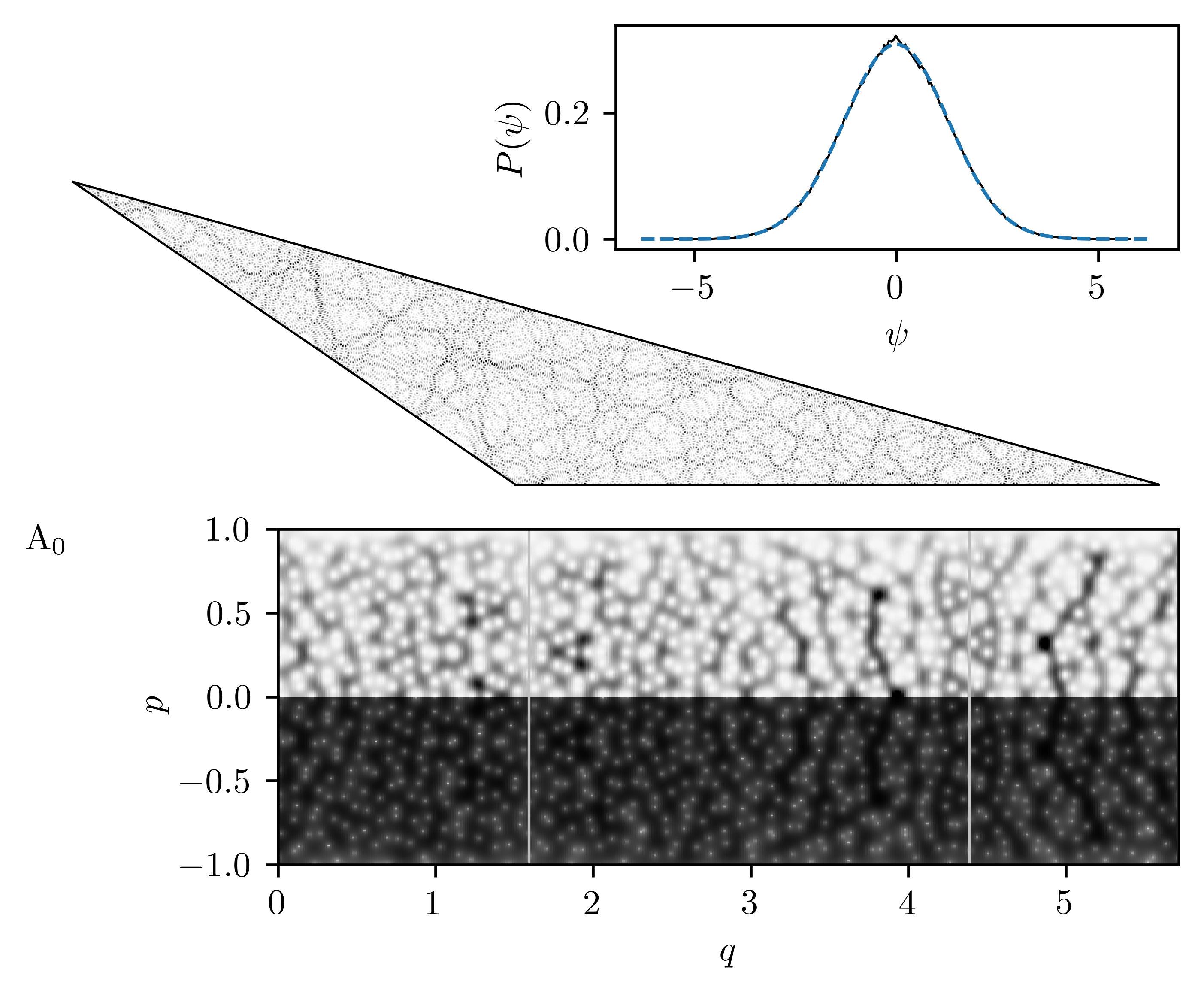}
  \caption{Uniformly ergodic state of the $\mathrm{A_0}$ triangle billiard, with $k = 593.6805$ and $l^{1} =0.693 $, $l^{2} =0.558 $. See Fig. \ref{fig:WfA0_typ} for description.} 
  \label{fig:WfA0_erg}
\end{figure*}

\begin{figure*}
  \centering
  \includegraphics[]{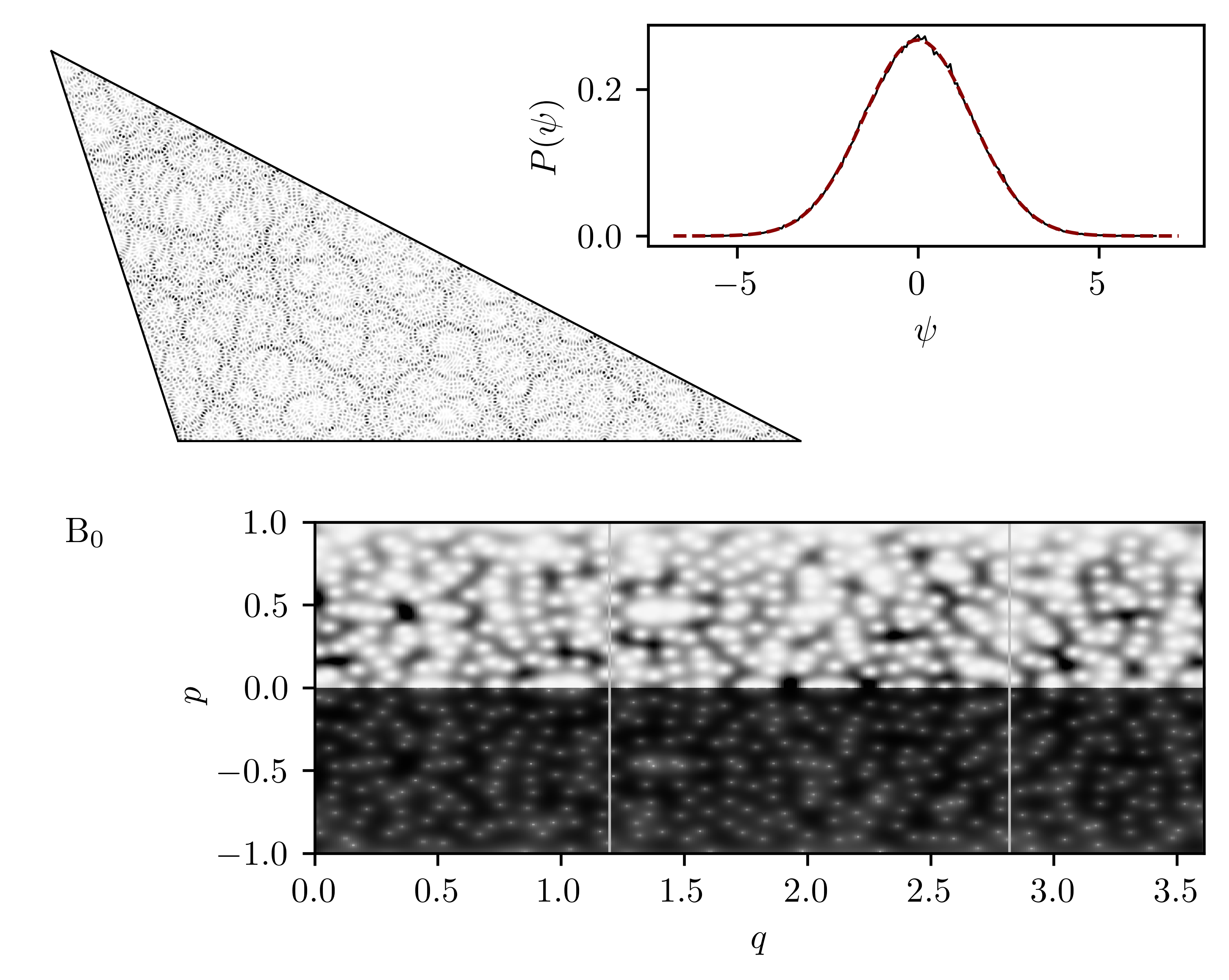}
  \caption{Uniformly ergodic state of the $\mathrm{B_0}$ triangle billiard, with $k =564.1286 $ and $l^{1} =0.696 $, $l^{2} =0.560 $. See Fig. \ref{fig:WfA0_typ} for description.} 
  \label{fig:WfB0_erg}
\end{figure*}

\begin{figure*}
  \centering
  \includegraphics[]{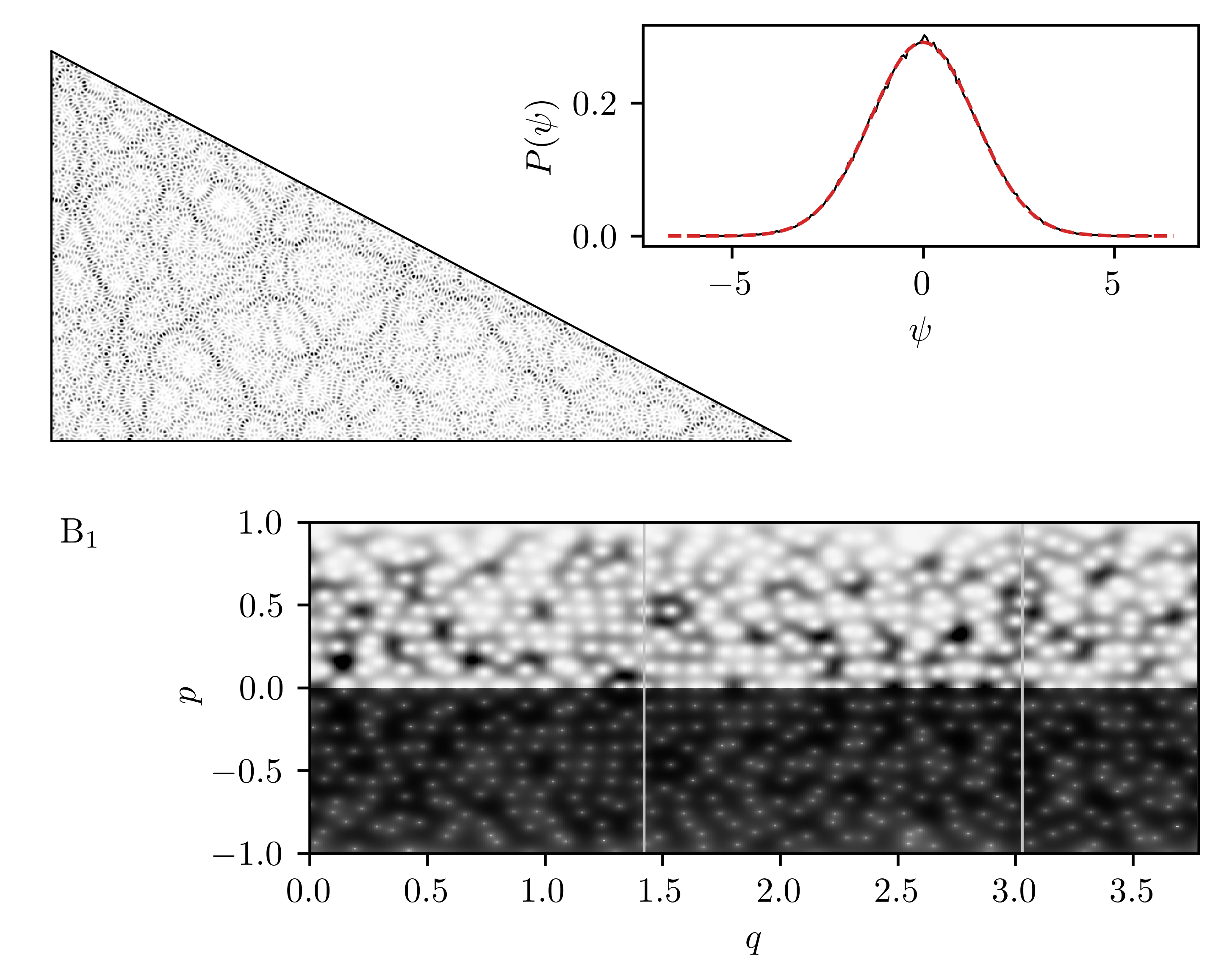}
  \caption{Uniformly ergodic state of the $\mathrm{B_1}$ triangle billiard, with $k = 504.6291$ and $l^{1} =0.699 $, $l^{2} = 0.559 $. See Fig. \ref{fig:WfA0_typ} for description.} 
  \label{fig:WfB1_erg}
\end{figure*}

Let us consider the wavefunction in the triangle billiards $\psi(\textbf{r})$, $\textbf{r} \in \mathcal{B}$ and the probability $|\psi|^2$ in the configuration space. The typical probability distributions should show the distinctive uniform nodal patterns similar to random plane waves. The distribution of the wavefunction values should be close to Gaussian. When scarring occurs, the probability is enhanced along the path of a classical unstable periodic orbit (or familiy of orbits). The semiclassical interpretation of the states is even easier, when we consider the states in the Husimi representation. Any scarring is easily observable as an enchancement (or localization) of the PH function near the classical periodic orbits. Some further information about the states may be gained also by observing the iso-contour patterns of the PH functions. Due to a deep link of Husimi functions to square-moduli of complex analytic functions of $z=q+ip$ (the so-called Bargmann representaiton of quantum states), the set of zeros -- points in phase space --- is completely characterizing the quantum state and resembles a star field, thus it is referred to as the {\em stellar representation} of quantum states \cite{tualle1995}. The number of zeros is proportional to the sequential mode number of the state, and their distributions show contrasting behaviors for regular and chaotic systems~\cite{prosen1996}. Similarly to SOS plots in classical dynamical systems, the nodes of the Husimi functions form regular ordered patterns along lines of invariant curves in regular systems and distribute themselves in more disordered patterns in chaotic systems.
We will now show a small selection of triangle eigenstates and discuss their properties. By considering the ELMs, discussed in the previous section, interesting states are easily identified. We will first show some states from the bulk of the ELM distributions, which we consider typical. These are presented in Figs. \ref{fig:WfA0_typ}, \ref{fig:WfA1_typ} and \ref{fig:WfB1_typ}. The states are mostly extended but some, scarring is visible in all three examples. The distribution $P(\psi)$ is very close to Gaussian in all three cases, with some small peaks near $\psi=0$. These small enhancements may be attributed partly to boundary effects and partly to scarring. The PH functions are mostly extended over the whole phase space, with some enchantment along the periodic orbits causing the scarring. The zeros in the stellar representation show similarly disordered patterns for each triangle, but depend on the underlying features of the classical dynamics. In Figs. \ref{fig:WfA0_scarr}, \ref{fig:WfB0_scarr} and \ref{fig:WfB1_scarr} we show some of the most heavily scarred states we observed.  We see that even in the most strongly-mixing triangle $\mathrm{A_0}$, scarring is still observable. The other two examples show states that are so severely scarred that they are localized on the classical bouncing ball invariant manifolds. In fact, these types of superscars are found in pseudointegrable systems, like for instance the right-triangles with rational corners and barrier billiards \cite{bogomolny2004, bogomolny2006barrier, bogomolny2021} and have been conjectured to exist in more general polygonal billiards. Here, we present some clear numerical evidence that they exist also in triangles with only one rational angle. Lastly, we show some uniformly ergodic extended states in Figs. \ref{fig:WfA0_erg}, \ref{fig:WfB0_erg} and \ref{fig:WfB1_erg}. These have been selected on the grounds of having ELMs above the typical values, i.e. from the right tails of the distributions. The PH functions are as close to uniformly distributed as possible for eigenstates. However, we may observe that the zeros in the stellar representation in the triangles $\mathrm{B_0}$ and $\mathrm{B_1}$ tend to order themselves in the $q$ direction (with fixed $p$). Let us remark that we have chosen to show states with relatively low energies purely for graphical convenience, and equivalent examples may be found higher in the spectrum.         

\section{Discussion and conclusions}\label{sec:Conclusion} 

We have presented an extensive numerical study of quantum chaos in two classes of triangle billiards. The first class (A) consist of generic triangles where all angles have irrational ratios with $\pi$ (with any pair of angles mutually incommensurate) and in the second class (B) exactly one of the angles has a rational ratio with $\pi$. 
While the classical dynamics of billiards in the class A is observed to be ergodic and mixing~\cite{CasPro1999tri}, the billiards in the class (B) with the right angle ($\pi/2$), specifically $B_1$ and $B_2$, are probably not even ergodic~\cite{wang2014nonergodicity}.
We examined the spectral statistics and the localization properties and the structure of the eigenstates. Our numerical study clearly confirms the long held belief that a dynamical system that has ergodic and mixing properties but is nevertheless not chaotic may still exhibit random matrix theory spectral statistics when quantized. The main example presented in this work is the generic (ergodic and mixing) triangle labeled $\mathrm{A_0}$. Both the short range statistics like the mode-fluctuations, the level spacing and the spacing ratio distributions, and the medium-to-long range statistics such as the number variance and spectral form factor are consistent with the GOE statistics, extending the quantum chaos conjecture. The eigenstates are consistent with quantum ergodicity, as shown by considering the localization measures in the Poincaré-Husimi representation. However, some small scaring effects are still observable and scarred bouncing ball states may be found even in the deep semiclassical limit. There is little doubt remaining that the triangle $\mathrm{A_0}$ is quantum ergodic.
There is the question remaining about the rate at which ergodicity manifests itself as a function of the energy. An illumination of this question might be gained by studying the matrix-elements of a characteristic function on a part of the triangle, akin to the study performed by Barnett for the Sinai billiard in Ref. \cite{barnett2006asymptotic} and is left for future work. 

The other generic triangles considered in this work, $\mathrm{A_1}$, $\mathrm{A_2}$, have been chosen to have less irrational angles (in the sense of a continued fraction expansion). The spectral statistics in these two cases are still very close to RMT. The distributions of the mode-fluctuations are Gaussian, but scaling of their variance indicates the presence of scarred bouncing ball states. The level spacing distributions are close to the asymptotic GOE prediction, in fact closer than the Wigner-Dyson approximation, but are lager than purely statistical deviations at the considered sample sizes. The number variance follows GOE for short ranges, but reveals a linear long-range regime, which is also consistent with the short time behavior of the connected spectral form factor, as it shows a small spectral compressibility. Qualitatively, we may believe that small deviations from GOE in the spectral statistics of generic triangles are mainly linked to the existence of scarred bouncing ball states, while more significant deviations in right triangle billiards are potentially linked to observed lack of ergodicity (pseudo-ergodicity), or ultra slow ergodicity. The localization measures reveal that the deviations from the empirical beta distribution in the lower tail are consistently greater in the "less irrational" triangles, owing to the existence of shorter time periodic orbits that produce a greater scarring effect. Again, it is hard to predict how deep into the semiclassical limit these scarring effects persist, but they are certainly present at $e = 2 \cdot 10^6$ which was the highest unfolded energy considered.

The situation is very similar in the class (B) triangles, with one rational angle. Indeed, the spectral statistics of the $\mathrm{B_0}$ triangle (where the rational angle is not $\pi/2$) are very similar to those of $\mathrm{A_1}$ and particularly $\mathrm{A_2}$. However, in $\mathrm{B_0}$ the scarring because of the short time periodic orbits is even more prominent as seen in the distributions of the ELMs and in particular the IPRs, which are more sensitive to scarred states. We have found examples of very severely scarred states, localized almost exclusively on the classical bouncing ball regions, that we consider strong candidates for super-scarred states. 
The heavily scarred states are even more noteworthy in the two right-triangles $\mathrm{B_1}$ and $\mathrm{B_2}$. In these two examples, one may easily find states that appear regular in almost the entire configuration space, with the exception of the small areas around the two irrational corners. The deviations from GOE spectral statistics are also considerably larger than in all the other examples. The right triangles are also the only cases where we observed deviations from the Gausssian distribution of the mode-fluctuations. The energy scaling of the variance of the mode-fluctuations is dominated by the bouncing ball states and is close to the square-root scaling expected in regular systems. The deviations of the level spacing distributions are about an order of magnitude larger than in the $\mathrm{B_0}$ case. The deviation is visible also in the averaged level spacing ratio, which gives results consistent with GOE in all other cases. The number variance only follows the GOE curve for short correlation lengths and exhibits linear regimes with varying spectral compressibility. Consistently, the SFF shows a much larger deviation as $t \rightarrow 0$ and connects with the GOE result at $t\approx0.4$. This indicates a much longer classical transport time and is consistent with the observed wider distributions of localization measures. The contributions of the bouncing ball states to the mode-fluctuations have been studied in the stadium billiard \cite{aurich1997,backer1997number}. These states introduce additional oscillatory terms in the mean of the mode fluctuations. Special unfolding procedures may then be used to compensate for these terms, however this requires an intimate knowledge of the contributing periodic orbits. The approach might be feasibly applied to the right triangles, and is left for further consideration in the future.           

Let us make some comments on the technical aspects of the analysis. Firstly, we are convinced the tiny number of missing and spurious levels is completely negligible and has no effect on the statistical results. To make sure, we have tested the robustness of the spectral statistics with regard to missing levels by uniformly randomly removing some (up to $0.1\%$, vastly more than the number of missing levels) of the levels from the spectra and computing the results.  We have found no significant changes in the spectral statistics. We also removed the first $10^5$ levels from the spectral samples to check for any non-universal contributions of the low-lying states, and found no significant change in the spectral statistics. With regard to the long-range statistics, we make the following observations.  The expectation that the number variance converges to the semiclassical limit up to $L_\mathrm{max}\propto \sqrt{e}$ is in practice hampered by the fact that the proportionality coefficient may be arbitrarily small. The SFF also requires a large spectral sample, in our experience at least $10^5$ states, to ensure that the time smoothing sufficiently eliminates the fluctuations. We tested alternative definitions of the SFF involving an additional spectral filtering by using Gaussian functions of various widths to weigh the spectra (akin to those used in Ref. \cite{suntajs2020}). These all produced identical results after the removal of the short time disconnected part of the SFF and the time smoothing. The only difference was in the amplitude of the fluctuations, which is smallest with no filtering, since the sample size is effectively larger. Regarding the distributions of localization measures, we may comment that the results are consistent with those found in chaotic billiards and ergodic components of mixed-type billiards without stickiness. The empirical beta distributions seem to provide good descriptions of the results also in the triangle billiards, with qualitatively well-understood deviations. However, the origin of this distribution is still unknown and lacks a theoretical foundation, which would facilitate a more quantitative analysis.

Lastly, we will shortly discuss some referential numerical studies preformed as part of the investigation of quantum chaos in triangular billiards but not included in the paper. Since some deviations from the RMT statistics were observed, we attempted to produce a minimal example that would eliminate them. We thus computed the spectra of the triangle billiards, $\mathrm{A_1}$, $\mathrm{A_2}$, $\mathrm{B_1}$, $\mathrm{B_2}$, with rounded corners. To elaborate, we take each triangular billiard and cut out a circular arc with radius $R=0.1$ in each of the acute angles (leaving the obtuse angle as is). The circular arcs form a right angle with the two shorter sides of the triangle and are smoothly connected with the long side. This is mainly due to technical issues with the basis employed in the numerical method. The rounded corners cause the billiard to become chaotic due to the defocusing mechanism. Numerically, no islands of stability were found. The quality of the results is much worse than the triangle study, with many more missing levels (up to $0.1\%$) but some conclusions may still be made. The rounding restores the SFF to the GOE result in "triangles" $\mathrm{A_1}$, $\mathrm{A_2}$, but a less prominent deviation remains in $\mathrm{B_1}$ and $\mathrm{B_2}$. Similarly, the NV is restored to the GOE result in $\mathrm{A_1}$, but the linear regime remains in the other cases. To explain these findings, we must consider how the rounding of the corners affects the periodic orbits in the systems. Let us consider the simplest periodic orbits that are the cause of the bouncing ball modes. We start a trajectory in the perpendicular direction from one of the edges and follow it. A bounce on the straight edge can easily be visualized by reflecting the billiard table (or unfolding) and continuing the trajectory until the next collision and repeating the process. If the collision is perpendicular, we have found a periodic orbit. Once we find a periodic orbit, it is easy to see that a whole family of bouncing ball orbits exists in the vicinity. The termination condition is when the orbit hits one of the corners, which are the source of instability. By rounding the corners we greatly increase the area of instability from a single point to a whole section of the boundary and thus eliminate many of the periodic orbits. However, in the right triangles it is easy to see that the family of bouncing ball orbits remains as long there is still a straight segment left (like in the stadium billiard). These orbits continue to cause the scarring in the eigenstates, seen in the localization measure distributions, and deviations from GOE statistics, even though this is now a chaotic billiard. 

\section{Acknowledgements}

We thank Eugene Bogomolny, Marko Robnik, and Barbara Dietz for inspiring discussions. This work has been supported by the European Research Council (ERC) under the Advanced Grant No.\ 694544 -- OMNES, and by the Slovenian Research Agency (ARRS)
under the Program P1-0402.

\bibliographystyle{apsrev4-1} 
\bibliography{references} 

\providecommand{\noopsort}[1]{}\providecommand{\singleletter}[1]{#1}%
\begin{thebibliography}{70}%
\makeatletter
\providecommand \@ifxundefined [1]{%
 \@ifx{#1\undefined}
}%
\providecommand \@ifnum [1]{%
 \ifnum #1\expandafter \@firstoftwo
 \else \expandafter \@secondoftwo
 \fi
}%
\providecommand \@ifx [1]{%
 \ifx #1\expandafter \@firstoftwo
 \else \expandafter \@secondoftwo
 \fi
}%
\providecommand \natexlab [1]{#1}%
\providecommand \enquote  [1]{``#1''}%
\providecommand \bibnamefont  [1]{#1}%
\providecommand \bibfnamefont [1]{#1}%
\providecommand \citenamefont [1]{#1}%
\providecommand \href@noop [0]{\@secondoftwo}%
\providecommand \href [0]{\begingroup \@sanitize@url \@href}%
\providecommand \@href[1]{\@@startlink{#1}\@@href}%
\providecommand \@@href[1]{\endgroup#1\@@endlink}%
\providecommand \@sanitize@url [0]{\catcode `\\12\catcode `\$12\catcode
  `\&12\catcode `\#12\catcode `\^12\catcode `\_12\catcode `\%12\relax}%
\providecommand \@@startlink[1]{}%
\providecommand \@@endlink[0]{}%
\providecommand \url  [0]{\begingroup\@sanitize@url \@url }%
\providecommand \@url [1]{\endgroup\@href {#1}{\urlprefix }}%
\providecommand \urlprefix  [0]{URL }%
\providecommand \Eprint [0]{\href }%
\providecommand \doibase [0]{http://dx.doi.org/}%
\providecommand \selectlanguage [0]{\@gobble}%
\providecommand \bibinfo  [0]{\@secondoftwo}%
\providecommand \bibfield  [0]{\@secondoftwo}%
\providecommand \translation [1]{[#1]}%
\providecommand \BibitemOpen [0]{}%
\providecommand \bibitemStop [0]{}%
\providecommand \bibitemNoStop [0]{.\EOS\space}%
\providecommand \EOS [0]{\spacefactor3000\relax}%
\providecommand \BibitemShut  [1]{\csname bibitem#1\endcsname}%
\let\auto@bib@innerbib\@empty
\bibitem [{\citenamefont {Bohigas}\ \emph {et~al.}(1984)\citenamefont
  {Bohigas}, \citenamefont {Giannoni},\ and\ \citenamefont {Schmit}}]{BGS}%
  \BibitemOpen
  \bibfield  {author} {\bibinfo {author} {\bibfnamefont {O.}~\bibnamefont
  {Bohigas}}, \bibinfo {author} {\bibfnamefont {M.-J.}\ \bibnamefont
  {Giannoni}}, \ and\ \bibinfo {author} {\bibfnamefont {C.}~\bibnamefont
  {Schmit}},\ }\href@noop {} {\bibfield  {journal} {\bibinfo  {journal}
  {Physical review letters}\ }\textbf {\bibinfo {volume} {52}},\ \bibinfo
  {pages} {1} (\bibinfo {year} {1984})}\BibitemShut {NoStop}%
\bibitem [{\citenamefont {Casati}\ \emph {et~al.}(1980)\citenamefont {Casati},
  \citenamefont {Valz-Gris},\ and\ \citenamefont {Guarnieri}}]{CVG}%
  \BibitemOpen
  \bibfield  {author} {\bibinfo {author} {\bibfnamefont {G.}~\bibnamefont
  {Casati}}, \bibinfo {author} {\bibfnamefont {F.}~\bibnamefont {Valz-Gris}}, \
  and\ \bibinfo {author} {\bibfnamefont {I.}~\bibnamefont {Guarnieri}},\
  }\href@noop {} {\bibfield  {journal} {\bibinfo  {journal} {Lettere al Nuovo
  Cimento}\ }\textbf {\bibinfo {volume} {28}},\ \bibinfo {pages} {279}
  (\bibinfo {year} {1980})}\BibitemShut {NoStop}%
\bibitem [{\citenamefont {Sieber}\ and\ \citenamefont
  {Richter}(2001)}]{Sieber2001}%
  \BibitemOpen
  \bibfield  {author} {\bibinfo {author} {\bibfnamefont {M.}~\bibnamefont
  {Sieber}}\ and\ \bibinfo {author} {\bibfnamefont {K.}~\bibnamefont
  {Richter}},\ }\href@noop {} {\bibfield  {journal} {\bibinfo  {journal} {Phys.
  Scr.}\ }\textbf {\bibinfo {volume} {T90}},\ \bibinfo {pages} {128} (\bibinfo
  {year} {2001})}\BibitemShut {NoStop}%
\bibitem [{\citenamefont {M\"uller}\ \emph {et~al.}(2004)\citenamefont
  {M\"uller}, \citenamefont {Heusler}, \citenamefont {Braun}, \citenamefont
  {Haake},\ and\ \citenamefont {Altland}}]{Mueller2004a}%
  \BibitemOpen
  \bibfield  {author} {\bibinfo {author} {\bibfnamefont {S.}~\bibnamefont
  {M\"uller}}, \bibinfo {author} {\bibfnamefont {S.}~\bibnamefont {Heusler}},
  \bibinfo {author} {\bibfnamefont {P.}~\bibnamefont {Braun}}, \bibinfo
  {author} {\bibfnamefont {F.}~\bibnamefont {Haake}}, \ and\ \bibinfo {author}
  {\bibfnamefont {A.}~\bibnamefont {Altland}},\ }\href@noop {} {\bibfield
  {journal} {\bibinfo  {journal} {Physical Review Letters}\ }\textbf {\bibinfo
  {volume} {93}},\ \bibinfo {pages} {014103} (\bibinfo {year}
  {2004})}\BibitemShut {NoStop}%
\bibitem [{\citenamefont {Heusler}\ \emph {et~al.}(2004)\citenamefont
  {Heusler}, \citenamefont {M\"uller}, \citenamefont {Braun},\ and\
  \citenamefont {Haake}}]{Mueller2004b}%
  \BibitemOpen
  \bibfield  {author} {\bibinfo {author} {\bibfnamefont {S.}~\bibnamefont
  {Heusler}}, \bibinfo {author} {\bibfnamefont {S.}~\bibnamefont {M\"uller}},
  \bibinfo {author} {\bibfnamefont {P.}~\bibnamefont {Braun}}, \ and\ \bibinfo
  {author} {\bibfnamefont {F.}~\bibnamefont {Haake}},\ }\href@noop {}
  {\bibfield  {journal} {\bibinfo  {journal} {Journal of Physics A:
  Mathematical and General}\ }\textbf {\bibinfo {volume} {37}},\ \bibinfo
  {pages} {L31} (\bibinfo {year} {2004})}\BibitemShut {NoStop}%
\bibitem [{\citenamefont {M\"uller}\ \emph {et~al.}(2005)\citenamefont
  {M\"uller}, \citenamefont {Heusler}, \citenamefont {Braun}, \citenamefont
  {Haake},\ and\ \citenamefont {Altland}}]{Mueller2005}%
  \BibitemOpen
  \bibfield  {author} {\bibinfo {author} {\bibfnamefont {S.}~\bibnamefont
  {M\"uller}}, \bibinfo {author} {\bibfnamefont {S.}~\bibnamefont {Heusler}},
  \bibinfo {author} {\bibfnamefont {P.}~\bibnamefont {Braun}}, \bibinfo
  {author} {\bibfnamefont {F.}~\bibnamefont {Haake}}, \ and\ \bibinfo {author}
  {\bibfnamefont {A.}~\bibnamefont {Altland}},\ }\href@noop {} {\bibfield
  {journal} {\bibinfo  {journal} {Physical Review E}\ }\textbf {\bibinfo
  {volume} {72}},\ \bibinfo {pages} {046207} (\bibinfo {year}
  {2005})}\BibitemShut {NoStop}%
\bibitem [{\citenamefont {Berry}\ and\ \citenamefont
  {Tabor}(1977)}]{BerryTabor}%
  \BibitemOpen
  \bibfield  {author} {\bibinfo {author} {\bibfnamefont {M.~V.}\ \bibnamefont
  {Berry}}\ and\ \bibinfo {author} {\bibfnamefont {M.}~\bibnamefont {Tabor}},\
  }\href@noop {} {\bibfield  {journal} {\bibinfo  {journal} {Proceedings of the
  Royal Society of London. A. Mathematical and Physical Sciences}\ }\textbf
  {\bibinfo {volume} {356}},\ \bibinfo {pages} {375} (\bibinfo {year}
  {1977})}\BibitemShut {NoStop}%
\bibitem [{\citenamefont {Berkovitz}\ \emph {et~al.}(2006)\citenamefont
  {Berkovitz}, \citenamefont {Frigg},\ and\ \citenamefont {Kronz}}]{EH}%
  \BibitemOpen
  \bibfield  {author} {\bibinfo {author} {\bibfnamefont {J.}~\bibnamefont
  {Berkovitz}}, \bibinfo {author} {\bibfnamefont {R.}~\bibnamefont {Frigg}}, \
  and\ \bibinfo {author} {\bibfnamefont {F.}~\bibnamefont {Kronz}},\
  }\href@noop {} {\bibfield  {journal} {\bibinfo  {journal} {Studies in History
  and Philosophy of Science Part B: Studies in History and Philosophy of Modern
  Physics}\ }\textbf {\bibinfo {volume} {37}},\ \bibinfo {pages} {661}
  (\bibinfo {year} {2006})}\BibitemShut {NoStop}%
\bibitem [{\citenamefont {Gutkin}(1986)}]{gutkin1986}%
  \BibitemOpen
  \bibfield  {author} {\bibinfo {author} {\bibfnamefont {E.}~\bibnamefont
  {Gutkin}},\ }\href@noop {} {\bibfield  {journal} {\bibinfo  {journal}
  {Physica D: Nonlinear Phenomena}\ }\textbf {\bibinfo {volume} {19}},\
  \bibinfo {pages} {311} (\bibinfo {year} {1986})}\BibitemShut {NoStop}%
\bibitem [{\citenamefont {Gutkin}(1996)}]{gutkin1996}%
  \BibitemOpen
  \bibfield  {author} {\bibinfo {author} {\bibfnamefont {E.}~\bibnamefont
  {Gutkin}},\ }\href@noop {} {\bibfield  {journal} {\bibinfo  {journal}
  {Journal of statistical physics}\ }\textbf {\bibinfo {volume} {83}},\
  \bibinfo {pages} {7} (\bibinfo {year} {1996})}\BibitemShut {NoStop}%
\bibitem [{\citenamefont {Schwartz}(2009)}]{schwartz2009}%
  \BibitemOpen
  \bibfield  {author} {\bibinfo {author} {\bibfnamefont {R.~E.}\ \bibnamefont
  {Schwartz}},\ }\href@noop {} {\bibfield  {journal} {\bibinfo  {journal}
  {Experimental Mathematics}\ }\textbf {\bibinfo {volume} {18}},\ \bibinfo
  {pages} {137} (\bibinfo {year} {2009})}\BibitemShut {NoStop}%
\bibitem [{\citenamefont {Casati}\ and\ \citenamefont
  {Prosen}(1999)}]{CasPro1999tri}%
  \BibitemOpen
  \bibfield  {author} {\bibinfo {author} {\bibfnamefont {G.}~\bibnamefont
  {Casati}}\ and\ \bibinfo {author} {\bibfnamefont {T.}~\bibnamefont
  {Prosen}},\ }\href@noop {} {\bibfield  {journal} {\bibinfo  {journal}
  {Physical Review Letters}\ }\textbf {\bibinfo {volume} {83}},\ \bibinfo
  {pages} {4729} (\bibinfo {year} {1999})}\BibitemShut {NoStop}%
\bibitem [{\citenamefont {Artuso}\ \emph {et~al.}(1997)\citenamefont {Artuso},
  \citenamefont {Casati},\ and\ \citenamefont {Guarneri}}]{ArtCasGua1997}%
  \BibitemOpen
  \bibfield  {author} {\bibinfo {author} {\bibfnamefont {R.}~\bibnamefont
  {Artuso}}, \bibinfo {author} {\bibfnamefont {G.}~\bibnamefont {Casati}}, \
  and\ \bibinfo {author} {\bibfnamefont {I.}~\bibnamefont {Guarneri}},\
  }\href@noop {} {\bibfield  {journal} {\bibinfo  {journal} {Physical Review
  E}\ }\textbf {\bibinfo {volume} {55}},\ \bibinfo {pages} {6384} (\bibinfo
  {year} {1997})}\BibitemShut {NoStop}%
\bibitem [{\citenamefont {Wang}\ \emph {et~al.}(2014)\citenamefont {Wang},
  \citenamefont {Casati},\ and\ \citenamefont
  {Prosen}}]{wang2014nonergodicity}%
  \BibitemOpen
  \bibfield  {author} {\bibinfo {author} {\bibfnamefont {J.}~\bibnamefont
  {Wang}}, \bibinfo {author} {\bibfnamefont {G.}~\bibnamefont {Casati}}, \ and\
  \bibinfo {author} {\bibfnamefont {T.}~\bibnamefont {Prosen}},\ }\href@noop {}
  {\bibfield  {journal} {\bibinfo  {journal} {Physical Review E}\ }\textbf
  {\bibinfo {volume} {89}},\ \bibinfo {pages} {042918} (\bibinfo {year}
  {2014})}\BibitemShut {NoStop}%
\bibitem [{\citenamefont {Huang}\ and\ \citenamefont {Zhao}(2017)}]{huang}%
  \BibitemOpen
  \bibfield  {author} {\bibinfo {author} {\bibfnamefont {J.}~\bibnamefont
  {Huang}}\ and\ \bibinfo {author} {\bibfnamefont {H.}~\bibnamefont {Zhao}},\
  }\href {\doibase 10.1103/PhysRevE.95.032209} {\bibfield  {journal} {\bibinfo
  {journal} {Phys. Rev. E}\ }\textbf {\bibinfo {volume} {95}},\ \bibinfo
  {pages} {032209} (\bibinfo {year} {2017})}\BibitemShut {NoStop}%
\bibitem [{Note1()}]{Note1}%
  \BibitemOpen
  \bibinfo {note} {We define the system as pseudo-ergodic, if a fixed typical
  trajectory (for almost any initial condition) comes arbitrary close to any
  point in phase space, but the corresponding phase-space measure is not flat
  (as observed in ~\cite {wang2014nonergodicity}).}\BibitemShut {Stop}%
\bibitem [{\citenamefont {Bogomolny}\ and\ \citenamefont
  {Schmit}(2004)}]{bogomolny2004}%
  \BibitemOpen
  \bibfield  {author} {\bibinfo {author} {\bibfnamefont {E.}~\bibnamefont
  {Bogomolny}}\ and\ \bibinfo {author} {\bibfnamefont {C.}~\bibnamefont
  {Schmit}},\ }\href@noop {} {\bibfield  {journal} {\bibinfo  {journal}
  {Physical Review Letters}\ }\textbf {\bibinfo {volume} {92}},\ \bibinfo
  {pages} {244102} (\bibinfo {year} {2004})}\BibitemShut {NoStop}%
\bibitem [{\citenamefont {Bogomolny}(2021)}]{bogomolny2021}%
  \BibitemOpen
  \bibfield  {author} {\bibinfo {author} {\bibfnamefont {E.}~\bibnamefont
  {Bogomolny}},\ }\href@noop {} {\bibfield  {journal} {\bibinfo  {journal}
  {Journal of Physics Communications}\ }\textbf {\bibinfo {volume} {5}},\
  \bibinfo {pages} {055010} (\bibinfo {year} {2021})}\BibitemShut {NoStop}%
\bibitem [{\citenamefont {Richens}\ and\ \citenamefont
  {Berry}(1981)}]{richens1981}%
  \BibitemOpen
  \bibfield  {author} {\bibinfo {author} {\bibfnamefont {P.}~\bibnamefont
  {Richens}}\ and\ \bibinfo {author} {\bibfnamefont {M.}~\bibnamefont
  {Berry}},\ }\href@noop {} {\bibfield  {journal} {\bibinfo  {journal} {Physica
  D: Nonlinear Phenomena}\ }\textbf {\bibinfo {volume} {2}},\ \bibinfo {pages}
  {495} (\bibinfo {year} {1981})}\BibitemShut {NoStop}%
\bibitem [{\citenamefont {Vergini}\ and\ \citenamefont
  {Saraceno}(1995)}]{VerSer1995}%
  \BibitemOpen
  \bibfield  {author} {\bibinfo {author} {\bibfnamefont {E.}~\bibnamefont
  {Vergini}}\ and\ \bibinfo {author} {\bibfnamefont {M.}~\bibnamefont
  {Saraceno}},\ }\href@noop {} {\bibfield  {journal} {\bibinfo  {journal}
  {Phys. Rev. E}\ }\textbf {\bibinfo {volume} {52}},\ \bibinfo {pages} {2204}
  (\bibinfo {year} {1995})}\BibitemShut {NoStop}%
\bibitem [{\citenamefont {Barnett}(2001)}]{BarnettPHD}%
  \BibitemOpen
  \bibfield  {author} {\bibinfo {author} {\bibfnamefont {A.}~\bibnamefont
  {Barnett}},\ }\emph {\bibinfo {title} {Dissipation in Deforming Chaotic
  Billiards}},\ \href@noop {} {Ph.D. thesis},\ \bibinfo  {school} {Harvard
  University} (\bibinfo {year} {2001})\BibitemShut {NoStop}%
\bibitem [{\citenamefont {Barnett}\ and\ \citenamefont
  {Betcke}(2007)}]{BarBet2007}%
  \BibitemOpen
  \bibfield  {author} {\bibinfo {author} {\bibfnamefont {A.~H.}\ \bibnamefont
  {Barnett}}\ and\ \bibinfo {author} {\bibfnamefont {T.}~\bibnamefont
  {Betcke}},\ }\href@noop {} {\bibfield  {journal} {\bibinfo  {journal} {Chaos:
  An Interdisciplinary Journal of Nonlinear Science}\ }\textbf {\bibinfo
  {volume} {17}},\ \bibinfo {pages} {043125} (\bibinfo {year}
  {2007})}\BibitemShut {NoStop}%
\bibitem [{\citenamefont {Lozej}\ \emph {et~al.}()\citenamefont {Lozej},
  \citenamefont {Batisti{\'c}},\ and\ \citenamefont
  {Lukman}}]{QuantumBilliards}%
  \BibitemOpen
  \bibfield  {author} {\bibinfo {author} {\bibfnamefont {{\v C}.}~\bibnamefont
  {Lozej}}, \bibinfo {author} {\bibfnamefont {B.}~\bibnamefont {Batisti{\'c}}},
  \ and\ \bibinfo {author} {\bibfnamefont {D.}~\bibnamefont {Lukman}},\
  }\href@noop {} {\enquote {\bibinfo {title} {Quantum billiards},}\ }\bibinfo
  {note} {Available at
  https://github.com/clozej/quantum-billiards/tree/crt{\_}public.}\BibitemShut
  {Stop}%
\bibitem [{\citenamefont {Lima}\ \emph {et~al.}(2013)\citenamefont {Lima},
  \citenamefont {Rodr{\'\i}guez-P{\'e}rez},\ and\ \citenamefont
  {de~Aguiar}}]{lima2013}%
  \BibitemOpen
  \bibfield  {author} {\bibinfo {author} {\bibfnamefont {T.~A.}\ \bibnamefont
  {Lima}}, \bibinfo {author} {\bibfnamefont {S.}~\bibnamefont
  {Rodr{\'\i}guez-P{\'e}rez}}, \ and\ \bibinfo {author} {\bibfnamefont
  {F.}~\bibnamefont {de~Aguiar}},\ }\href@noop {} {\bibfield  {journal}
  {\bibinfo  {journal} {Physical Review E}\ }\textbf {\bibinfo {volume} {87}},\
  \bibinfo {pages} {062902} (\bibinfo {year} {2013})}\BibitemShut {NoStop}%
\bibitem [{\citenamefont {Baltes}\ and\ \citenamefont {Hilf}(1976)}]{Hilf}%
  \BibitemOpen
  \bibfield  {author} {\bibinfo {author} {\bibfnamefont {H.~P.}\ \bibnamefont
  {Baltes}}\ and\ \bibinfo {author} {\bibfnamefont {E.~R.}\ \bibnamefont
  {Hilf}},\ }\href@noop {} {\emph {\bibinfo {title} {Spectra of finite
  systems}}}\ (\bibinfo {year} {Mannheim: BI-Wissenschaftsverlag,
  1976})\BibitemShut {NoStop}%
\bibitem [{\citenamefont {Aurich}\ \emph {et~al.}(1994)\citenamefont {Aurich},
  \citenamefont {Bolte},\ and\ \citenamefont {Steiner}}]{aurich1994}%
  \BibitemOpen
  \bibfield  {author} {\bibinfo {author} {\bibfnamefont {R.}~\bibnamefont
  {Aurich}}, \bibinfo {author} {\bibfnamefont {J.}~\bibnamefont {Bolte}}, \
  and\ \bibinfo {author} {\bibfnamefont {F.}~\bibnamefont {Steiner}},\
  }\href@noop {} {\bibfield  {journal} {\bibinfo  {journal} {Physical Review
  Letters}\ }\textbf {\bibinfo {volume} {73}},\ \bibinfo {pages} {1356}
  (\bibinfo {year} {1994})}\BibitemShut {NoStop}%
\bibitem [{\citenamefont {Aurich}\ \emph {et~al.}(1997)\citenamefont {Aurich},
  \citenamefont {B{\"a}cker},\ and\ \citenamefont {Steiner}}]{aurich1997}%
  \BibitemOpen
  \bibfield  {author} {\bibinfo {author} {\bibfnamefont {R.}~\bibnamefont
  {Aurich}}, \bibinfo {author} {\bibfnamefont {A.}~\bibnamefont {B{\"a}cker}},
  \ and\ \bibinfo {author} {\bibfnamefont {F.}~\bibnamefont {Steiner}},\
  }\href@noop {} {\bibfield  {journal} {\bibinfo  {journal} {International
  Journal of Modern Physics B}\ }\textbf {\bibinfo {volume} {11}},\ \bibinfo
  {pages} {805} (\bibinfo {year} {1997})}\BibitemShut {NoStop}%
\bibitem [{\citenamefont {Berry}(1985)}]{berry1985}%
  \BibitemOpen
  \bibfield  {author} {\bibinfo {author} {\bibfnamefont {M.~V.}\ \bibnamefont
  {Berry}},\ }\href@noop {} {\bibfield  {journal} {\bibinfo  {journal}
  {Proceedings of the Royal Society of London. A. Mathematical and Physical
  Sciences}\ }\textbf {\bibinfo {volume} {400}},\ \bibinfo {pages} {229}
  (\bibinfo {year} {1985})}\BibitemShut {NoStop}%
\bibitem [{\citenamefont {B{\"a}cker}\ and\ \citenamefont
  {Schubert}(2002)}]{backer2002autocorrelation}%
  \BibitemOpen
  \bibfield  {author} {\bibinfo {author} {\bibfnamefont {A.}~\bibnamefont
  {B{\"a}cker}}\ and\ \bibinfo {author} {\bibfnamefont {R.}~\bibnamefont
  {Schubert}},\ }\href@noop {} {\bibfield  {journal} {\bibinfo  {journal}
  {Journal of Physics A: Mathematical and General}\ }\textbf {\bibinfo {volume}
  {35}},\ \bibinfo {pages} {539} (\bibinfo {year} {2002})}\BibitemShut
  {NoStop}%
\bibitem [{\citenamefont {Alt}\ \emph {et~al.}(1998)\citenamefont {Alt},
  \citenamefont {B{\"a}cker}, \citenamefont {Dembowski}, \citenamefont
  {Gr{\"a}f}, \citenamefont {Hofferbert}, \citenamefont {Rehfeld},\ and\
  \citenamefont {Richter}}]{alt1998mode}%
  \BibitemOpen
  \bibfield  {author} {\bibinfo {author} {\bibfnamefont {H.}~\bibnamefont
  {Alt}}, \bibinfo {author} {\bibfnamefont {A.}~\bibnamefont {B{\"a}cker}},
  \bibinfo {author} {\bibfnamefont {C.}~\bibnamefont {Dembowski}}, \bibinfo
  {author} {\bibfnamefont {H.-D.}\ \bibnamefont {Gr{\"a}f}}, \bibinfo {author}
  {\bibfnamefont {R.}~\bibnamefont {Hofferbert}}, \bibinfo {author}
  {\bibfnamefont {H.}~\bibnamefont {Rehfeld}}, \ and\ \bibinfo {author}
  {\bibfnamefont {A.}~\bibnamefont {Richter}},\ }\href@noop {} {\bibfield
  {journal} {\bibinfo  {journal} {Physical Review E}\ }\textbf {\bibinfo
  {volume} {58}},\ \bibinfo {pages} {1737} (\bibinfo {year}
  {1998})}\BibitemShut {NoStop}%
\bibitem [{Note2()}]{Note2}%
  \BibitemOpen
  \bibinfo {note} {Thereby generalizing the term `bouncing ball' which is
  usually used only in reference to orbits bouncing between parallel
  walls}\BibitemShut {NoStop}%
\bibitem [{\citenamefont {Batisti{\'c}}\ and\ \citenamefont
  {Robnik}(2013)}]{batistic2013dynamical}%
  \BibitemOpen
  \bibfield  {author} {\bibinfo {author} {\bibfnamefont {B.}~\bibnamefont
  {Batisti{\'c}}}\ and\ \bibinfo {author} {\bibfnamefont {M.}~\bibnamefont
  {Robnik}},\ }\href@noop {} {\bibfield  {journal} {\bibinfo  {journal}
  {Journal of Physics A: Mathematical and Theoretical}\ }\textbf {\bibinfo
  {volume} {46}},\ \bibinfo {pages} {315102} (\bibinfo {year}
  {2013})}\BibitemShut {NoStop}%
\bibitem [{\citenamefont {Dietz}\ and\ \citenamefont
  {Haake}(1990)}]{dietz1990taylor}%
  \BibitemOpen
  \bibfield  {author} {\bibinfo {author} {\bibfnamefont {B.}~\bibnamefont
  {Dietz}}\ and\ \bibinfo {author} {\bibfnamefont {F.}~\bibnamefont {Haake}},\
  }\href@noop {} {\bibfield  {journal} {\bibinfo  {journal} {Zeitschrift
  f{\"u}r Physik B Condensed Matter}\ }\textbf {\bibinfo {volume} {80}},\
  \bibinfo {pages} {153} (\bibinfo {year} {1990})}\BibitemShut {NoStop}%
\bibitem [{\citenamefont {B{\"a}cker}\ \emph {et~al.}(1997)\citenamefont
  {B{\"a}cker}, \citenamefont {Schubert},\ and\ \citenamefont
  {Stifter}}]{backer1997number}%
  \BibitemOpen
  \bibfield  {author} {\bibinfo {author} {\bibfnamefont {A.}~\bibnamefont
  {B{\"a}cker}}, \bibinfo {author} {\bibfnamefont {R.}~\bibnamefont
  {Schubert}}, \ and\ \bibinfo {author} {\bibfnamefont {P.}~\bibnamefont
  {Stifter}},\ }\href@noop {} {\bibfield  {journal} {\bibinfo  {journal}
  {Journal of Physics A: Mathematical and General}\ }\textbf {\bibinfo {volume}
  {30}},\ \bibinfo {pages} {6783} (\bibinfo {year} {1997})}\BibitemShut
  {NoStop}%
\bibitem [{\citenamefont {Atas}\ \emph {et~al.}(2013)\citenamefont {Atas},
  \citenamefont {Bogomolny}, \citenamefont {Giraud},\ and\ \citenamefont
  {Roux}}]{atas2013}%
  \BibitemOpen
  \bibfield  {author} {\bibinfo {author} {\bibfnamefont {Y.~Y.}\ \bibnamefont
  {Atas}}, \bibinfo {author} {\bibfnamefont {E.}~\bibnamefont {Bogomolny}},
  \bibinfo {author} {\bibfnamefont {O.}~\bibnamefont {Giraud}}, \ and\ \bibinfo
  {author} {\bibfnamefont {G.}~\bibnamefont {Roux}},\ }\href@noop {} {\bibfield
   {journal} {\bibinfo  {journal} {Physical Review Letters}\ }\textbf {\bibinfo
  {volume} {110}},\ \bibinfo {pages} {084101} (\bibinfo {year}
  {2013})}\BibitemShut {NoStop}%
\bibitem [{\citenamefont {B{\"a}cker}\ \emph {et~al.}(1995)\citenamefont
  {B{\"a}cker}, \citenamefont {Steiner},\ and\ \citenamefont
  {Stifter}}]{backer1995spectral}%
  \BibitemOpen
  \bibfield  {author} {\bibinfo {author} {\bibfnamefont {A.}~\bibnamefont
  {B{\"a}cker}}, \bibinfo {author} {\bibfnamefont {F.}~\bibnamefont {Steiner}},
  \ and\ \bibinfo {author} {\bibfnamefont {P.}~\bibnamefont {Stifter}},\
  }\href@noop {} {\bibfield  {journal} {\bibinfo  {journal} {Physical Review
  E}\ }\textbf {\bibinfo {volume} {52}},\ \bibinfo {pages} {2463} (\bibinfo
  {year} {1995})}\BibitemShut {NoStop}%
\bibitem [{\citenamefont {Heusler}\ \emph {et~al.}(2007)\citenamefont
  {Heusler}, \citenamefont {M\"uller}, \citenamefont {Altland}, \citenamefont
  {Braun},\ and\ \citenamefont {Haake}}]{Heustler2007}%
  \BibitemOpen
  \bibfield  {author} {\bibinfo {author} {\bibfnamefont {S.}~\bibnamefont
  {Heusler}}, \bibinfo {author} {\bibfnamefont {S.}~\bibnamefont {M\"uller}},
  \bibinfo {author} {\bibfnamefont {A.}~\bibnamefont {Altland}}, \bibinfo
  {author} {\bibfnamefont {P.}~\bibnamefont {Braun}}, \ and\ \bibinfo {author}
  {\bibfnamefont {F.}~\bibnamefont {Haake}},\ }\href@noop {} {\bibfield
  {journal} {\bibinfo  {journal} {Physical Review Letters}\ }\textbf {\bibinfo
  {volume} {98}},\ \bibinfo {pages} {044103} (\bibinfo {year}
  {2007})}\BibitemShut {NoStop}%
\bibitem [{\citenamefont {M\"uller}\ \emph {et~al.}(2009)\citenamefont
  {M\"uller}, \citenamefont {Heusler}, \citenamefont {Altland}, \citenamefont
  {Braun},\ and\ \citenamefont {Haake}}]{Mueller2009}%
  \BibitemOpen
  \bibfield  {author} {\bibinfo {author} {\bibfnamefont {S.}~\bibnamefont
  {M\"uller}}, \bibinfo {author} {\bibfnamefont {S.}~\bibnamefont {Heusler}},
  \bibinfo {author} {\bibfnamefont {A.}~\bibnamefont {Altland}}, \bibinfo
  {author} {\bibfnamefont {P.}~\bibnamefont {Braun}}, \ and\ \bibinfo {author}
  {\bibfnamefont {F.}~\bibnamefont {Haake}},\ }\href@noop {} {\bibfield
  {journal} {\bibinfo  {journal} {New Journal of Physics}\ }\textbf {\bibinfo
  {volume} {11}},\ \bibinfo {pages} {103025} (\bibinfo {year}
  {2009})}\BibitemShut {NoStop}%
\bibitem [{\citenamefont {Bertini}\ \emph {et~al.}(2018)\citenamefont
  {Bertini}, \citenamefont {Kos},\ and\ \citenamefont
  {Prosen}}]{bertini2018exact}%
  \BibitemOpen
  \bibfield  {author} {\bibinfo {author} {\bibfnamefont {B.}~\bibnamefont
  {Bertini}}, \bibinfo {author} {\bibfnamefont {P.}~\bibnamefont {Kos}}, \ and\
  \bibinfo {author} {\bibfnamefont {T.}~\bibnamefont {Prosen}},\ }\href@noop {}
  {\bibfield  {journal} {\bibinfo  {journal} {Physical Review Letters}\
  }\textbf {\bibinfo {volume} {121}},\ \bibinfo {pages} {264101} (\bibinfo
  {year} {2018})}\BibitemShut {NoStop}%
\bibitem [{\citenamefont {Bertini}\ \emph {et~al.}(2021)\citenamefont
  {Bertini}, \citenamefont {Kos},\ and\ \citenamefont
  {Prosen}}]{bertini2021cmp}%
  \BibitemOpen
  \bibfield  {author} {\bibinfo {author} {\bibfnamefont {B.}~\bibnamefont
  {Bertini}}, \bibinfo {author} {\bibfnamefont {P.}~\bibnamefont {Kos}}, \ and\
  \bibinfo {author} {\bibfnamefont {T.}~\bibnamefont {Prosen}},\ }\href@noop {}
  {\bibfield  {journal} {\bibinfo  {journal} {Communications in Mathematical
  Physics}\ ,\ \bibinfo {pages} {1}} (\bibinfo {year} {2021})}\BibitemShut
  {NoStop}%
\bibitem [{\citenamefont {Prange}(1997)}]{prange1997spectral}%
  \BibitemOpen
  \bibfield  {author} {\bibinfo {author} {\bibfnamefont {R.}~\bibnamefont
  {Prange}},\ }\href@noop {} {\bibfield  {journal} {\bibinfo  {journal}
  {Physical Review Letters}\ }\textbf {\bibinfo {volume} {78}},\ \bibinfo
  {pages} {2280} (\bibinfo {year} {1997})}\BibitemShut {NoStop}%
\bibitem [{\citenamefont {Kaplan}\ and\ \citenamefont
  {Heller}(1998)}]{kaplan1998}%
  \BibitemOpen
  \bibfield  {author} {\bibinfo {author} {\bibfnamefont {L.}~\bibnamefont
  {Kaplan}}\ and\ \bibinfo {author} {\bibfnamefont {E.~J.}\ \bibnamefont
  {Heller}},\ }\href@noop {} {\bibfield  {journal} {\bibinfo  {journal}
  {Physica D: Nonlinear Phenomena}\ }\textbf {\bibinfo {volume} {121}},\
  \bibinfo {pages} {1} (\bibinfo {year} {1998})}\BibitemShut {NoStop}%
\bibitem [{\citenamefont {Horvat}\ \emph {et~al.}(2009)\citenamefont {Horvat},
  \citenamefont {Degli~Esposti}, \citenamefont {Isola}, \citenamefont
  {Prosen},\ and\ \citenamefont {Bunimovich}}]{horvat}%
  \BibitemOpen
  \bibfield  {author} {\bibinfo {author} {\bibfnamefont {M.}~\bibnamefont
  {Horvat}}, \bibinfo {author} {\bibfnamefont {M.}~\bibnamefont
  {Degli~Esposti}}, \bibinfo {author} {\bibfnamefont {S.}~\bibnamefont
  {Isola}}, \bibinfo {author} {\bibfnamefont {T.}~\bibnamefont {Prosen}}, \
  and\ \bibinfo {author} {\bibfnamefont {L.}~\bibnamefont {Bunimovich}},\
  }\href@noop {} {\bibfield  {journal} {\bibinfo  {journal} {Physica D:
  Nonlinear Phenomena}\ }\textbf {\bibinfo {volume} {238}},\ \bibinfo {pages}
  {395} (\bibinfo {year} {2009})}\BibitemShut {NoStop}%
\bibitem [{\citenamefont {Gutkin}(2003)}]{gutkin2003review}%
  \BibitemOpen
  \bibfield  {author} {\bibinfo {author} {\bibfnamefont {E.}~\bibnamefont
  {Gutkin}},\ }\href@noop {} {\bibfield  {journal} {\bibinfo  {journal}
  {Regular and chaotic dynamics}\ }\textbf {\bibinfo {volume} {8}},\ \bibinfo
  {pages} {1} (\bibinfo {year} {2003})}\BibitemShut {NoStop}%
\bibitem [{\citenamefont {Shnirelman}(1974)}]{shnirel1974ergodic}%
  \BibitemOpen
  \bibfield  {author} {\bibinfo {author} {\bibfnamefont {A.~I.}\ \bibnamefont
  {Shnirelman}},\ }\href@noop {} {\bibfield  {journal} {\bibinfo  {journal}
  {Uspekhi Matematicheskikh Nauk}\ }\textbf {\bibinfo {volume} {29}},\ \bibinfo
  {pages} {181} (\bibinfo {year} {1974})}\BibitemShut {NoStop}%
\bibitem [{\citenamefont {Zelditch}\ and\ \citenamefont
  {Zworski}(1996)}]{zelditch1996ergodicity}%
  \BibitemOpen
  \bibfield  {author} {\bibinfo {author} {\bibfnamefont {S.}~\bibnamefont
  {Zelditch}}\ and\ \bibinfo {author} {\bibfnamefont {M.}~\bibnamefont
  {Zworski}},\ }\href@noop {} {\bibfield  {journal} {\bibinfo  {journal}
  {Communications in mathematical physics}\ }\textbf {\bibinfo {volume}
  {175}},\ \bibinfo {pages} {673} (\bibinfo {year} {1996})}\BibitemShut
  {NoStop}%
\bibitem [{\citenamefont {Barnett}(2006)}]{barnett2006asymptotic}%
  \BibitemOpen
  \bibfield  {author} {\bibinfo {author} {\bibfnamefont {A.~H.}\ \bibnamefont
  {Barnett}},\ }\href@noop {} {\bibfield  {journal} {\bibinfo  {journal}
  {Communications on Pure and Applied Mathematics: A Journal Issued by the
  Courant Institute of Mathematical Sciences}\ }\textbf {\bibinfo {volume}
  {59}},\ \bibinfo {pages} {1457} (\bibinfo {year} {2006})}\BibitemShut
  {NoStop}%
\bibitem [{\citenamefont {Berry}(1977)}]{Berry1977wf}%
  \BibitemOpen
  \bibfield  {author} {\bibinfo {author} {\bibfnamefont {M.~V.}\ \bibnamefont
  {Berry}},\ }\href@noop {} {\bibfield  {journal} {\bibinfo  {journal} {Journal
  of Physics A: Mathematical and General}\ }\textbf {\bibinfo {volume} {12}},\
  \bibinfo {pages} {2083} (\bibinfo {year} {1977})}\BibitemShut {NoStop}%
\bibitem [{\citenamefont {Heller}(1984)}]{Heller1984}%
  \BibitemOpen
  \bibfield  {author} {\bibinfo {author} {\bibfnamefont {E.~J.}\ \bibnamefont
  {Heller}},\ }\href {\doibase 10.1103/PhysRevLett.53.1515} {\bibfield
  {journal} {\bibinfo  {journal} {Phys. Rev. Lett.}\ }\textbf {\bibinfo
  {volume} {53}},\ \bibinfo {pages} {1515} (\bibinfo {year}
  {1984})}\BibitemShut {NoStop}%
\bibitem [{\citenamefont {Lozej}(2020)}]{LozejPHD}%
  \BibitemOpen
  \bibfield  {author} {\bibinfo {author} {\bibfnamefont {{\v C}.}~\bibnamefont
  {Lozej}},\ }\emph {\bibinfo {title} {Transport and Localizaion in Classical
  and Quantum Billiards}},\ \href@noop {} {Ph.D. thesis},\ \bibinfo  {school}
  {University of Maribor} (\bibinfo {year} {2020})\BibitemShut {NoStop}%
\bibitem [{\citenamefont {Wigner}(1932)}]{Wig1932}%
  \BibitemOpen
  \bibfield  {author} {\bibinfo {author} {\bibfnamefont {E.}~\bibnamefont
  {Wigner}},\ }\href@noop {} {\bibfield  {journal} {\bibinfo  {journal}
  {Physical Review}\ }\textbf {\bibinfo {volume} {40}},\ \bibinfo {pages} {749}
  (\bibinfo {year} {1932})}\BibitemShut {NoStop}%
\bibitem [{\citenamefont {Hillery}\ \emph {et~al.}(1984)\citenamefont
  {Hillery}, \citenamefont {O'Connell}, \citenamefont {Scully},\ and\
  \citenamefont {Wigner}}]{Hillery1984}%
  \BibitemOpen
  \bibfield  {author} {\bibinfo {author} {\bibfnamefont {M.~O. S.~M.}\
  \bibnamefont {Hillery}}, \bibinfo {author} {\bibfnamefont {R.~F.}\
  \bibnamefont {O'Connell}}, \bibinfo {author} {\bibfnamefont {M.~O.}\
  \bibnamefont {Scully}}, \ and\ \bibinfo {author} {\bibfnamefont {E.~P.}\
  \bibnamefont {Wigner}},\ }\href@noop {} {\bibfield  {journal} {\bibinfo
  {journal} {Physics reports}\ }\textbf {\bibinfo {volume} {106}},\ \bibinfo
  {pages} {121} (\bibinfo {year} {1984})}\BibitemShut {NoStop}%
\bibitem [{\citenamefont {Husimi}(1940)}]{Hus1940}%
  \BibitemOpen
  \bibfield  {author} {\bibinfo {author} {\bibfnamefont {K.}~\bibnamefont
  {Husimi}},\ }\href@noop {} {\bibfield  {journal} {\bibinfo  {journal} {Proc.
  Phys. Math. Soc. Jpn.}\ }\textbf {\bibinfo {volume} {22}},\ \bibinfo {pages}
  {264} (\bibinfo {year} {1940})}\BibitemShut {NoStop}%
\bibitem [{\citenamefont {Tualle}\ and\ \citenamefont
  {Voros}(1995)}]{tualle1995}%
  \BibitemOpen
  \bibfield  {author} {\bibinfo {author} {\bibfnamefont {J.-M.}\ \bibnamefont
  {Tualle}}\ and\ \bibinfo {author} {\bibfnamefont {A.}~\bibnamefont {Voros}},\
  }\href@noop {} {\bibfield  {journal} {\bibinfo  {journal} {Chaos, Solitons \&
  Fractals}\ }\textbf {\bibinfo {volume} {5}},\ \bibinfo {pages} {1085}
  (\bibinfo {year} {1995})}\BibitemShut {NoStop}%
\bibitem [{\citenamefont {B\"acker}\ \emph {et~al.}(2004)\citenamefont
  {B\"acker}, \citenamefont {F\"urstberger},\ and\ \citenamefont
  {Schubert}}]{Baecker2004hus}%
  \BibitemOpen
  \bibfield  {author} {\bibinfo {author} {\bibfnamefont {A.}~\bibnamefont
  {B\"acker}}, \bibinfo {author} {\bibfnamefont {S.}~\bibnamefont
  {F\"urstberger}}, \ and\ \bibinfo {author} {\bibfnamefont {R.}~\bibnamefont
  {Schubert}},\ }\href@noop {} {\bibfield  {journal} {\bibinfo  {journal}
  {Physical Review E}\ }\textbf {\bibinfo {volume} {70}},\ \bibinfo {pages}
  {036204} (\bibinfo {year} {2004})}\BibitemShut {NoStop}%
\bibitem [{\citenamefont {Batisti{\'c}}\ \emph {et~al.}(2020)\citenamefont
  {Batisti{\'c}}, \citenamefont {Lozej},\ and\ \citenamefont
  {Robnik}}]{batistic2020distribution}%
  \BibitemOpen
  \bibfield  {author} {\bibinfo {author} {\bibfnamefont {B.}~\bibnamefont
  {Batisti{\'c}}}, \bibinfo {author} {\bibfnamefont {{\v{C}}.}~\bibnamefont
  {Lozej}}, \ and\ \bibinfo {author} {\bibfnamefont {M.}~\bibnamefont
  {Robnik}},\ }\href@noop {} {\bibfield  {journal} {\bibinfo  {journal}
  {Nonlinear phenomena in complex systems}\ }\textbf {\bibinfo {volume} {23}},\
  \bibinfo {pages} {17} (\bibinfo {year} {2020})}\BibitemShut {NoStop}%
\bibitem [{\citenamefont {Batisti{\'c}}\ \emph {et~al.}(2019)\citenamefont
  {Batisti{\'c}}, \citenamefont {Lozej},\ and\ \citenamefont
  {Robnik}}]{BatLozRob2019}%
  \BibitemOpen
  \bibfield  {author} {\bibinfo {author} {\bibfnamefont {B.}~\bibnamefont
  {Batisti{\'c}}}, \bibinfo {author} {\bibfnamefont {{\v{C}}.}~\bibnamefont
  {Lozej}}, \ and\ \bibinfo {author} {\bibfnamefont {M.}~\bibnamefont
  {Robnik}},\ }\href@noop {} {\bibfield  {journal} {\bibinfo  {journal}
  {Physical Review E}\ }\textbf {\bibinfo {volume} {100}},\ \bibinfo {pages}
  {062208} (\bibinfo {year} {2019})}\BibitemShut {NoStop}%
\bibitem [{\citenamefont {Lozej}\ \emph
  {et~al.}(2021{\natexlab{a}})\citenamefont {Lozej}, \citenamefont {Lukman},\
  and\ \citenamefont {Robnik}}]{lozej2021effects}%
  \BibitemOpen
  \bibfield  {author} {\bibinfo {author} {\bibfnamefont {{\v{C}}.}~\bibnamefont
  {Lozej}}, \bibinfo {author} {\bibfnamefont {D.}~\bibnamefont {Lukman}}, \
  and\ \bibinfo {author} {\bibfnamefont {M.}~\bibnamefont {Robnik}},\
  }\href@noop {} {\bibfield  {journal} {\bibinfo  {journal} {Physical Review
  E}\ }\textbf {\bibinfo {volume} {103}},\ \bibinfo {pages} {012204} (\bibinfo
  {year} {2021}{\natexlab{a}})}\BibitemShut {NoStop}%
\bibitem [{\citenamefont {Izrailev}(1990)}]{Izr1990}%
  \BibitemOpen
  \bibfield  {author} {\bibinfo {author} {\bibfnamefont {F.~M.}\ \bibnamefont
  {Izrailev}},\ }\href@noop {} {\bibfield  {journal} {\bibinfo  {journal}
  {Phys. Rep.}\ }\textbf {\bibinfo {volume} {196}},\ \bibinfo {pages} {299}
  (\bibinfo {year} {1990})}\BibitemShut {NoStop}%
\bibitem [{\citenamefont {Lozej}\ \emph
  {et~al.}(2021{\natexlab{b}})\citenamefont {Lozej}, \citenamefont {Lukman},\
  and\ \citenamefont {Robnik}}]{lozej2021nostick}%
  \BibitemOpen
  \bibfield  {author} {\bibinfo {author} {\bibfnamefont {{\v{C}}.}~\bibnamefont
  {Lozej}}, \bibinfo {author} {\bibfnamefont {D.}~\bibnamefont {Lukman}}, \
  and\ \bibinfo {author} {\bibfnamefont {M.}~\bibnamefont {Robnik}},\
  }\href@noop {} {\bibfield  {journal} {\bibinfo  {journal} {Nonlinear
  phenomena in complex systems}\ }\textbf {\bibinfo {volume} {24}},\ \bibinfo
  {pages} {1} (\bibinfo {year} {2021}{\natexlab{b}})}\BibitemShut {NoStop}%
\bibitem [{\citenamefont {Wang}\ and\ \citenamefont
  {Robnik}(2020)}]{wang2020dicke}%
  \BibitemOpen
  \bibfield  {author} {\bibinfo {author} {\bibfnamefont {Q.}~\bibnamefont
  {Wang}}\ and\ \bibinfo {author} {\bibfnamefont {M.}~\bibnamefont {Robnik}},\
  }\href@noop {} {\bibfield  {journal} {\bibinfo  {journal} {Physical Review
  E}\ }\textbf {\bibinfo {volume} {102}},\ \bibinfo {pages} {032212} (\bibinfo
  {year} {2020})}\BibitemShut {NoStop}%
\bibitem [{\citenamefont {Gnutzmann}\ and\ \citenamefont
  {Zyczkowski}(2001)}]{gnutzmann2001renyi}%
  \BibitemOpen
  \bibfield  {author} {\bibinfo {author} {\bibfnamefont {S.}~\bibnamefont
  {Gnutzmann}}\ and\ \bibinfo {author} {\bibfnamefont {K.}~\bibnamefont
  {Zyczkowski}},\ }\href@noop {} {\bibfield  {journal} {\bibinfo  {journal}
  {Journal of Physics A: Mathematical and General}\ }\textbf {\bibinfo {volume}
  {34}},\ \bibinfo {pages} {10123} (\bibinfo {year} {2001})}\BibitemShut
  {NoStop}%
\bibitem [{\citenamefont {Villase{\~n}or}\ \emph {et~al.}(2021)\citenamefont
  {Villase{\~n}or}, \citenamefont {Pilatowsky-Cameo}, \citenamefont
  {Bastarrachea-Magnani}, \citenamefont {Lerma-Hern{\'a}ndez},\ and\
  \citenamefont {Hirsch}}]{villasenor2021quantum}%
  \BibitemOpen
  \bibfield  {author} {\bibinfo {author} {\bibfnamefont {D.}~\bibnamefont
  {Villase{\~n}or}}, \bibinfo {author} {\bibfnamefont {S.}~\bibnamefont
  {Pilatowsky-Cameo}}, \bibinfo {author} {\bibfnamefont {M.}~\bibnamefont
  {Bastarrachea-Magnani}}, \bibinfo {author} {\bibfnamefont {S.}~\bibnamefont
  {Lerma-Hern{\'a}ndez}}, \ and\ \bibinfo {author} {\bibfnamefont
  {J.}~\bibnamefont {Hirsch}},\ }\href@noop {} {\bibfield  {journal} {\bibinfo
  {journal} {Physical Review E}\ }\textbf {\bibinfo {volume} {103}},\ \bibinfo
  {pages} {052214} (\bibinfo {year} {2021})}\BibitemShut {NoStop}%
\bibitem [{\citenamefont {Pilatowsky-Cameo}\ \emph
  {et~al.}(2021{\natexlab{a}})\citenamefont {Pilatowsky-Cameo}, \citenamefont
  {Villase{\~n}or}, \citenamefont {Bastarrachea-Magnani}, \citenamefont
  {Lerma-Hern{\'a}ndez}, \citenamefont {Santos},\ and\ \citenamefont
  {Hirsch}}]{pilatowsky2021ubiquitous}%
  \BibitemOpen
  \bibfield  {author} {\bibinfo {author} {\bibfnamefont {S.}~\bibnamefont
  {Pilatowsky-Cameo}}, \bibinfo {author} {\bibfnamefont {D.}~\bibnamefont
  {Villase{\~n}or}}, \bibinfo {author} {\bibfnamefont {M.~A.}\ \bibnamefont
  {Bastarrachea-Magnani}}, \bibinfo {author} {\bibfnamefont {S.}~\bibnamefont
  {Lerma-Hern{\'a}ndez}}, \bibinfo {author} {\bibfnamefont {L.~F.}\
  \bibnamefont {Santos}}, \ and\ \bibinfo {author} {\bibfnamefont {J.~G.}\
  \bibnamefont {Hirsch}},\ }\href@noop {} {\bibfield  {journal} {\bibinfo
  {journal} {Nature communications}\ }\textbf {\bibinfo {volume} {12}},\
  \bibinfo {pages} {1} (\bibinfo {year} {2021}{\natexlab{a}})}\BibitemShut
  {NoStop}%
\bibitem [{\citenamefont {Pilatowsky-Cameo}\ \emph
  {et~al.}(2021{\natexlab{b}})\citenamefont {Pilatowsky-Cameo}, \citenamefont
  {Villase{\~n}or}, \citenamefont {Bastarrachea-Magnani}, \citenamefont
  {Lerma-Hern{\'a}ndez}, \citenamefont {Santos},\ and\ \citenamefont
  {Hirsch}}]{pilatowsky2021identification}%
  \BibitemOpen
  \bibfield  {author} {\bibinfo {author} {\bibfnamefont {S.}~\bibnamefont
  {Pilatowsky-Cameo}}, \bibinfo {author} {\bibfnamefont {D.}~\bibnamefont
  {Villase{\~n}or}}, \bibinfo {author} {\bibfnamefont {M.~A.}\ \bibnamefont
  {Bastarrachea-Magnani}}, \bibinfo {author} {\bibfnamefont {S.}~\bibnamefont
  {Lerma-Hern{\'a}ndez}}, \bibinfo {author} {\bibfnamefont {L.~F.}\
  \bibnamefont {Santos}}, \ and\ \bibinfo {author} {\bibfnamefont {J.~G.}\
  \bibnamefont {Hirsch}},\ }\href@noop {} {\bibfield  {journal} {\bibinfo
  {journal} {arXiv preprint arXiv:2107.06894}\ } (\bibinfo {year}
  {2021}{\natexlab{b}})}\BibitemShut {NoStop}%
\bibitem [{\citenamefont {Urbina}\ and\ \citenamefont
  {Richter}(2013)}]{urbina2013random}%
  \BibitemOpen
  \bibfield  {author} {\bibinfo {author} {\bibfnamefont {J.-D.}\ \bibnamefont
  {Urbina}}\ and\ \bibinfo {author} {\bibfnamefont {K.}~\bibnamefont
  {Richter}},\ }\href@noop {} {\bibfield  {journal} {\bibinfo  {journal}
  {Advances in Physics}\ }\textbf {\bibinfo {volume} {62}},\ \bibinfo {pages}
  {363} (\bibinfo {year} {2013})}\BibitemShut {NoStop}%
\bibitem [{\citenamefont {Srednicki}(1994)}]{srednicki1994eth}%
  \BibitemOpen
  \bibfield  {author} {\bibinfo {author} {\bibfnamefont {M.}~\bibnamefont
  {Srednicki}},\ }\href@noop {} {\bibfield  {journal} {\bibinfo  {journal}
  {Physical Review E}\ }\textbf {\bibinfo {volume} {50}},\ \bibinfo {pages}
  {888} (\bibinfo {year} {1994})}\BibitemShut {NoStop}%
\bibitem [{\citenamefont {Prosen}(1996)}]{prosen1996}%
  \BibitemOpen
  \bibfield  {author} {\bibinfo {author} {\bibfnamefont {T.}~\bibnamefont
  {Prosen}},\ }\href@noop {} {\bibfield  {journal} {\bibinfo  {journal}
  {Physica D: Nonlinear Phenomena}\ }\textbf {\bibinfo {volume} {91}},\
  \bibinfo {pages} {244} (\bibinfo {year} {1996})}\BibitemShut {NoStop}%
\bibitem [{\citenamefont {Bogomolny}\ \emph {et~al.}(2006)\citenamefont
  {Bogomolny}, \citenamefont {Dietz}, \citenamefont {Friedrich}, \citenamefont
  {Miski-Oglu}, \citenamefont {Richter}, \citenamefont {Sch{\"a}fer},\ and\
  \citenamefont {Schmit}}]{bogomolny2006barrier}%
  \BibitemOpen
  \bibfield  {author} {\bibinfo {author} {\bibfnamefont {E.}~\bibnamefont
  {Bogomolny}}, \bibinfo {author} {\bibfnamefont {B.}~\bibnamefont {Dietz}},
  \bibinfo {author} {\bibfnamefont {T.}~\bibnamefont {Friedrich}}, \bibinfo
  {author} {\bibfnamefont {M.}~\bibnamefont {Miski-Oglu}}, \bibinfo {author}
  {\bibfnamefont {A.}~\bibnamefont {Richter}}, \bibinfo {author} {\bibfnamefont
  {F.}~\bibnamefont {Sch{\"a}fer}}, \ and\ \bibinfo {author} {\bibfnamefont
  {C.}~\bibnamefont {Schmit}},\ }\href@noop {} {\bibfield  {journal} {\bibinfo
  {journal} {Physical Review Letters}\ }\textbf {\bibinfo {volume} {97}},\
  \bibinfo {pages} {254102} (\bibinfo {year} {2006})}\BibitemShut {NoStop}%
\bibitem [{\citenamefont {{\v{S}}untajs}\ \emph {et~al.}(2020)\citenamefont
  {{\v{S}}untajs}, \citenamefont {Bon{\v{c}}a}, \citenamefont {Prosen},\ and\
  \citenamefont {Vidmar}}]{suntajs2020}%
  \BibitemOpen
  \bibfield  {author} {\bibinfo {author} {\bibfnamefont {J.}~\bibnamefont
  {{\v{S}}untajs}}, \bibinfo {author} {\bibfnamefont {J.}~\bibnamefont
  {Bon{\v{c}}a}}, \bibinfo {author} {\bibfnamefont {T.}~\bibnamefont {Prosen}},
  \ and\ \bibinfo {author} {\bibfnamefont {L.}~\bibnamefont {Vidmar}},\
  }\href@noop {} {\bibfield  {journal} {\bibinfo  {journal} {Physical Review
  E}\ }\textbf {\bibinfo {volume} {102}},\ \bibinfo {pages} {062144} (\bibinfo
  {year} {2020})}\BibitemShut {NoStop}%
\end{thebibliography}%

\end{document}